\documentclass[11pt,a4paper]{article}
\pdfoutput=1
\usepackage{a4wide}
\usepackage{amsfonts}
\usepackage{amssymb}
\usepackage{amsmath}
\usepackage{graphicx}
\usepackage{booktabs}
\usepackage{array}
\usepackage{color}
\usepackage{ulem}
\usepackage[small,bf]{caption}
\usepackage{cite}
\usepackage[usenames,dvipsnames]{xcolor}
\usepackage{subfigure}
\usepackage{verbatim}
\usepackage{dsfont}
\usepackage{multirow}
\usepackage{url}

\def\1{\mathbf{1}}
\def\3{\mathbf{3}}
\def\2{\mathbf{2}}

\allowdisplaybreaks[1]

\numberwithin{equation}{section}

\newcounter{mysubequation}[equation]

\definecolor{pink}{rgb}{1.,.2,.8}

\renewcommand{\emph}[1]{{\it #1}}

\begin{document}

\begin{titlepage}

\vspace*{-15mm}
\begin{flushright}
TTP16-022\\
MPP-2016-127
\end{flushright}
\vspace*{0.7cm}

\begin{center}
{ \bf\LARGE Predictivity of \\\vspace{0.12cm} Neutrino Mass Sum Rules}
\\[8mm]
Julia Gehrlein$^{\, a,}$\footnote{E-mail: \texttt{julia.gehrlein@student.kit.edu}},
Alexander Merle$^{\, b,}$\footnote{E-mail: \texttt{amerle@mpp.mpg.de}},
Martin Spinrath$^{\, a,}$\footnote{E-mail: \texttt{martin.spinrath@kit.edu}}
\\[1mm]
\end{center}
\vspace*{0.50cm}
\centerline{$^a$ \it Institut f\"ur Theoretische Teilchenphysik, Karlsruhe Institute of Technology,}
\centerline{\it Engesserstra\ss{}e 7, D-76131 Karlsruhe, Germany}
\centerline{$^b$ \it Max-Planck-Institut f\"ur Physik (Werner-Heisenberg-Institut),}
\centerline{\it F\"ohringer Ring 6, D-80805 M\"unchen, Germany}
\vspace*{0.2cm}

\vspace*{1.20cm}

\begin{abstract}
\noindent
Correlations between light neutrino observables are arguably the strongest predictions of lepton flavour models based on (discrete) symmetries, except for the very few cases which unambiguously predict the full set of leptonic mixing angles. A subclass of these correlations are neutrino mass sum rules, which connect the three (complex) light neutrino mass eigenvalues among each other. This connection constrains both the light neutrino mass scale and the Majorana phases, so that mass sum rules generically lead to a nonzero value of the lightest neutrino mass and to distinct predictions for the effective mass probed in neutrinoless double beta decay. However, in nearly all cases known, the neutrino mass sum rules are not exact and receive corrections from various sources. We introduce a formalism to handle these corrections perturbatively in a model-independent manner, which overcomes issues present in earlier approaches. Our ansatz allows us to quantify the modification of the predictions derived from neutrino mass sum rules. We show that, in most cases, the predictions are fairly stable: while small quantitative changes can appear, they are generally rather mild. We therefore establish the predictivity of neutrino mass sum rules on a level far more general than previously known.
\end{abstract}

\end{titlepage}
\setcounter{footnote}{0}

\section{Introduction}

Neutrinos keep on surprising us when it comes to experiments. We are still puzzled by their curious properties which are experimentally established but which we still do not understand on a fundamental level. For example, the neutrino mass is extremely small compared to any other fermionic mass we know, namely below $1$~eV (or possibly even $0.1$~eV). This we know from several experiments and observations, either in the lab by kinematical determinations of the neutrino mass~\cite{Neutrino-Kin} and from the hunt for neutrinoless double beta decay~\cite{0nbb-Exp}, or in space by the time-of-flight measurements of supernova neutrinos~\cite{Pagliaroli:2010ik} or from cosmological considerations~\cite{Cosmo-Nu}. Similarly, the leptonic mixing angles are a mystery to us, for we have measured their values to be fairly large (see \url{nu-fit.org}~\cite{Gonzalez-Garcia:2014bfa}), but yet we have absolutely no theoretical understanding of these numbers.

Although far from perfect, still the best idea we have to explain leptonic mixing is to motivate the values of the mixing angles by relating them to discrete flavour symmetries, see~\cite{flavour-reviews} for recent reviews.\footnote{For alternative ideas, one could for example consider anarchy~\cite{Haba:2000be} or radiative transmission~\cite{Adulpravitchai:2009re}.} However, flavour models based on discrete symmetries generically face one big problem: they often do not give us any testable prediction beyond fitting the known mixing angles within their experimental ranges (note that, if a model did not, it would in any case be discarded). On the other hand, at least some groups of flavour models are more powerful than that, in the sense that they predict certain correlations between different observable quantities. Among these correlations, the most popular ones discussed in the literature are mixing angle sum rules~\cite{Mixing-SR}; but, when looking at the total mass matrix of neutrinos, a second class of correlations arises, \emph{neutrino mass sum rules}.

It is these mass sum rules (in the following referred to by ``SRs'') that we will investigate in this text. Basically, what they do is to connect the three \emph{complex} neutrino mass eigenvalues $\tilde m_i$ in a simple relation, where all three contributions sum to zero. For example, $\tilde m_1 + \tilde m_2 - \tilde m_3 = 0$ would be a valid SR, as well as $ \tilde m_1^{-1} + \tilde m_2^{-1} - \tilde m_3^{-1} = 0$. Several studies of SRs have been presented earlier. Among the first works investigating SRs were Refs.~\cite{Altarelli:2008bg,Hirsch:2008rp,Bazzocchi:2009da}. However, these actually did not mention the name ``sum rules'' at all. That term came up only later, in Refs.~\cite{Altarelli:2009kr,Chen:2009um,Barry:2010yk,SR11}. The probably most comprehensive study of SRs, which includes all known cases we are aware of, had been presented in Ref.~\cite{King:2013psa}. Based on this study, Ref.~\cite{SR-GERDA} has shown that, indeed, with realistic assumptions on the experimental side one could truly distinguish at least some classes of models with near-future data -- as long as our current understanding of the nuclear physics aspects of neutrinoless double beta decay is not totally flawed.

However, it is clear in nearly all cases that sum rules, even though predicted, are in fact \emph{not} exact.\footnote{See Ref.~\cite{Cooper:2012bd} for a notable exception, where the SR at least holds to next-to-leading order.} These can arise, e.g., from higher-order correction terms arising from flavour symmetry breaking, or from corrections transmitted from the charged lepton sector. A further correction that we had investigated earlier are those arising from renormalisation group running, see Ref.~\cite{Gehrlein:2015ena}. We have indeed been able to show that at least these corrections do not change the predictions from SRs significantly, but we cannot claim that this particular type of corrections would be the most general one.

In this manuscript, we will try to close this gap by computing the effect of nonexact SRs on the predictions in a very general framework, based on a perturbative approach. The only attempt to investigate approximate sum rules that we are aware of had been presented in Ref.~\cite{Barry:2010yk}, however, it had been pointed out~\cite{King:2013psa,Gehrlein:2015ena} that the approach in that reference is likely to be insufficient because the corrections have been attributed to one particular mass eigenvalue only, namely $m_3$. However, this may create problems both due to measure dependence of the perturbations and due to the fact that $m_3$ plays a very different role for the two mass orderings. This is another shortcoming which we will at least partially overcome with our approach, by describing a procedure by which the dependence on a choice of mass eigenvalue can be minimised. Using the techniques developed, we will investigate the predictions of all sum rules we found in the literature. As we will see, while the predictions are of course changed for approximate SRs, in most cases these changes are comparatively mild, thereby keeping the predictivity of the SRs alive. Only in one case, namely SR~10, a qualitative change does happen which could in fact be visible in an experiment.

This work is organised as follows. We  start in Sec.~\ref{sec:par} by reintroducing the parametrisation of SRs used by us, and we will also visualise how to interpret neutrino mass sum rules, no matter if exact or not, in a geometrical manner. Sec.~\ref{sec:corr} is dedicated to a discussion of the possible origins of the various corrections, to clarify which cases are covered by our formalism. Our numerical results, along with detailed plots for each sum rule, are presented in Sec.~\ref{sec:num}. We conclude in Sec.~\ref{sec:conc}. Technical details on how to derive the physical leptonic mixing parameters from the charged lepton and neutrino sectors are summarised in Appendix~\ref{sec:parametrisations}.

\section{\label{sec:par}Parametrisation and geometrical interpretation}

\begin{table}
\centering
\begin{tabular}{lcc} 
\toprule
Parameter & Best fit ($\pm 1\sigma$) & $ 3\sigma$ range\\ 
\midrule 
$\theta_{12}$ in $^{\circ}$ & $ 33.48^{+0.78}_{-0.75}$& $31.29\rightarrow 35.91$\\[0.5 pc]
$\theta_{13}$ in $^{\circ}$ & $ 8.50^{+0.20}_{-0.21}\oplus 8.51^{+0.20}_{-0.21} $& $7.85\rightarrow 9.10 \oplus 7.87\rightarrow 9.11$\\[0.5 pc]
$\theta_{23}$ in $^{\circ}$ & $ 42.3^{+3.0}_{-1.6}\oplus 49.5^{+1.5}_{-2.2}$ & $38.2\rightarrow 53.3 \oplus 38.6\rightarrow 53.3$\\[0.5 pc]
$\delta$  in $^{\circ}$&$251^{+67}_{-59}$&$0\rightarrow 360$\\
\midrule
$\Delta m_{21}^{2}$ in $10^{-5}$~eV$^2$ & $7.50^{+0.19}_{-0.17}$ & $7.02\rightarrow 8.09$\\[0,5 pc]
$\Delta m_{31}^{2}$ in $10^{-3}$~eV$^2$~(NO) &$2.457^{+0.047}_{-0.047}$&$2.317\rightarrow 2.607$\\[0,5 pc]
$\Delta m_{32}^{2}$ in $10^{-3}$~eV$^2$~(IO) &$-2.449^{+0.048}_{-0.047}$&$-2.590\rightarrow -2.307$\\
\bottomrule
\end{tabular}
\caption{\label{tab:exp_parameters}The best-fit values and the 3$\sigma$ ranges for the parameters taken from~\cite{Gonzalez-Garcia:2014bfa}, {\tt v2.0}. The two minima for both $\theta_{13}$ and $\theta_{23}$ correspond to normal and inverted mass ordering, respectively.}
\end{table}

To start off, let us define our conventions. First of all, we parametrise the leptonic mixing matrix, the so-called Pontecorvo-Maki-Nakagawa-Sakata (PMNS) matrix, by using  the standard parametrisation suggested by the \emph{Particle Data Group} (PDG)~\cite{pdg}
\begin{align}
U_\text{PMNS}&=R_{23}U_{13}R_{12}P_{0}\nonumber\\&=
\begin{pmatrix}
c_{12}c_{13}&s_{12}c_{13}&s_{13}\mathrm{e}^{-\mathrm{i}\delta}\\
-s_{12}c_{23}-c_{12}s_{23}s_{13}\mathrm{e}^{\mathrm{i}\delta}&c_{12}c_{23}-s_{12}s_{23}s_{13}\mathrm{e}^{\mathrm{i}\delta}&s_{23}c_{13}\\
s_{12}s_{23}-c_{12}c_{23}s_{13}\mathrm{e}^{\mathrm{i}\delta}&-c_{12}s_{23}-s_{12}c_{23}s_{13}\mathrm{e}^{\mathrm{i}\delta}&c_{23}c_{13}\\
\end{pmatrix}
P_{0},
\label{eq:U2}
\end{align}
where $\delta$ is the Dirac CP-phase and $P_{0}$=diag$(\mathrm{e}^{-\mathrm{i}\phi_1/2},\mathrm{e}^{-\mathrm{i}\phi_2/2},1)$ is a diagonal matrix containing the two Majorana phases $\phi_{1,2}$. However, note that our definition of the Majorana phases is different compared to what the PDG uses~\cite{pdg} in their Eq.~(14.78): $\phi_1 = -\alpha_{31}$ and $\phi_2 = \alpha_{21}-\alpha_{31}$. As for the mixing parameters, we have used the {\tt v2.0} version from \url{nu-fit.org}, as reported in Tab.~\ref{tab:exp_parameters}.\footnote{Note that, just while this work was in its final stages, a new version of mixing parameters {\tt v2.1} was released. We have verified that our results are not significantly changed for some example cases, however, given the time-consuming numerics behind this manuscript we have decided against rerunning all of our code and have instead decided to consistently present the results obtained for {\tt v2.0}.}

With that said, let us next introduce our formalism to treat neutrino mass sum rules (SRs). In~\cite{Gehrlein:2015ena} we have investigated the effect of renormalisation group corrections to neutrino mass SRs. In this paper we want to examine the impact of next-to-leading-order (NLO) corrections to neutrino mass matrices on mass SRs.

A general exact SR can be parametrised according to~\cite{Gehrlein:2015ena}:
\begin{align}
 s(m_1,m_2,&m_3,\phi_1,\phi_2;c_1,c_2,d,\Delta \chi_{13},\Delta \chi_{23}) \equiv \nonumber\\
 &c_1 \left(m_1 \text{e}^{-\text{i}\phi_{1}}\right)^d 
\text{e}^{\text{i}\Delta \chi_{13}}+
c_2 \left(m_2 \text{e}^{-\text{i}\phi_{2}}\right)^d 
\text{e}^{\text{i}\Delta \chi_{23}}
+m_3^d~ \stackrel{!}{=} 0 \;,
\label{eq:parametrisation_SR}
\end{align}
where $\phi_i$, $i=1,~2$ are the Majorana phases. The quantities $c_1,~c_2,~ d,~\Delta \chi_{13}$, and $\Delta \chi_{23}$ are parameters which characterise the SR, e.g., SR~1, $\tilde m_1 + \tilde m_2 = \tilde m_3$, is characterised by $(c_1, c_2, d, \Delta \chi_{13}, \Delta \chi_{23}) = (1, 1, 1, \pi, \pi)$, while SR~7, $\tilde m_1^{-1} = 2 \tilde m_2^{-1} + \tilde m_3^{-1}$, is characterised by $(c_1, c_2, d, \Delta \chi_{13}, \Delta \chi_{23}) = (1, 2, -1, \pi, 0)$, see Tab.~\ref{tab:overview_SR} for a summary. Note that, in this notation, $\tilde m_i$ are the \emph{complex} mass eigenvalues, i.e., with the phase information included. In Tab.~\ref{tab:overview_SR} we have collected all the SRs we found in the literature with their parameters $c_1,~c_2,~ d,~\Delta \chi_{13}$, and $\Delta \chi_{23}$.

\begin{table}
\centering
\begin{tabular}{c c c c c c c}
\toprule
Sum rule & References & $c_1$&$c_2$&$d$&$\Delta \chi_{13}$&$\Delta \chi_{23}$ \\
\midrule
	1&~\cite{Barry:2010yk,Bazzocchi:2009da,Ding:2010pc,Ma:2005sha,Ma:2006wm,Honda:2008rs,Brahmachari:2008fn,Kang:2015xfa,SR1} &$1$&$1$&$1$&$\pi$&$\pi$\\
	2&~\cite{SR2} &$1$&$2$&$1$&$\pi$&$\pi$\\
	3&~\cite{Barry:2010yk,Ma:2005sha,Ma:2006wm,Honda:2008rs,Brahmachari:2008fn,Altarelli:2005yx,Chen:2009um,Chen:2009gy,Kang:2015xfa,SR3} &$1$&$2$&$1$&$\pi$&$0$\\
	4&~\cite{SR4} &$1/2$&$1/2$&$1$&$\pi$&$\pi$\\
	5&~\cite{SR5} &$\tfrac{2}{\sqrt{3}+1}$&$\tfrac{\sqrt{3}-1}{\sqrt{3}+1}$&$1$&$0$&$\pi$\\
	6&~\cite{Barry:2010yk,Bazzocchi:2009da,Ding:2010pc,Cooper:2012bd,SR6,Gehrlein:2014wda} &$1$&$1$&$-1$&$\pi$&$\pi$\\
	7&~\cite{Barry:2010yk,Altarelli:2005yx,Chen:2009um,Chen:2009gy,Altarelli:2009kr,SR7,Altarelli:2008bg} &$1$&$2$&$-1$&$\pi$&$0$\\
	8&~\cite{SR8} &$1$&$2$&$-1$&$0$&$\pi$\\
	9&~\cite{SR9} &$1$&$2$&$-1$&$\pi$&$\pi/2,3\pi/2$\\
	10&~\cite{SR10,Hirsch:2008rp} &$1$&$2$&$1/2$&$\pi,0,\pi/2$ &$0,\pi,\pi/2$ \\
	11&~\cite{SR11} &$1/3$&$1$&$1/2$&$\pi$&$0$\\
	12&~\cite{SR12} &$1/2$&$1/2$&$-1/2$&$\pi$&$\pi$\\
	\bottomrule
\end{tabular}
\caption{\label{tab:overview_SR}Summary table of the SRs we will analyse in the following. The parameters $c_1, c_2, d, \Delta \chi_{13}$, and $\Delta \chi_{23}$ that characterise them are defined in Eq.~\eqref{eq:parametrisation_SR}. In SRs~9 and~10, two possible signs appear which lead to two possible values of $\Delta \chi_{i3}$.
}
\end{table}

A complex perturbation governed by the complex parameter $\delta m_i~\text{e}^{\text{i}\delta \phi_i}$ to the neutrino mass matrix shifts its complex eigenvalues to
\begin{align} 
\tilde{m}_i=m_i \text{e}^{-\text{i}\phi_i}=m_i^{(0)} \text{e}^{-\text{i}\phi_i^ {(0)}}+ \delta m_i~\text{e}^{\text{i}\delta \phi_i}~.
\label{eq:corrections_m}
\end{align}
Thus, explicitly, the corrections are connected to the physical parameters as follows:
\begin{align}
m_i&\approx m_i^{(0)} \left( 1 + \frac{\delta m_i}{m_i^{(0)}} \cos(\delta \phi_i-\phi_i^{(0)}) \right) ~,\\
\phi_i&\approx -\arctan\left(\frac{-m_i^{(0)}\sin (\phi_i^{(0)})+\delta m_i \sin (\delta \phi_i)}{m_i^{(0)}\cos (\phi_i^{(0)})+\delta m_i \cos (\delta \phi_i)}\right) ~.
\end{align}
Thus, e.g.\ $\delta \phi_i$ is \emph{not} the correction to the phase of the complex mass. We assume that, in general, the correction to each mass can have a different phase $\delta \phi_i$ than that of the zeroth-order SR, i.e., $\delta \phi_i$ need \emph{not} be a small number. We furthermore assume that $\delta m_i/m_i^ {(0)}\ll 1$, with $\delta m_i > 0$ without loss of generality. So we can expand Eq.~\eqref{eq:parametrisation_SR} in the small parameters $\delta m_i$ to obtain the deviation from the exact SR. This results in
\begin{align}
0 \neq s \approx s^{(0)}+\delta s \;,
\label{eq:s}
\end{align}
where 
\begin{align}
s^{(0)}= c_1 \left(m_1^{(0)} \text{e}^{-\text{i}\phi_{1}^{(0)}}\right)^d 
\text{e}^{\text{i}\Delta \chi_{13}}+
c_2 \left(m_2^{(0)} \text{e}^{-\text{i}\phi_{2}^{(0)}}\right)^d 
\text{e}^{\text{i}\Delta \chi_{23}}
+\left(m_3^{(0)}\right)^d~
\label{eq:s0}
\end{align}
and
\begin{align}
\delta s &= d \Big[ c_1 \left(m_1^{(0)} \text{e}^{-\text{i}\phi_1^{(0)}}\right)^d \text{e}^{\text{i}(\Delta \chi_{13}+\phi_1^ {(0)})} \frac{\delta m_1 \text{e}^{\text{i}\delta \phi_1} }{m_1^ {(0)}} \nonumber\\
&+c_2 \left(m_2^{(0)} \text{e}^{-\text{i}\phi_2^{(0)}}\right)^d \text{e}^{\text{i}(\Delta \chi_{23}+\phi_2^{(0)})} \frac{\delta m_2 \text{e}^{\text{i}\delta \phi_2} }{m_2^ {(0)}} + \left(m_3^{(0)}\right)^{d}  \frac{\delta m_3 \text{e}^{\text{i}\delta \phi_3} }{m_3^ {(0)}} \Big]~.
\label{eq:ds}
\end{align}
With the leading-order (LO) expressions for the masses the SR is exactly fulfilled (i.e., $s^{(0)} = 0$), in case the SR does allow for the mass ordering under consideration. Note that the SR is complex and hence the correction $\delta s$ will in general be complex as well.

\begin{figure}
\centering
\includegraphics[scale=0.8]{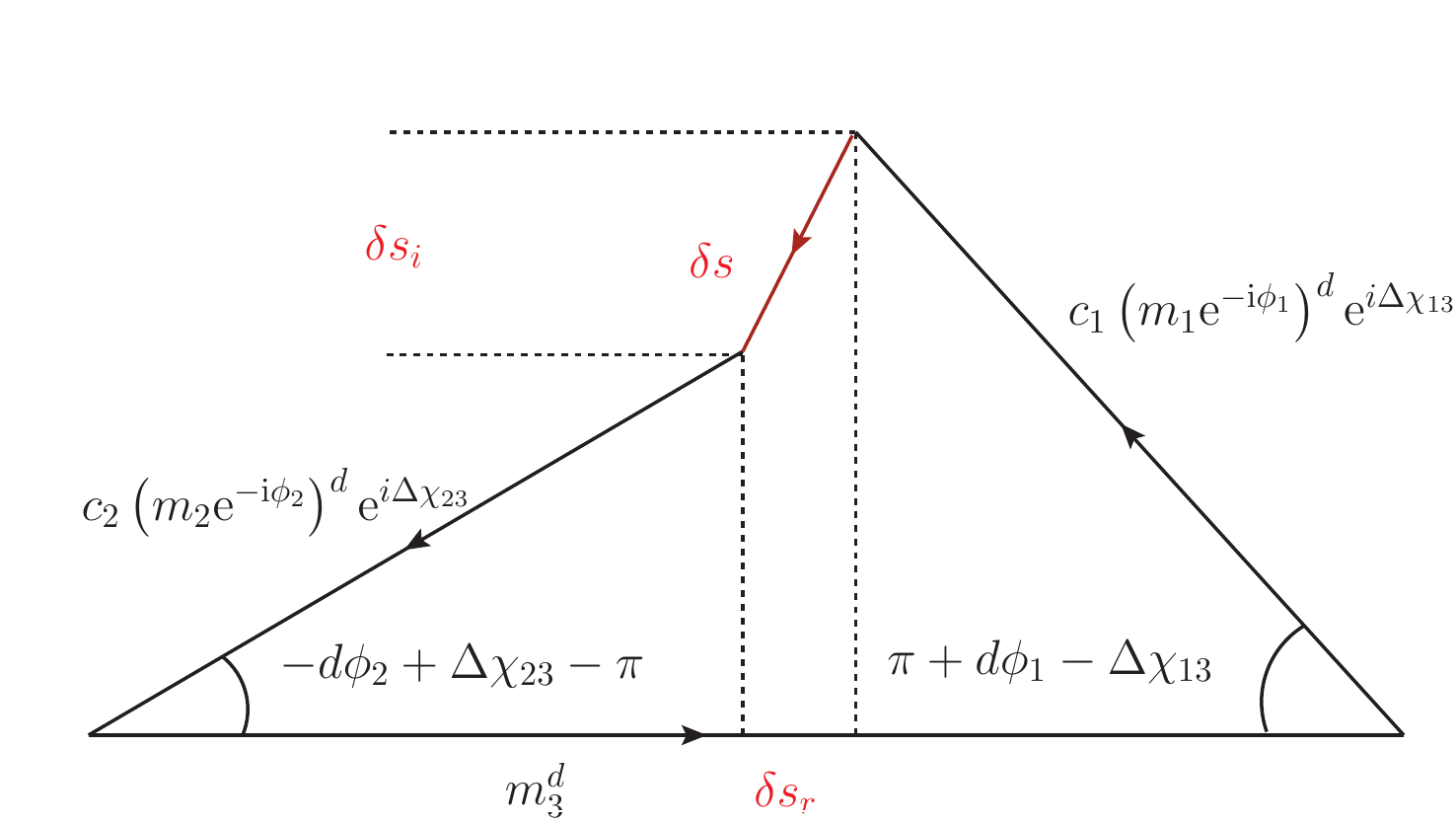}
\caption{\label{fig:triangle}Definition of the parameters $\delta s_r$ and $\delta s_i$ used to measure the corrections to the SRs. Note that these parameters are dimensionful. We normalise $\delta s$ according to Eq.~\eqref{eq:mn} such that we obtain dimensionless quantities. The parameters $m_i$ and $\phi_i$ are already corrected. }
\end{figure}

Graphically the deviation from a SR corresponds to an ``incomplete'' triangle in the complex plane, as illustrated in Fig.~\ref{fig:triangle}. Note that the parameters which appear in the triangle are now the \emph{corrected} masses and phases, which are complicated functions of the leading-order parameters and of the corrections. We will use the parameters $\delta s_r$ and $\delta s_i$ to measure the effect of the perturbation, where $\delta s_r$ ($\delta s_i$) corresponds to the real (imaginary) part of the deviation.

 At this point we also want to note that $\delta s$ is a dimensionful quantity, and hence not well suited to express corrections since the notion of small and big is not meaningful. We therefore introduce the normalised hatted quantities
\begin{equation}
 \hat{s} \equiv \frac{ s}{m_n^d} \text{ and } \delta \hat{s} \equiv \frac{\delta s}{m_n^d} \;, \label{eq:dsh}~
\end{equation}
where $m_n$ is chosen in such a way that the coefficients in front of $\delta m_i/m_i^{(0)}$ in Eq.~\eqref{eq:ds} are not artificially enhanced by $m_i/m_j \gg 1$. Explicitly, that is:
\begin{equation}
 m_n = \begin{cases}
        m_3 & \text{for } d>0 \text{ and NO,} \\
        m_2 & \text{for } d>0 \text{ and IO,} \\
        m_1 & \text{for } d<0 \text{ and NO,} \\
        m_3 & \text{for } d<0 \text{ and IO,} \\
       \end{cases} \label{eq:mn}
\end{equation}
where NO (IO) stands for normal (inverted) ordering.
With this choice, $|\delta \hat{s}|$ should be much smaller than one -- if we want to talk about small corrections and one of the sides of the triangle in Fig.~\ref{fig:triangle} has a length of $\mathcal{O}(1)$.

Before going on, it is important to realise that Eq.~\eqref{eq:mn} implies that the quantity $\delta \hat{s} < b$ can in fact have a slightly different meaning even for one and the same bound $b$, depending on the mass ordering and on the sign of $d$. This is, however, not so much an inconsistency than simply a convenient approach to use the same formalism for all cases considered. The decisive point is that, in any case, the two mass orderings are physically different, and so are cases with a different sign of $d$, so that it is simply impossible to put all SRs on the same footing for all cases. Thus, a bound such as, e.g., $\delta \hat{s} < 0.1$ may be more or less restrictive, depending on the actual case under consideration. However, as we will see, the difference induced by this subtlety is not really decisive and will thus not be a major concern for the remainder of this manuscript.

\section{\label{sec:corr}Main origin of corrections}

In this section, we will give examples for a possible origin of deviations from exact SRs. Note that, in principle, no matter where the corrections arise from, they will always be covered by our formalism, cf.\ section~\ref{sec:par}. However, the important restriction is that our Eq.~\eqref{eq:ds} relies on the assumption that one can expand the full SR $s$ in the small quantities $\delta m_i/m_i^ {(0)}$. If this is not possible for some reason, our formalism will not apply.

Keeping this in mind, we will now discuss three possible origins for deviations to SRs. The first origin is higher-dimensional operators, Sec.~\ref{sec:example_corr_sr}, which typically arise from including suppressed terms that ultimately arise from the flavour symmetry being broken. The next possibility to modify SRs is to have corrections from the charged-lepton mass matrix. These can arise if the mass matrix in the charged lepton sector is not diagonal, but has to be diagonalised to arrive at the standard definition of leptonic mixing angles; this is discussed in Sec.~\ref{sec:CL-corrections}. Finally, as discussed in our earlier reference~\cite{Gehrlein:2015ena}, renormalisation group evolution (RGE) corrections can also lead to modifications. This possibility, to be introduced in Sec.~\ref{sec:RGEs}, is discussed in some more detail in what follows, for the simple reason that we can give a detailed comparison to the previous results.

\subsection{\label{sec:example_corr_sr}Higher-dimensional operators}

In many models, the mass matrices have a leading-order structure which is supposed here to give one of the mass sum rules. But then due to some higher-dimensional operators this leading structure gets disturbed. As an example for a possible higher order correction we want to study the $A_{5}\times SU(5)$ model proposed in~\cite{Gehrlein:2014wda,Gehrlein:2015dxa,Gehrlein:2015dza} where a correction to the leading Yukawa  matrix is introduced to account for the observed baryon asymmetry of the universe via the leptogenesis mechanism. In order to simplify the notation, we introduce the dimensionless matrices
\begin{align}
 \hat{y}&=
 \begin{pmatrix}
 1 & 0 & 0 \\
 0 & 0 & 1 \\
 0 & 1 & 0
 \end{pmatrix} \text{ and }
 \delta \hat{y}= \begin{pmatrix}
 0&1&0\\
 -1&0&0\\
 0&0&0
 \end{pmatrix} ~.
\end{align}
The neutrino Yukawa coupling in this model is then proportional to $\hat{y} + c \text{ e}^{\text{i } \gamma}\delta\hat{y}$, with $c \ll 1$ and $\gamma$ being an arbitrary phase coming from a higher-dimensional operator.

The mass matrix for the right-handed neutrinos is proportional to (see also~\cite{Cooper:2012bd}):
\begin{align}
 \hat{m}_{\text{RR}} &= \begin{pmatrix}
\frac{X}{\sqrt{6}}+\frac{Y \text{e}^{\text{i}\chi}}{\sqrt{30}}&-\frac{Y \text{e}^{\text{i}\chi}}{\sqrt{15}}& -\frac{Y \text{e}^{\text{i}\chi}}{\sqrt{15}}\\
-\frac{Y \text{e}^{\text{i}\chi}}{\sqrt{15}}&\frac{15 X-\sqrt{5}Y\text{e}^{\text{i}\chi}}{10\sqrt{6}}&\frac{-5 X-\sqrt{5}Y\text{e}^{\text{i}\chi}}{10\sqrt{6}}\\
-\frac{Y \text{e}^{\text{i}\chi}}{\sqrt{15}}&\frac{-5 X-\sqrt{5}Y\text{e}^{\text{i}\chi}}{10\sqrt{6}}&\frac{15 X-\sqrt{5}Y\text{e}^{\text{i}\chi}}{10\sqrt{6}}\\
\end{pmatrix} \; .
\label{eq:massmatrix_righthanded}
\end{align}

The light neutrino mass matrix is then generated via the type~I seesaw mechanism~\cite{seesaw}. It is up to $\mathcal{O}(c)$ given by
\begin{align}
m_\nu=-\hat{y}~ \hat{m}_{\text{RR}} ^{-1}~ \hat{y}-c~\text{e}^ {\text{i}\gamma} \left(\delta \hat{y}~ \hat{m}_{\text{RR}}^{-1}~ \hat{y}+\hat{y}~ \hat{m}_{\text{RR}} ^{-1}~ \delta \hat{y}\right) \;,
\end{align}
where we have absorbed all coefficients into $c$ and $\hat{m}_{\text{RR}}^{-1}$, and we thus have only five effective parameters. These are the dimensionful $X,~Y$; the phases $\chi$ and $\gamma$, and the small parameter $c$. One can easily map our simplified notation here to the original notation used in~\cite{Gehrlein:2015dxa,Gehrlein:2015dza}, by rewriting these parameters with the respective prefactors.

Since the leading-order neutrino mass matrix ($c\equiv 0$) depends only on two (complex) parameters, we find a mass SR which corresponds to SR~6 from Tab.~\ref{tab:overview_SR}:
\begin{align}
 \frac{\text{e}^{\text{i}\phi_1}}{m^{(0)}_1}+\frac{\text{e}^{\text{i}\phi_2}}{m^{(0)}_2}-\frac{1}{m_3^{(0)}}=0~.
\end{align}
 The corrections to the complex masses from Eq.~\eqref{eq:corrections_m} are up to order $c$  
\begin{align}
\delta m_1 &= \sqrt{6 (3 + \sqrt{5})}\frac{Y \sin \gamma}{X \cos \chi} \sqrt{\frac{1 }{ (X^2+Y^2+2 XY \cos \chi)}} \, c~,
\label{eq:delta_m1}\\
\delta m_2&= \sqrt{6 (3 - \sqrt{5})}\frac{Y \sin \gamma}{X \cos \chi} \sqrt{\frac{1 }{ (X^2+Y^2-2 XY \cos \chi)}} \, c~
\label{eq:delta_m2}\\
\delta m_3&=0~,\\
\delta\phi_1&=\arctan\left(\frac{1}{\tan \chi}+\frac{X}{Y}\frac{1}{\sin \chi}\right)~,\\
\delta \phi_2&=\arctan\left(\frac{1}{\tan \chi}-\frac{X}{Y}\frac{1}{\sin \chi}\right)~,\\
\delta \phi_3&=0~.
\end{align}
Note that for $\sin \gamma = 0$, $\delta m_1$ and $\delta m_2$ get corrected only at $\mathcal{O}(c^2)$ and $\tilde{m}_3$ does not get a correction at $\mathcal{O}(c)$. The correction is enhanced by the ratio $Y/X$ and  $\chi\approx n \pi/2$ ($n \in \mathbb{Z}$).

As an example parameter point we consider normal ordering with $X=24.0,~Y=185.3,~\chi=0.4$ so that, for $c=0.05$ and $\gamma=0.2$, we obtain $\delta \hat{s}_r\approx-0.013 $ and $\delta \hat{s}_i\approx   0.13$, which corresponds to $\delta \hat{s}\approx 0.13$.

\subsection{\label{sec:CL-corrections}Charged lepton sector}

Despite the fact that the mass SR is a feature of the neutrino sector of a given model, it can be influenced by a nondiagonal charged lepton mass matrix. The phases which appear in the PMNS matrix (i.e., the Dirac CP-phase and the two Majorana phases) depend on the phases from the neutrino and the charged lepton mixing matrices, since $U_{\text{PMNS}}=U_e^\dagger U_{\nu}$. This leads to relations between the leptonic mixing angles and phases. In the derivation of the formulas, however, some subtleties can arise, since all unphysical phases have to be correctly extracted. In App.~\ref{sec:parametrisations} we present a comprehensive derivation of relations between the parameters in the PMNS matrix and the neutrino and charged lepton mixing parameters. Here we will quote the results for the approximate expressions for the PMNS parameters in terms of SRs of neutrino mixing angles and the charged lepton mixing angles. For $\theta_{13}^\nu\approx 0$ and $\theta_{ 23}^e\approx \theta_{13}^e\approx 0$, they read
\begin{align}
 c_{13}^{\text{PMNS}} c_{12}^{\text{PMNS}} \text{e}^{\text{i}(\eta_1-\phi_1/2)}&=c_{12}^e c_{12}^\nu \text{e}^{\text{i}\omega_1^\nu}+c_{23}^\nu s_{12}^e s_{12}^\nu\text{e}^{\text{i}(\omega_2^\nu+\delta_{12}^\nu-\delta_{12}^e)}~,\label{eq:sr1}\\
  s_{12}^{\text{PMNS}} c_{13}^{\text{PMNS}}\text{e}^{\text{i} (\eta_1-\phi_2/2)} &= s_{12}^{\nu}c_{12}^e \text{e}^{-\text{i}\delta_{12}^{\nu} }\text{e}^{\text{i}\omega_1^\nu}-s_{12}^{e}c_{23}^{\nu}c_{12}^{\nu}\text{e}^{-\text{i}\delta_{12}^e }\text{e}^{\text{i}\omega_2^\nu } \;,\label{eq:sr2}\\
   s_{13}^{\text{PMNS}}\text{e}^{\text{i}( \eta_1-\delta)} &=-s_{12}^e s_{23}^{\nu} \text{e}^{-\text{i}(\delta_{12}^{e}+\delta_{23}^\nu)}\text{e}^{\text{i}\omega_2^\nu}~.  \label{eq:sr3} 
\end{align}
The superscript $\nu$ denotes the neutrino mixing parameters, while the superscript $e$ signifies the charged lepton mixing parameters. The phases $\eta_i$ and $\omega_i^\nu$ are unphysical, but they nevertheless have to be treated with care in order to obtain the correct results for the phases. Together with
\begin{multline}
c_{12}^{\text{PMNS}} \left(c_{13}^{\text{PMNS}}\right)^2 c_{23}^{\text{PMNS}} s_{13}^{\text{PMNS}} \left(s_{12}^{\text{PMNS}}s_{23}^{\text{PMNS}}\text{e}^{-\text{i}\delta}-c_{12}^{\text{PMNS}}c_{23}^{\text{PMNS}}s_{13}^{\text{PMNS}}\right) = \\
\left(U_{11}^{\text{PMNS}}\right)^*U_{13}^{\text{PMNS}}U_{31}^{\text{PMNS}} \left(U_{33}^{\text{PMNS}}\right)^*~,
\label{eq:PMNS-rel}
\end{multline}
which we get if we exploit the structure of the PMNS matrix, we can close the system to solve the four equations~\eqref{eq:sr1} to~\eqref{eq:PMNS-rel} to determine the Majorana phases. In~\cite{King:2002nf,Antusch:2005kw,Antusch:2008yc} these expressions were also derived but their formulas apply directly to the case where the unphysical phases are taken correctly into account and then subsequently absorbed. From Eqs.~(\ref{eq:sr1}, \ref{eq:sr2}, \ref{eq:sr3}) we see that the Majorana phases indeed depend on the charged lepton phases.

As a concrete example, we consider again the $A_5\times SU(5)$ model proposed in~\cite{Gehrlein:2014wda,Gehrlein:2015dxa,Gehrlein:2015dza}, which features a nondiagonal charged lepton mass matrix, a vanishing reactor angle, and a maximal atmospheric neutrino mixing angle. We introduced only a small 1-2 mixing in the charged lepton sector. The parameters in the neutrino mass matrix are complex, and hence the $\delta_{ij}^{\nu}$ depend on these parameters. But, for simplicity, we take the neutrino mass matrix to be real ($\chi = c = 0$). Taking into account all phases in the PMNS matrix, we obtain for the physical Majorana phases
\begin{equation}
\phi_1=\sqrt{3+\sqrt{5}}\theta_{12}^e \sin \delta_{12}^e~\text{ and }~\phi_2=\pi-\frac{\sqrt{5}-1}{\sqrt{2}}\theta_{12}^e\sin \delta_{12}^e~.
\end{equation} 
This result tells us that $\delta_{12}^e\approx\pi/2$  leads to the maximal correction to the Majorana phases and $\phi_1$ gets more affected by the charged lepton phases than $\phi_2$. We see that the physical Majorana phases have a dependence on the charged lepton phases, and thus the Majorana phases which appear in the SR are in general \emph{not} equal to the physical Majorana phases one obtains in a model with a nondiagonal charged lepton mass matrix. For this reason, a SR can get destroyed in this type of models.

To demonstrate how powerful our formalism is, we will now express these corrections in terms of
$\delta \hat{s}$ for the given example. On leading order we find that 
\begin{align}
\phi_1^{(0)}=0,&~\phi_2^{(0)}=\pi \;,~\\
m_1 = m_1^{(0)}+\mathcal{O}(\delta m_1^2),~m_2 &= m_2^{(0)}+\mathcal{O}(\delta m_2^2),~m_3 = m_3^{(0)} + \mathcal{O}(\delta m_3^2)\;~.
\end{align}
Note that the physical neutrino masses will only get corrected at the order $\mathcal{O}(\delta m_i^2)$.
With 
\begin{align}
\delta m_1&=-\sqrt{3 + \sqrt{5}} \, \theta_{12}^e \, m_1^{(0)} \sin\delta_{12}^e~,\\
\delta \phi_1&=-\frac{1}{2} \left(\pi + \sqrt{3 + \sqrt{5}} \, \theta_{12}^e \, \sin\delta_{12}^e \right)~,\\
\delta m_2&= -\sqrt{3-\sqrt{5}} \, \theta_{12}^e m_2^{(0)}  \, \sin\delta_{12}^e~,\\
\delta \phi_2&=-\frac{1} {2}\left(\pi-\theta_{12}^e \sqrt{3-\sqrt{5}} \sin\delta_{12}^e \right)~,\\
\delta m_3 &= 0~,
\end{align}
we get from eq.~\eqref{eq:ds} normalised to $m_1^d$ in leading order in $\theta_{12}^e$
\begin{align}
\delta \hat{s}\approx - \text{i} \frac{ \sqrt{3 - \sqrt{5}} \, m_1 - \sqrt{3 + \sqrt{5}} \, m_2 }{m_2} \theta_{12}^e \sin\delta_{12}^e~,
\end{align}
where we have used that $m_i \approx m_i^{(0)}$ in this approximation.
With $\theta_{12}^e\approx 12^{\circ}$ and $\delta_{12}^e=\pi/2$ we obtain as the maximal correction $\delta \hat{s}\approx 0.36$ and the correction vanishes for $\delta_{12}^e=0$.

\subsection{\label{sec:RGEs}RGE corrections}

Finally, a very generic correction to the mass SR are renormalisation group effects on the masses and Majorana phases. In~\cite{Gehrlein:2015ena}, we have already investigated the effect of such most generic corrections on the predictions of SRs. Now we want to match the corrections to the complex masses from Eq.~\eqref{eq:corrections_m} to the corrections coming from the RGEs, as far as possible. We therefore extract the RGE corrections from the absolute values of the masses and from the phases, and we rewrite the corrected complex masses as
\begin{align}
m_i \text{e}^{-\text{i}\phi_i} = \left(m_i^{(0)}+m_i^{\text{RGE}}\right)\text{e}^{\text{i} (-\phi_i^{(0)}+\phi_i^{\text{RGE}})}~,
\end{align}
where the superscript $(0)$ denotes the LO masses and Majorana phases, and the superscript RGE labels the corrections from the renormalisation group running.
The connection to Eq.~\eqref{eq:corrections_m} is
\begin{align}
\delta m_i&=\sqrt{\left(m_i^{\text{RGE}}\right)^2 + \left(m_i^{(0)}\right)^2 \left(\phi_i^{\text{RGE}}\right)^2}~,
\label{eq:m_RGE}\\
\delta \phi_i&=-\phi_i^{(0)}+\text{arctan}\left(   \frac{m_i^{(0)}\phi_i^{\text{RGE}}}{m_i^{\text{RGE}}}\right)~.
\label{eq:phi_RGE}
\end{align}
Note that the index $i$ in Eq.~\eqref{eq:m_RGE} runs from 1~to~3, whereas $i$ in Eq.~\eqref{eq:phi_RGE} is either~1 or~2. The explicit formulas for $\phi_i^{\text{RGE}}$ and $\delta m_i^{\text{RGE}}$ can be found in~\cite{Antusch:2003kp}. The RGE corrections to the phases and masses have a dependence on the mass scale and in the minimal supersymmetric Standard Model (MSSM)  additionally on $\tan \beta$. For large mass scales and large $\tan \beta$, the effect of the corrections is enhanced. The $\beta$ functions of the Majorana phases also depend on the phases themselves which leads to a dependence of $\delta m_i$ and $\delta \phi_i$ on both  leading-order Majorana phases. The RGE correction to the masses is always positive in the MSSM and negative in the Standard Model (SM), and hence the sign of the contribution to $\phi_i^{(0)}$  in  Eq.~\eqref{eq:phi_RGE} is fixed. Except for  $\phi_1^{(0)}\approx \phi_2^{(0)}$, where the running of the phases is suppressed, the running of the Majorana phases is  stronger than the running of the masses, since the $\beta$-functions of the masses depend on the values of the masses themselves (see also the discussion in~\cite{Gehrlein:2015ena}). For  $\phi_1^{(0)}\neq \phi_2^{(0)}$,  the main contribution to $\delta m_i$ comes from the term $m_i^{(0)}\phi_i^{\text{RGE}}$, and the correction to $\phi_i^{(0)}$ is close to maximal (i.e.\ $\pi/2$) in Eq.~\eqref{eq:phi_RGE}.

In the next section we will discuss the effect of RGE corrections to various sum rules in more detail.

\section{\label{sec:num}Numerical results}

In this section, we will present our numerical results. We will answer the nontrivial question of whether it is possible to reconstitute forbidden mass orderings by corrections, since some SRs allow for only one of the two mass orderings. Furthermore, we will present our numerical results for the allowed ranges for the effective mass $|m_{ee}|$ and for the lightest neutrino mass eigenvalues $m$, which we obtain from the corrected SRs.

\subsection{\label{sec:forbidden}Reconstituting forbidden orderings}

Some SRs only allow for one particular mass ordering~\cite{Barry:2010yk,SR11,King:2013psa}, i.e., normal ($m_1<m_2<m_3$) or inverted ($m_3<m_1<m_2$). The question might arise if we can reconstitute those forbidden mass orderings by large enough corrections. In~\cite{Gehrlein:2015ena}, we did a similar study concerning the generic corrections arising from renormalisation group effects, where we concluded that this was \emph{not} possible.

For SRs 2, 3, 4, 5, 10 and 12 from Tab.~\ref{tab:overview_SR}, only one mass ordering is allowed. Hence, we have $s\neq 0$ for the forbidden orderings. In order to  obtain $s^{(0)}= 0$ we need the leading-order masses to respect the allowed ordering in these SRs, and due to corrections, the observed masses will obey  the forbidden ordering. In other words, the corrections to the masses have to alter the mass ordering. In principle the ordering of the leading-order masses is not restricted to be normal or inverted,  and one can also imagine having $m_1^{(0)}>m_3^{(0)}>m_2^{(0)}$ or other variations as long as they fulfill the leading-order SR. However, one would usually discard such cases, as they apparently do not correspond to reality.

To be more precise, we want to ask the question of how large $\delta \hat{s}$ has to be at least, such that $s^{(0)} = 0$ is fulfilled and the $m_i$ follow the ``forbidden'' mass ordering. This can be done most easily by considering again the geometrical interpretation of the SRs, as done in Fig.~\ref{fig:triangle}. We see that a SR without corrections is fulfilled if the values of $\delta s_r = \delta s_i =0$ (i.e., the triangle closes). Now, on the other hand, it can happen that -- for experimentally allowed values of the neutrino masses -- the triangle \emph{never} closes. For this case, we want to determine $\delta \hat{s}$ with $|\delta \hat{s}|$ being minimal. Obviously, $|\delta \hat{s}|$ is minimal for 
\begin{align}
 \delta \hat{s}_i &= 0 \;, \\
 \Delta \chi_{23} - d \phi_2 &= \pi \;, \\
  d \phi_1- \Delta \chi_{13} &= -\pi \;.
\end{align}
Then $|\delta \hat{s}| = |\delta \hat{s}_r|$ is minimal.

Plugging these expressions into the general formulas results in a rather long expressions, and hence we prefer to discuss what happens for the concrete example of SR~2, where $(c_1, c_2, d, \Delta \chi_{13}, \Delta \chi_{23}) = (1, 2, 1, \pi, \pi)$, and we want to focus on inverted mass ordering for the observed (corrected) masses. 
Since we have $\delta \hat{s}_i=0$, and the minimal value for $\delta \hat{s}_r$ corresponds to the situation where the sides of the triangle lie on the baseline, which is 
\begin{align}
\delta \hat{s}_r&= 2 -  \frac{m_3}{m_2} -  \frac{m_1}{m_2}.
\end{align}
For three different exemplary mass scales ($m_3 = 0.0,\ 0.001,\ 0.05$~eV), we obtain:
\begin{align}
\delta \hat{s}_r(m_3=0~\text{eV})&=1.02 \;,\\
  \delta \hat{s}_r(m_3=0.001~\text{eV})&=1.00 \;,\\
  \delta \hat{s}_r(m_3=0.05~\text{eV})&=0.30 \;.
\end{align}
The corrections  thus have to have at least this size to reconstitute inverted mass ordering.

\begin{table}
\centering
\begin{tabular}{c c c c c c c}
\toprule
Sum rule &  Forbidden ordering& \multicolumn{3}{c}{$\delta \hat{s}_r (m)$}\\
& &$m=0~\text{eV}$ & $m=0.001~\text{eV}$& $ m=0.05~\text{eV}$\\
\midrule	
	2& IO&1.02&1.00&0.30 \\
	3& IO&1.02&1.00&0.30 \\
	4&NO&0.92&0.90&0.30 \\
	5&NO& 0.95&0.94&0.30\\
	10& IO&1.01&0.87&0.16\\
	12&IO&1&0.86&0.16\\
	\bottomrule
\end{tabular}
\caption{\label{tab:forbidden_orderings}Summary table of the SRs which allow only one mass ordering and the minimal value of the correction $\delta \hat{s}_r$ that is needed to reconstitute the forbidden mass orderings for different mass scales.}
\end{table}

In Tab.~\ref{tab:forbidden_orderings} we have calculated the minimal values of $\delta \hat{s}_r$ for different mass scales, for a large mass scale, for a small mass scale and for a vanishing value of $m$ for the SRs which allow for only one mass ordering (SRs 2, 3, 4, 5, 10 and 12). We see that, in principle, we can reconstitute all forbidden orderings although we might need quite sizeable corrections. Especially for the case of very small neutrino masses, the corrections are so large that a perturbative approach is not suitable anymore, and it is simply not appropriate to talk of a neutrino mass SR at all.

Furthermore, the attentive reader might be surprised that we can suddenly reconstitute all forbidden orderings which was in no case possible for the RGE corrections only. We will show now that this is due to the fact that the correction to the SR had in nearly all cases a fixed sign pointing in the ``wrong'' direction, i.e., the RGE corrections -- although potentially sizeable -- tend to make the forbidden mass orderings even less likely than the LO SRs. In other words, instead of making the deviation of the SR smaller, RGE corrections increased it. Only for one case, the sign mentioned above was in principle suitable, but for that particular case, the corrections were way too small. Thus, our previous conclusions remain perfectly valid for the RGE-corrected case; however, as we will see, the general case actually \emph{can} make the otherwise forbidden orderings possible -- as to be expected.

As an example we will consider SR~2 in Tab.~\ref{tab:overview_SR}, where we try to reconstitute inverted mass ordering by RGE corrections. In order to do so, we will analyse the corrections to the lengths of the sides of the triangle. If these corrections are positive and large enough, they can close the open triangle. The nontrivial lengths of the triangle in case of SR~2 are $m_1/m_2$ and $m_3/m_2$ where we set 
\begin{align}
m_i=m_i^{(0)}+m_i^{\text{RGE}}~.
\label{eq:delta_msides}
\end{align}
 The RGE corrections for the masses are of the form
\begin{align}
m_i^{\text{RGE}}=\frac{1}{16 \pi^2}\left[\alpha_{\text{RGE}} \, m_i^{(0)}+F_i m_i^{(0)}\right] \log \left(\frac{\mu}{M_Z} \right)~,
\label{eq:m_RGE2}
\end{align}
where $\alpha_{\text{RGE}}\approx 3$ is a function which depends on the gauge and Yukawa couplings, $\mu>M_Z$ is the high energy scale and $F_i$ is a function which depends on the angles and on $\tan \beta$ in the MSSM. In the MSSM we expect the largest effect for $F_i>0$, whereas $F_i<0$ holds in the SM. We will focus on the MSSM case first. If we plug  Eq.~\eqref{eq:m_RGE2} into Eq.~\eqref{eq:delta_msides} and expand in $m_i^{\text{RGE}}$, we  see that the terms $\propto \alpha_{\text{RGE}}$ drop out since 
$\alpha_{\text{RGE}}$ is the same for all masses. We obtain for the corrections of the length of the sides: 
\begin{align}
\delta \left( \frac{m_1}{m_2}\right)&=\frac{m_1^ {(0)}}{m_2^ {(0)}}(F_1-F_2)~,\\
\delta \left( \frac{m_3}{m_2}\right)&=\frac{m_3^ {(0)}}{m_2^ {(0)}}(F_3-F_2)~.
\end{align}
In the $3 \sigma$ ranges  for the mixing angles from Tab.~\ref{tab:exp_parameters}, we have $F_1-F_2<0$ and $F_3-F_2>0$. Hence the length $m_3/m_2$ increases whereas the length $m_1/m_2$ decreases. If $m_3/m_2$ increases more strongly than $m_1/m_2$ decreases  we can hope to close the triangle. But for SR~2 this is not the case since $m_1/m_2$ decreases more strongly than $m_3/m_2$ increases. We conclude that the RGE corrections make the deviation from the SR  even larger. Hence the inverted ordering \emph{cannot} be reconstituted.
 
This statement can be transferred to SRs~3 and~10, where we also have to normalise the sides of the triangle by $m_2$. Also in these cases the corrections to the lengths of the sides are $\propto (F_1-F_2) $ and $\propto (F_3-F_2)$. In case of SRs~4 and~5, the sides of the triangle have to be normalised to $m_3$. The corrections to the length of the sides are $\propto (F_1-F_2)$ and $\propto (F_2-F_3)$. For the $3 \sigma $ ranges of the mixing angles, $F_1-F_2<0$ and $F_2-F_3<0$. Hence, the sides of the triangle both decrease due to RGE corrections. Also in these cases we \emph{cannot} reconstitute the forbidden orderings. Only in the case of SR~12 both sides of the triangle increase. In SR~12, $d<0$, and we have to normalise the sides  by $m_3^d$ to reconstitute normal ordering. This leads to terms $\propto(F_3-F_1)$ and $\propto (F_3-F_2)$ in the corrections to the lengths of the sides, which are both positive. However, numerically we would need a mass scale larger than $1$~eV for $\tan \beta=200$ to fulfil the sum rule, which is simply unrealistic.

In fact, for any SR that is not fulfilled with the low energy masses, the RGE effects in the MSSM have the wrong sign if $d>0$ for both orderings and if $d<0$  for normal ordering. In the SM, $\delta\hat{s}$ has the right sign to decrease the deviation from the SR but nevertheless the effects are too small to fulfil the sum rule.

In conclusion the fixed sign of the RGE corrections makes it barely possible to reconstitute forbidden orderings. Only in the case of $d<0$ for inverted ordering or in the SM case the sign was suitable, but the effects are nevertheless too small. Since, however, any corrections to the masses beyond those from RGEs do not have a fixed sign, it is nevertheless possible to reconstitute forbidden orderings in the general case.

In the following section we want to confirm these estimates also numerically.

\subsection{Effects on the lower bound of $\boldsymbol{m}$}

One major prediction of the SRs is the lower bound on the smallest neutrino mass eigenvalue $m$. The question arises as to how this bound changes in the light of corrections to the SR.

To answer this question, we will consider -- similarly as in the previous section -- the case where the sides of the triangle lie on the baseline, i.e., the case where the SR is just fulfilled. This clearly leads to the lower bound on the lightest mass $m$. In Tab.~\ref{tab:mlightest_change} we have summarised the results for the relative change of the mass scale in leading order in $\zeta\equiv \frac{\Delta m_{21}^2}{|\Delta m_ {32}^2|}\approx 0.03$. The results are presented in the form 
\begin{align}
\frac{m}{m^ {(0)}}=1-\kappa ~\delta \hat{s}~,
\end{align}
where $\kappa$ is a parameter that depends on the SR and on the mass ordering.

\begin{table}
\centering
\begin{tabular}{c c c }
\toprule
SR&Ordering& $\kappa$\\
\midrule 
\multirow{2}{*}{1}&NO& $4/3$ \\
&IO& $2/\zeta \approx 65.3$\\
\midrule 
2& NO&$9/8$ \\
\midrule 
3& NO& $9/8$\\
\midrule 
4&IO& $4/\zeta \approx 130.6$ \\
\midrule
5&IO& $\frac{33 +19 \sqrt{3}}{24}\approx 2.75 $ \\
\midrule
\multirow{2}{*}{6}&NO& $\sqrt[3]{2/(27\zeta)} \approx 1.34$\\
    &IO &$4/3$\\
    \midrule
  7 &NO& $4/3$\\
   &IO& $9/8$\\
  \midrule
  \multirow{2}{*}{8}&NO& $4/3$\\
   &IO& $9/8$\\
  \midrule
  \multirow{2}{*}{9}&NO& $4/3$\\
   &IO& $9/8$\\
  \midrule
  10&NO& $81/40\approx 2.03 $\\
  \midrule
 \multirow{2}{*}{11}&NO&$512/175\approx 2.93 $\\
  &IO&$243/65\approx 3.74$\\
  \midrule
  12&NO&2\\
\bottomrule
\end{tabular}
\caption{\label{tab:mlightest_change}Estimates for the relative change of the lower bound of the lightest mass, where $\zeta\equiv \frac{\Delta m_{21}^2}{|\Delta m_ {32}^2|}$. Please see the main text for more details.}
\end{table}

For SRs~1 and~4 in IO, the effect of the correction is enhanced because of the small $\zeta$ in the denominator. For the other SRs we find that the relative change is between $20\%$ and $80\%$ for $\delta \hat{s}=0.1$ and $0.3$. In the next section, we will verify these estimates numerically. The enhancement of the effect of the corrections for the mass scale for SR~1 and SR~4, for IO in both cases, is not visible when considering RGE corrections only, because  $\delta \hat{s}$ for a small mass scale is very small.

In the case of SR~7 for NO, we additionally get a correction to the upper mass bound:
\begin{align}
\frac{m}{m^ {(0)}}=1+\frac{4}{3} ~\delta \hat{s}~,
\end{align}
which is $1.4$ for $\delta \hat{s}=0.3$. Furthermore we obtain for $\delta\hat{s}\neq 0$ a new allowed mass region in NO. To open up this region we need for the lightest neutrino mass $m=(0.01, 0.05, 0.1)$~eV a $\delta \hat{s}= (-0.31, -0.27, -0.10)$. These values are well within our scan regions.

In the case of RGE effects only we encountered a parametric enhancement for the RGE effects for SR 1 and 4 in IO for a medium mass scale. This enhancement is \emph{not} present now because we parametrised the corrections differently, to avoid any artificial parametric enhancement.

\subsection{Neutrinoless double beta decay}
 
\begin{figure}
\begin{tabular}{lll}
\hspace{-1cm}
\includegraphics[width=5.4cm]{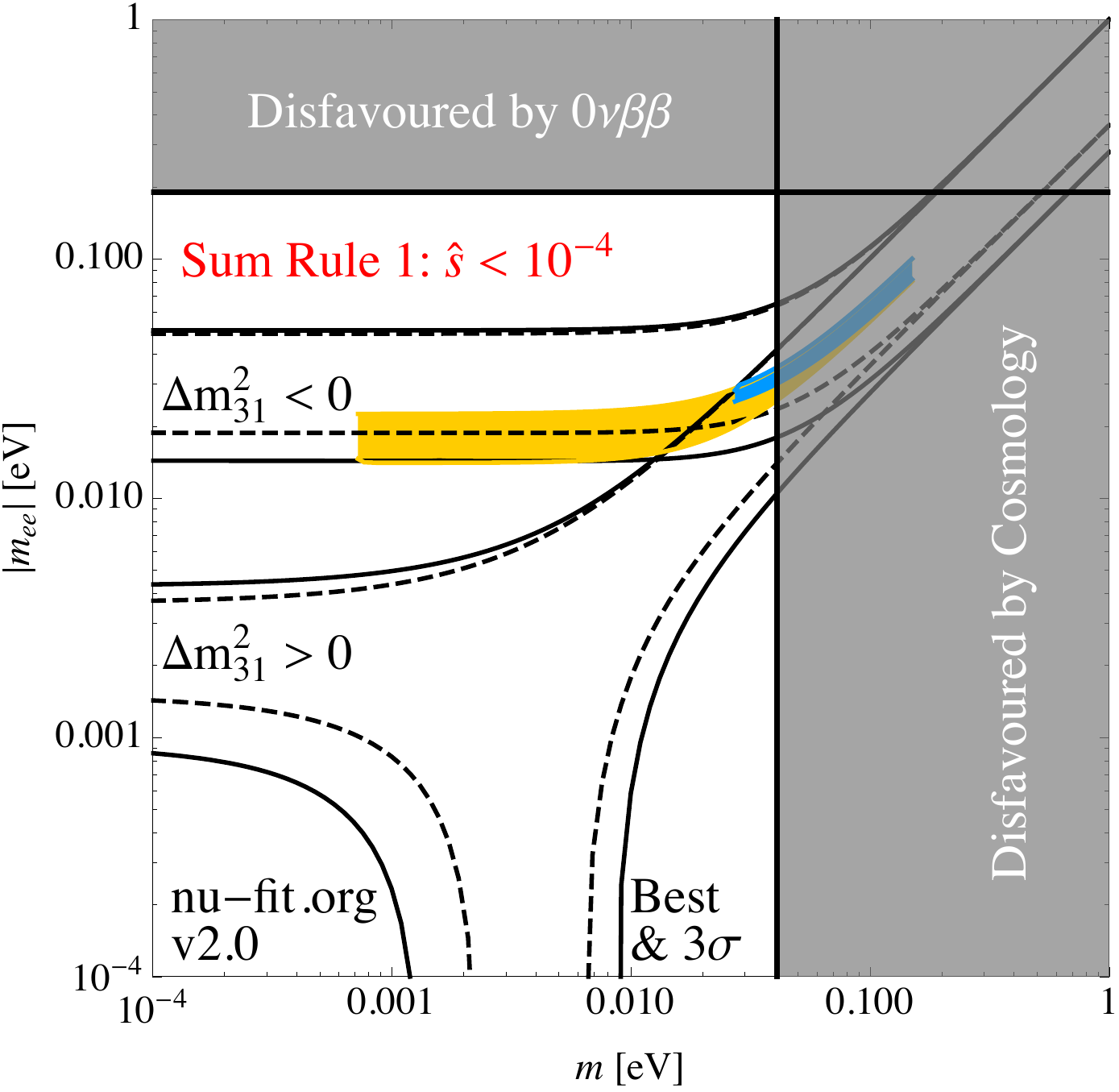} &
\includegraphics[width=5.4cm]{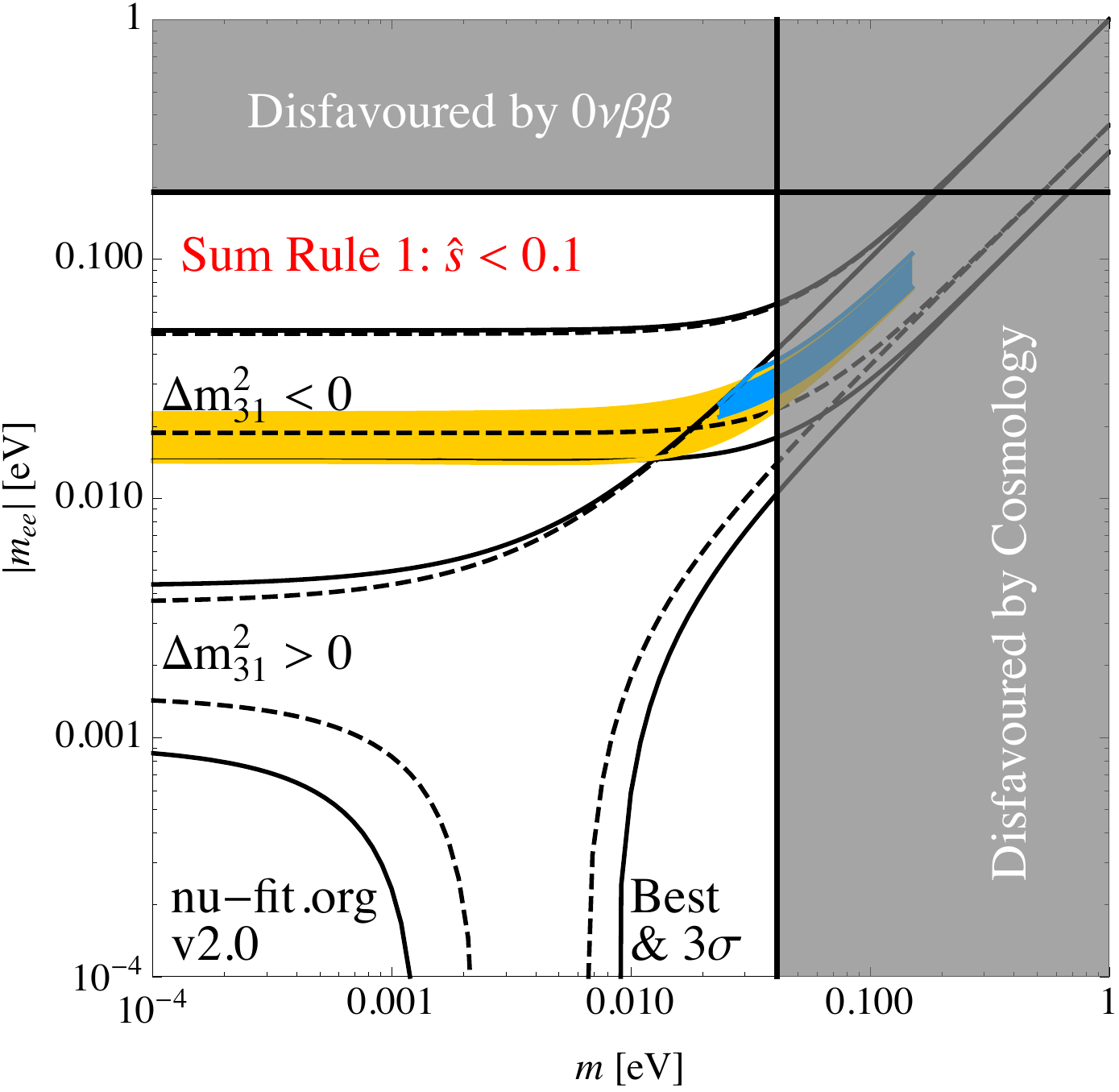} &
\includegraphics[width=5.4cm]{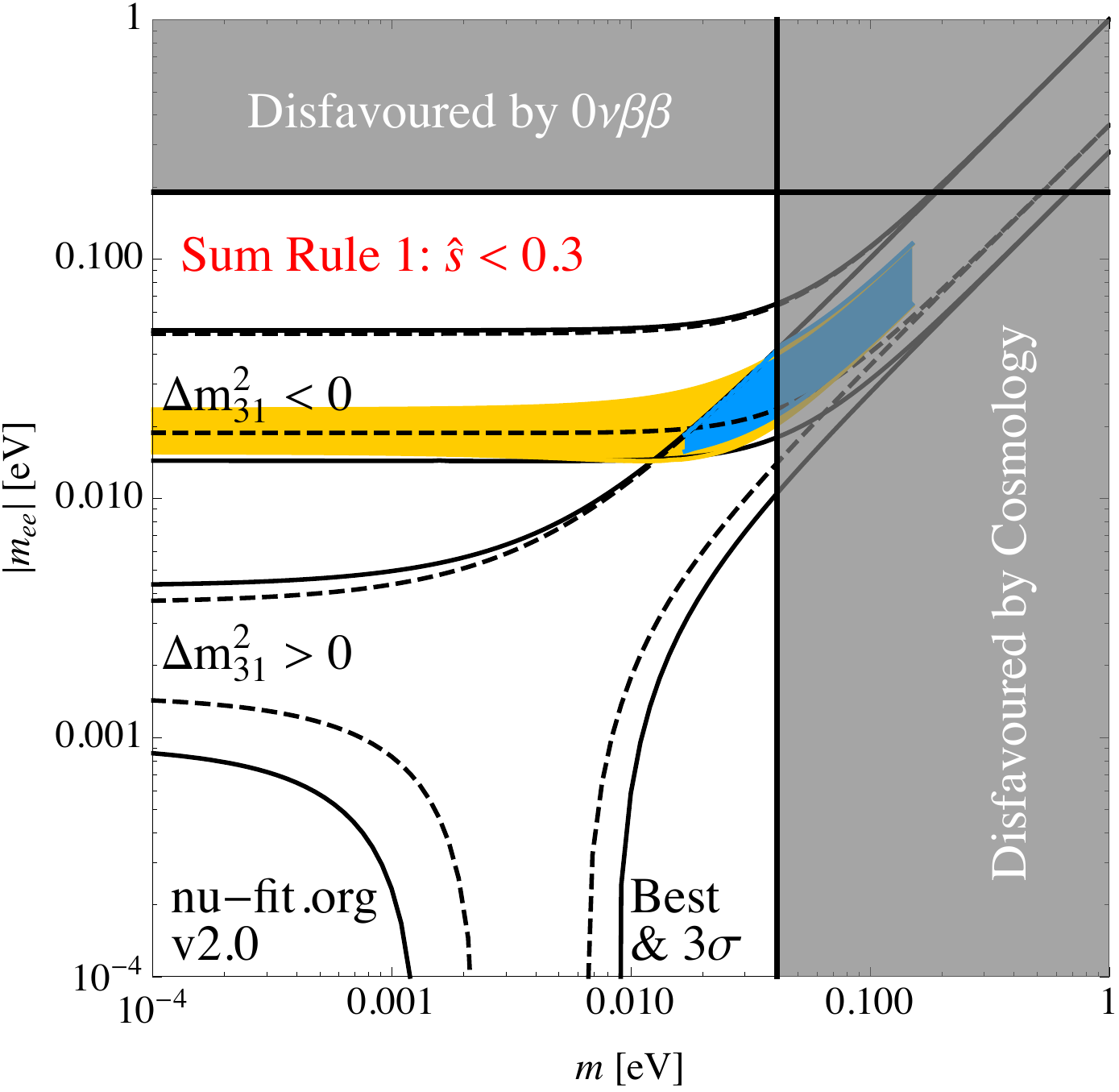}
\end{tabular}
\caption{\label{fig:SR1}Effective mass with SR~1 for $\hat{ s}=10^{-4}$, $0.1$, $0.3$.}
\end{figure}

\begin{figure}
\begin{tabular}{lll}
\hspace{-1cm}
\includegraphics[width=5.4cm]{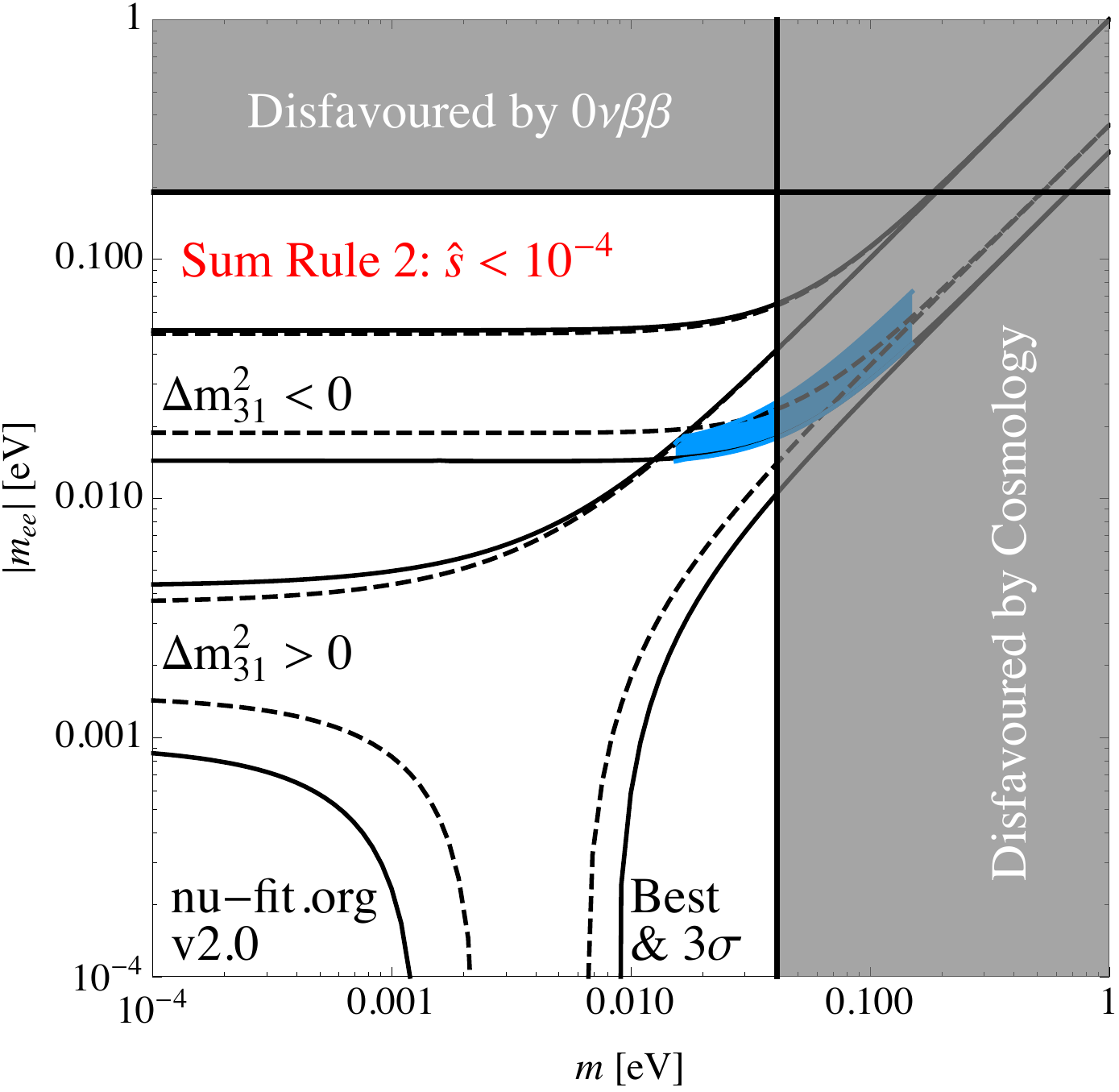} &
\includegraphics[width=5.4cm]{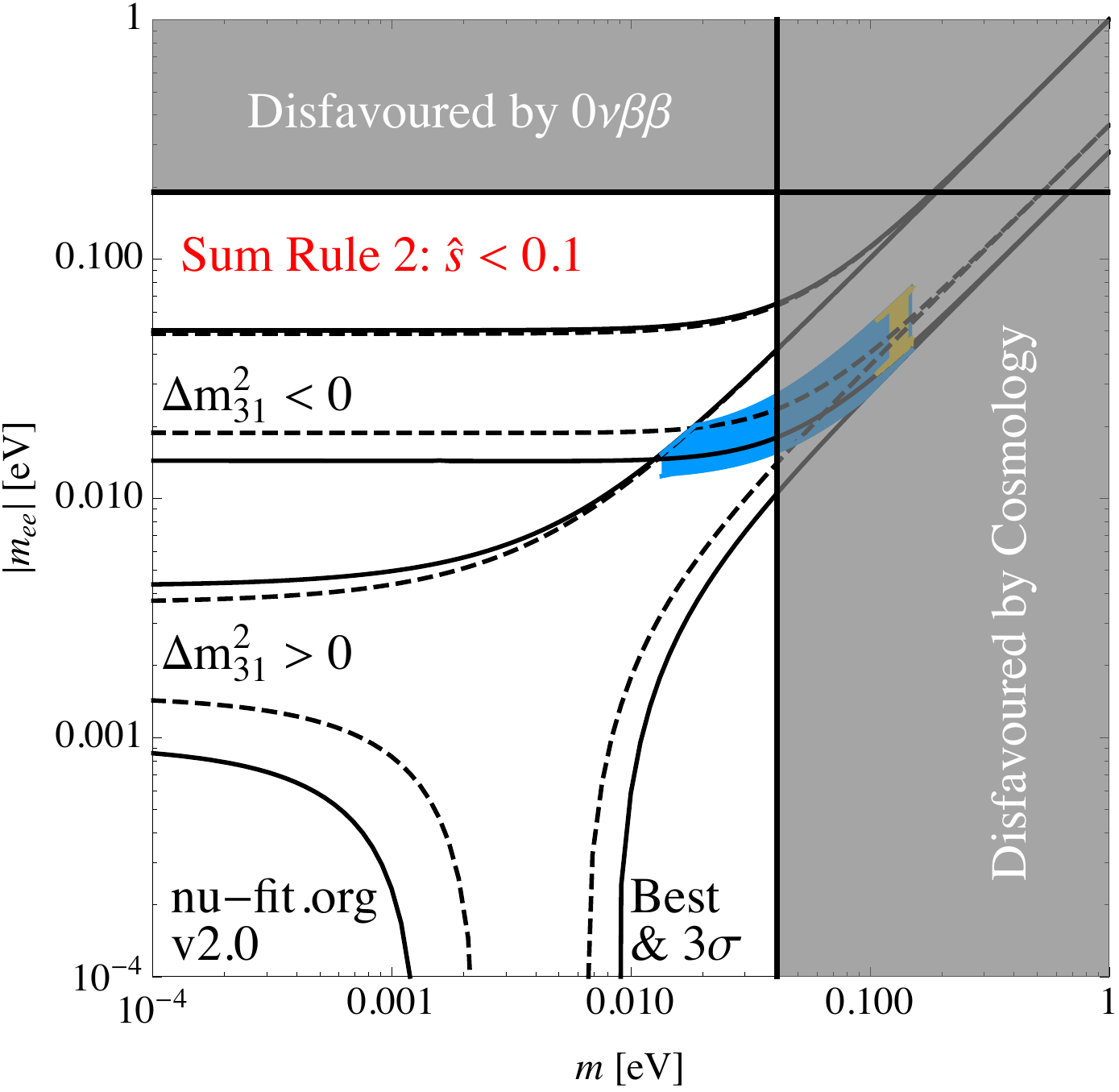} &
\includegraphics[width=5.4cm]{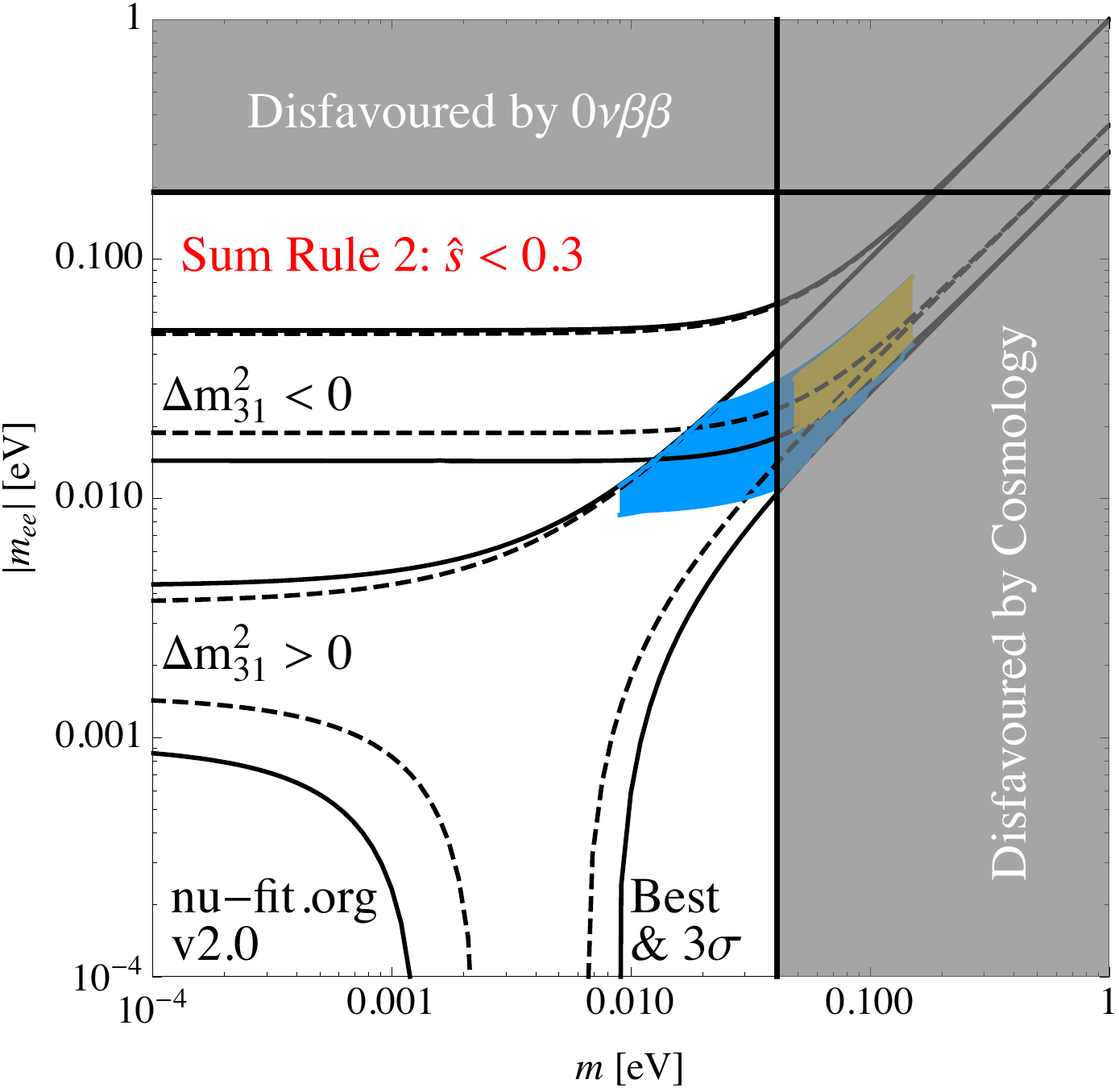}
\end{tabular}
\caption{\label{fig:SR2}Effective mass with SR~2 for $ \hat{s}=10^{-4}$, $0.1$, $0.3$.}
\end{figure}

\begin{figure}
\begin{tabular}{lll}
\hspace{-1cm}
\includegraphics[width=5.4cm]{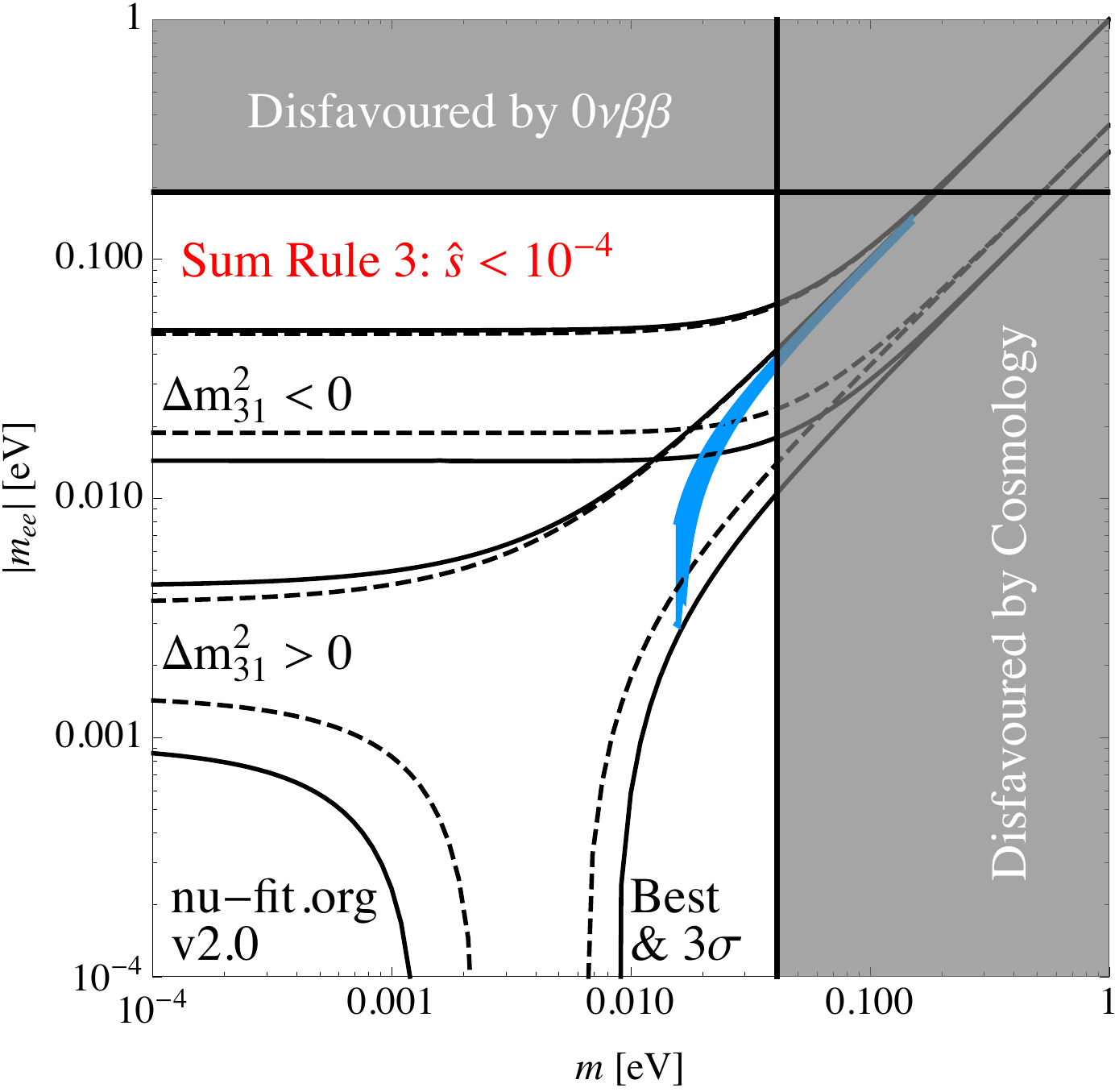} &
\includegraphics[width=5.4cm]{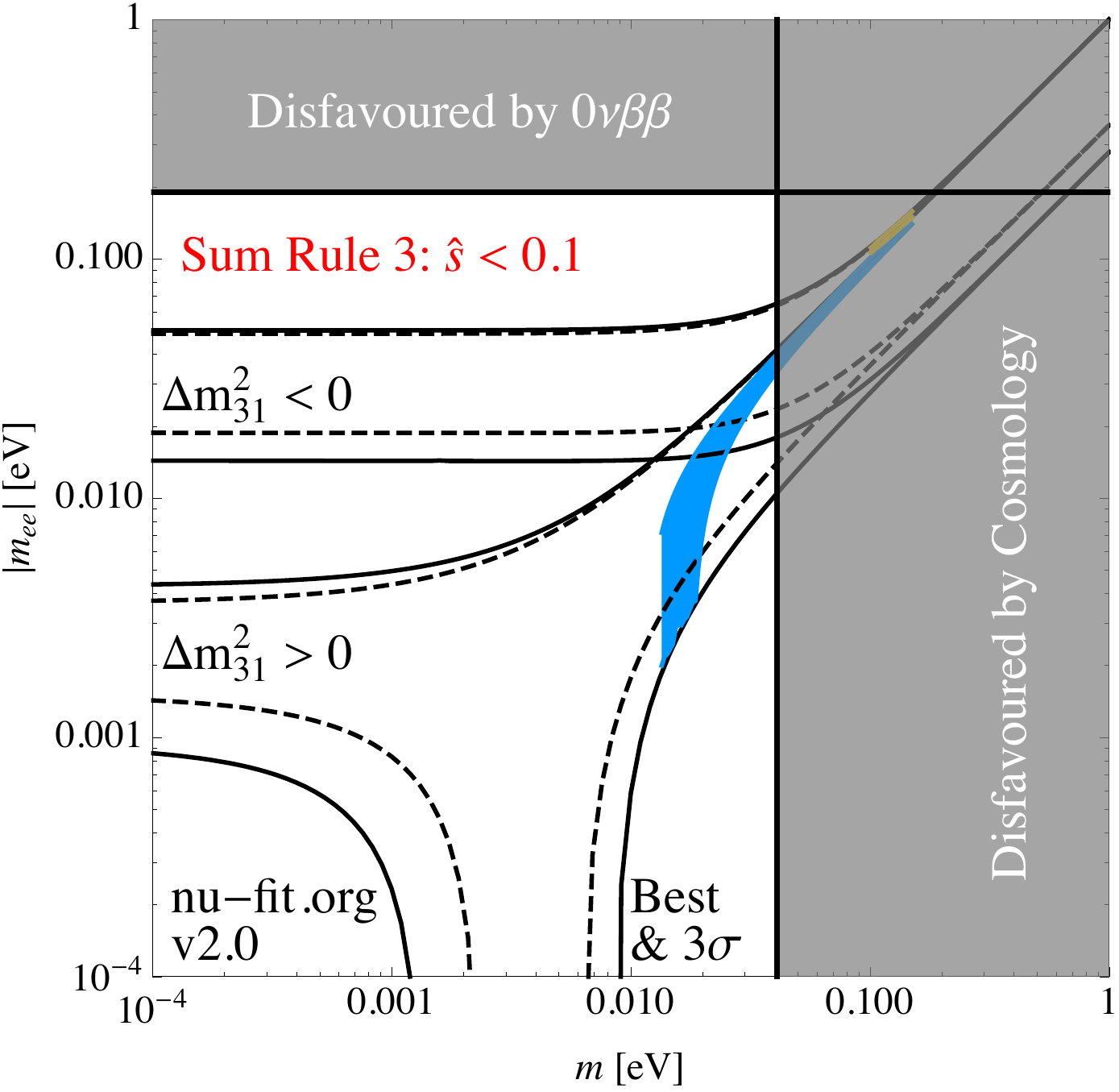} &
\includegraphics[width=5.4cm]{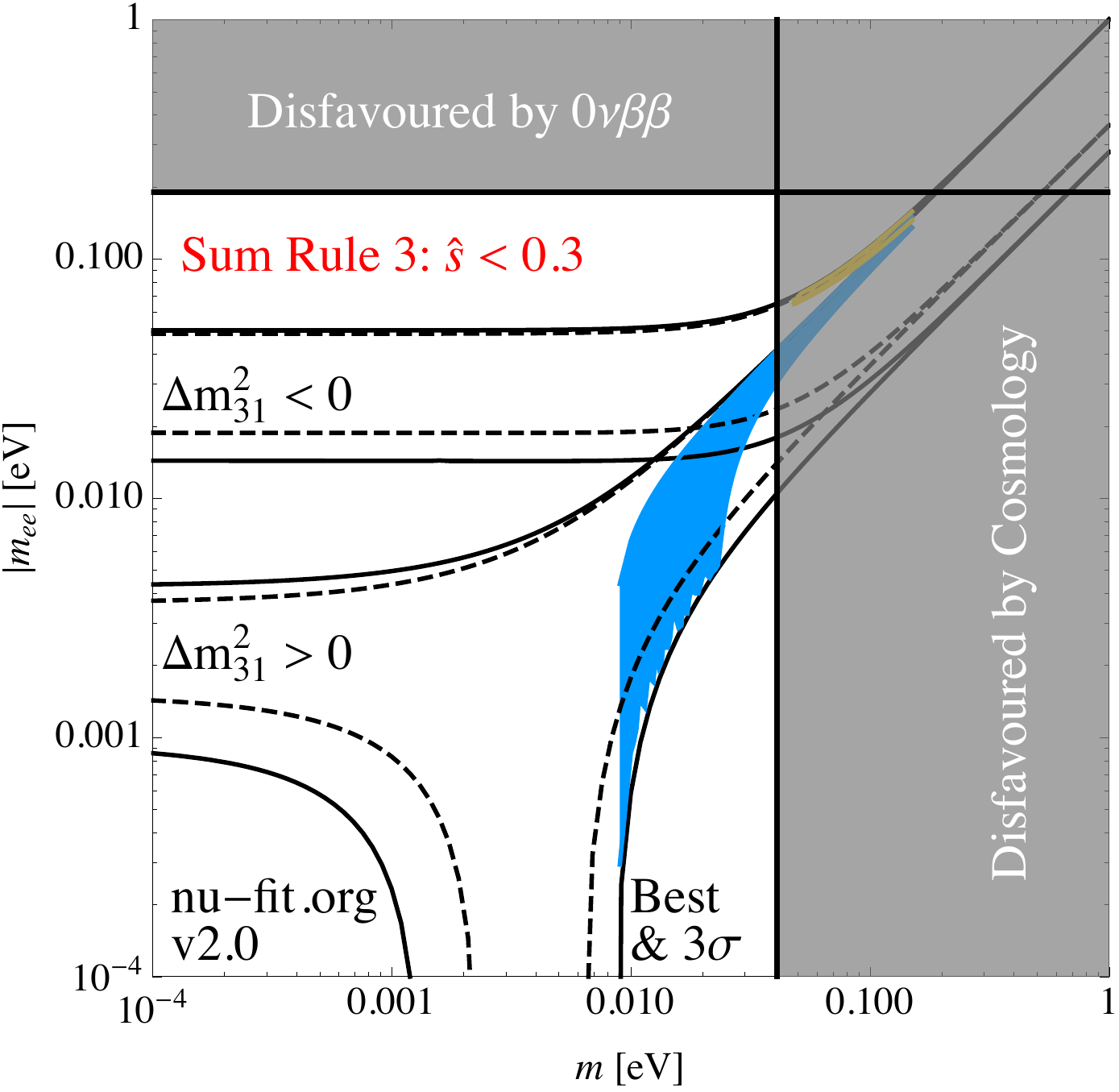}
\end{tabular}
\caption{\label{fig:SR3}Effective mass with SR~3 for $ \hat{s}=10^{-4}$, $0.1$, $0.3$.}
\end{figure}

\begin{figure}
\begin{tabular}{lll}
\hspace{-1cm}
\includegraphics[width=5.4cm]{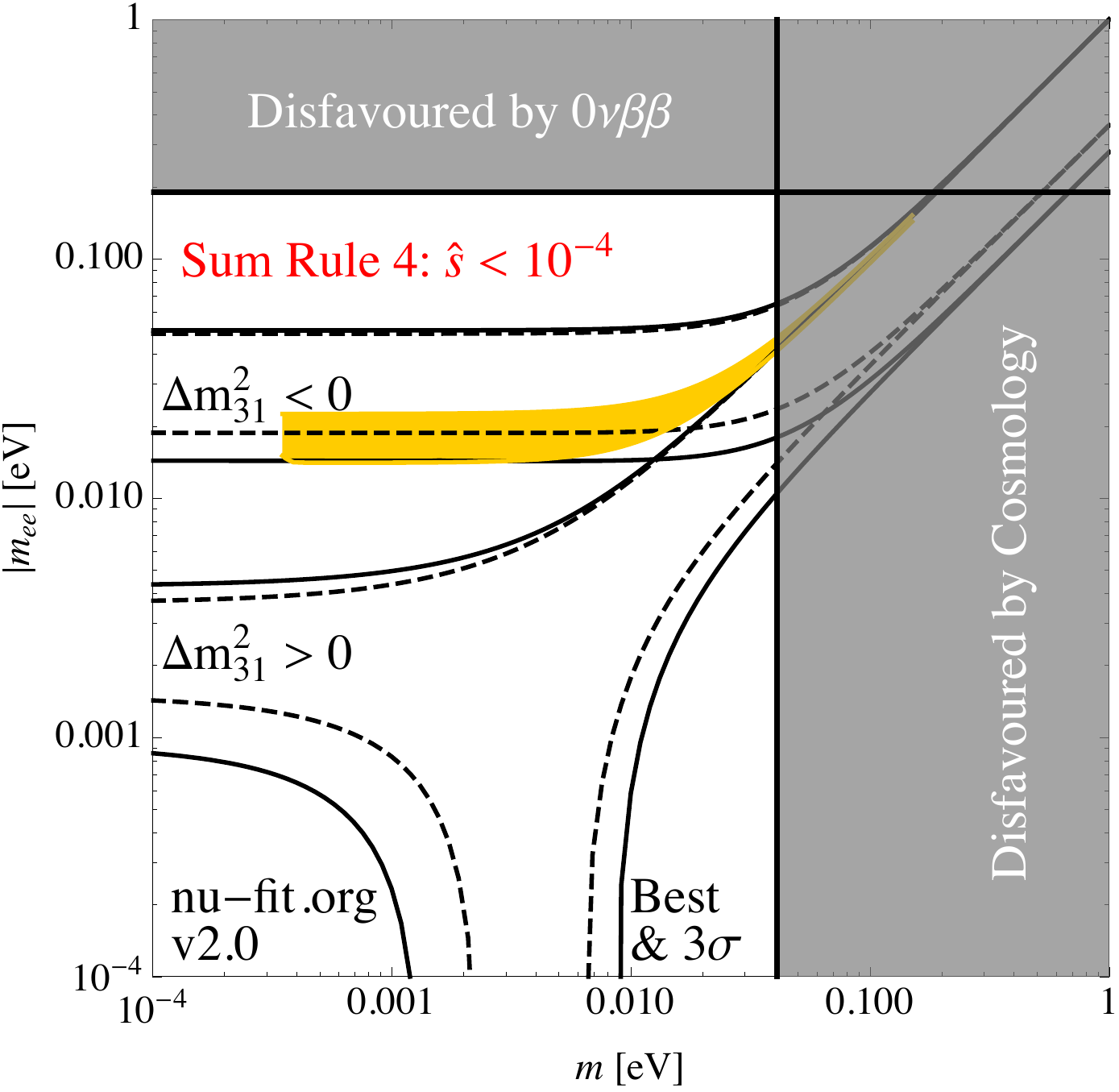} &
\includegraphics[width=5.4cm]{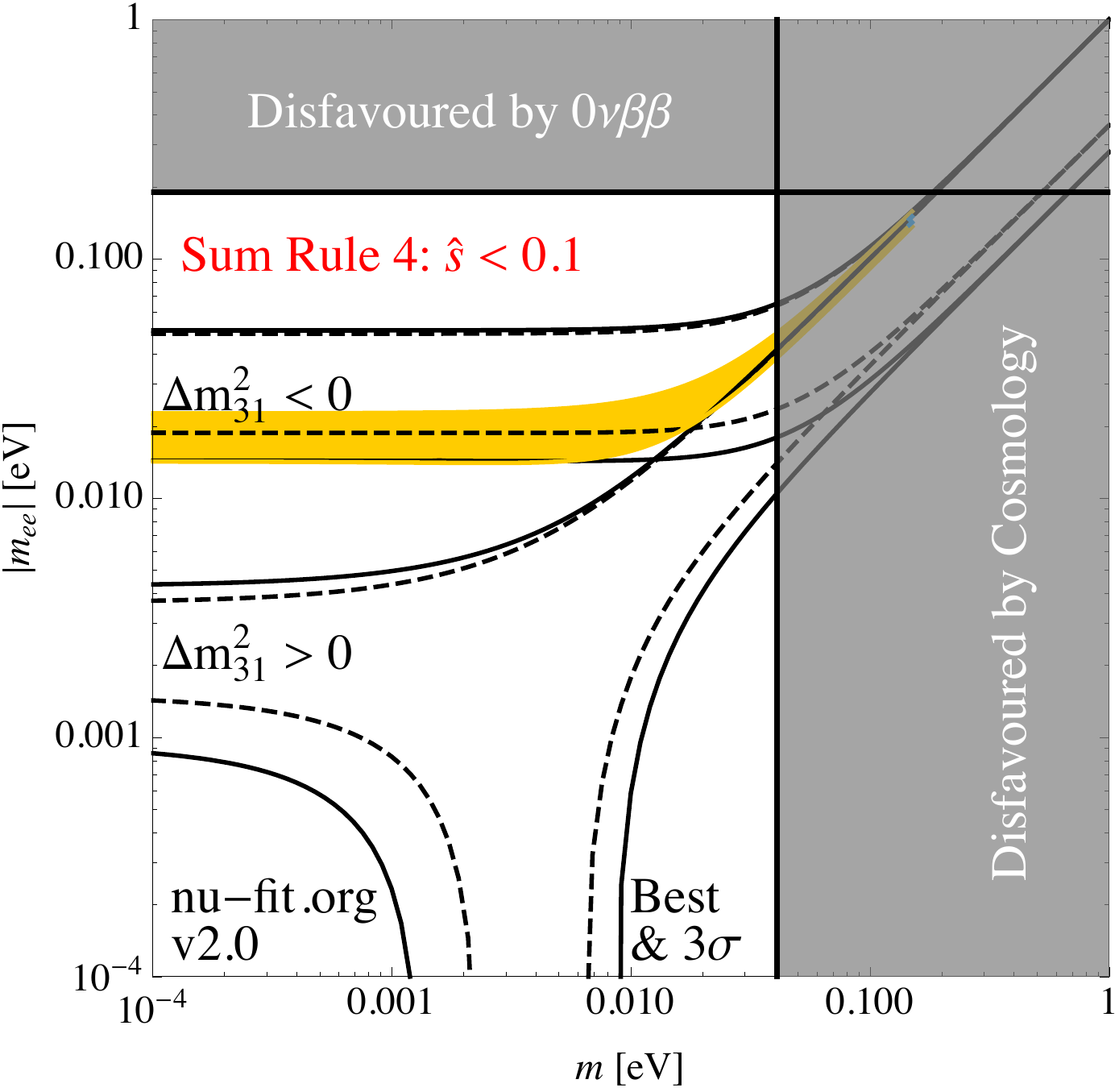} &
\includegraphics[width=5.4cm]{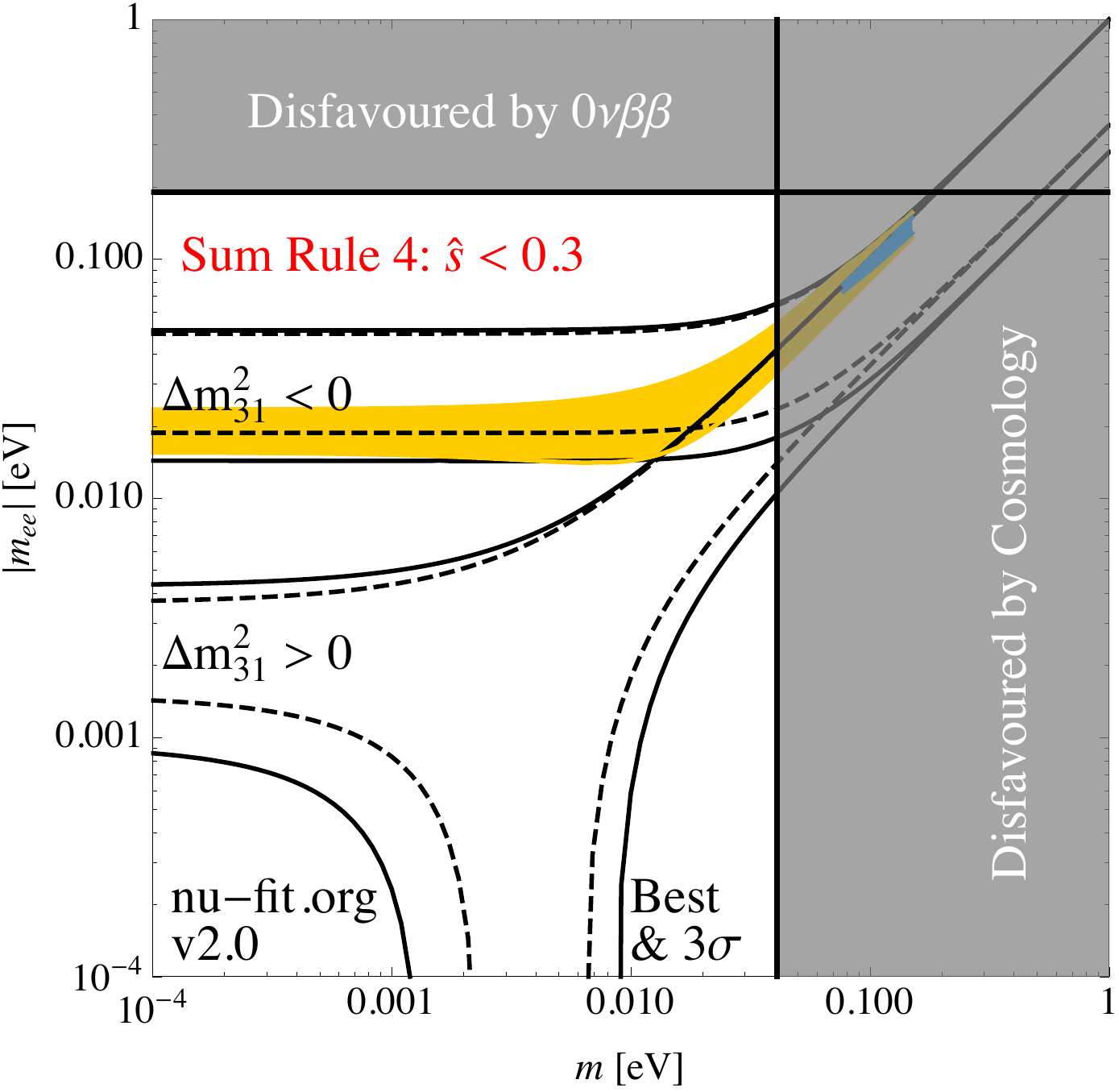}
\end{tabular}
\caption{\label{fig:SR4}Effective mass with SR~4 for $ \hat{s}=10^{-4}$, $0.1$, $0.3$.}
\end{figure}

\begin{figure}
\begin{tabular}{lll}
\hspace{-1cm}
\includegraphics[width=5.4cm]{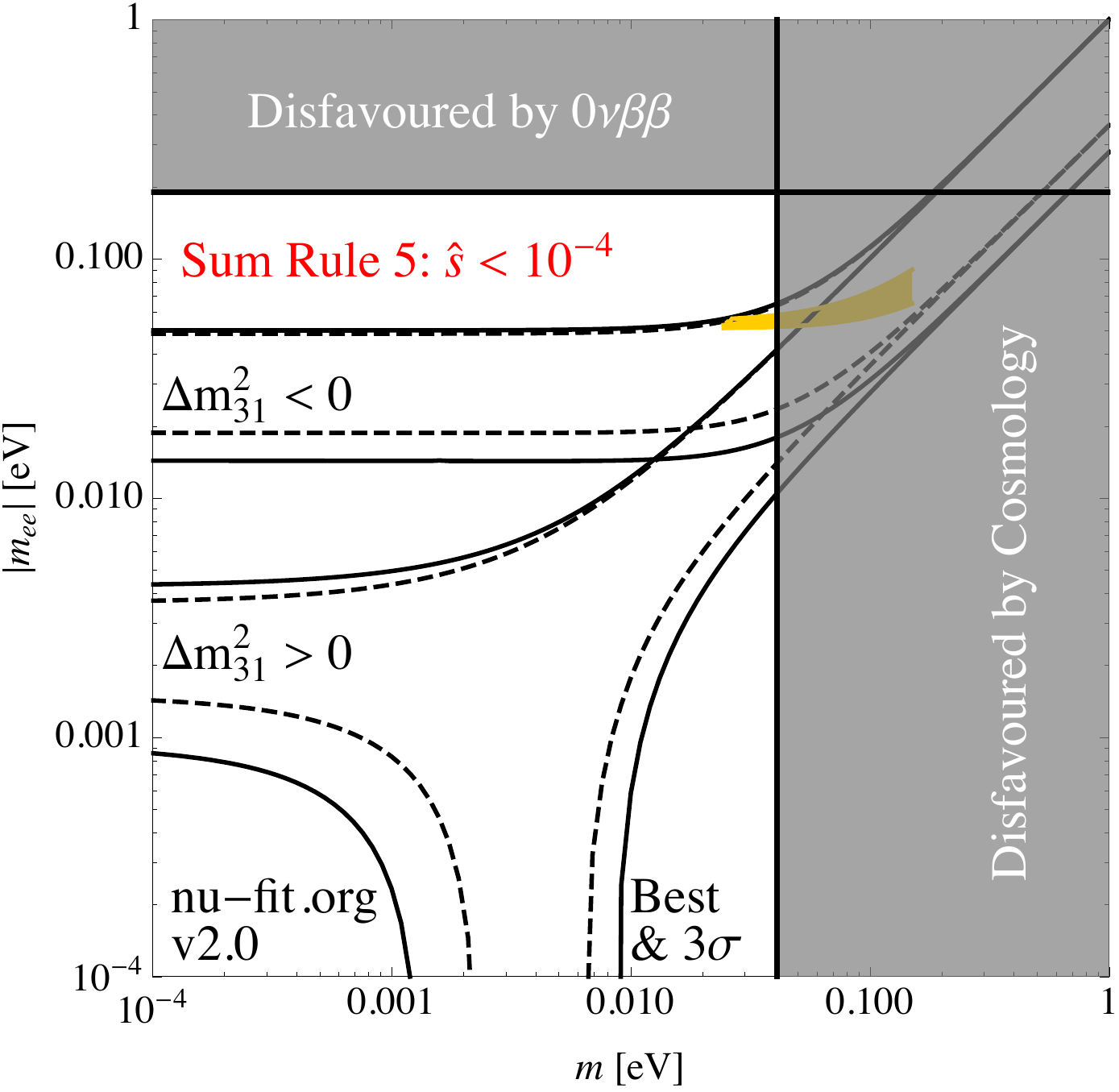} &
\includegraphics[width=5.4cm]{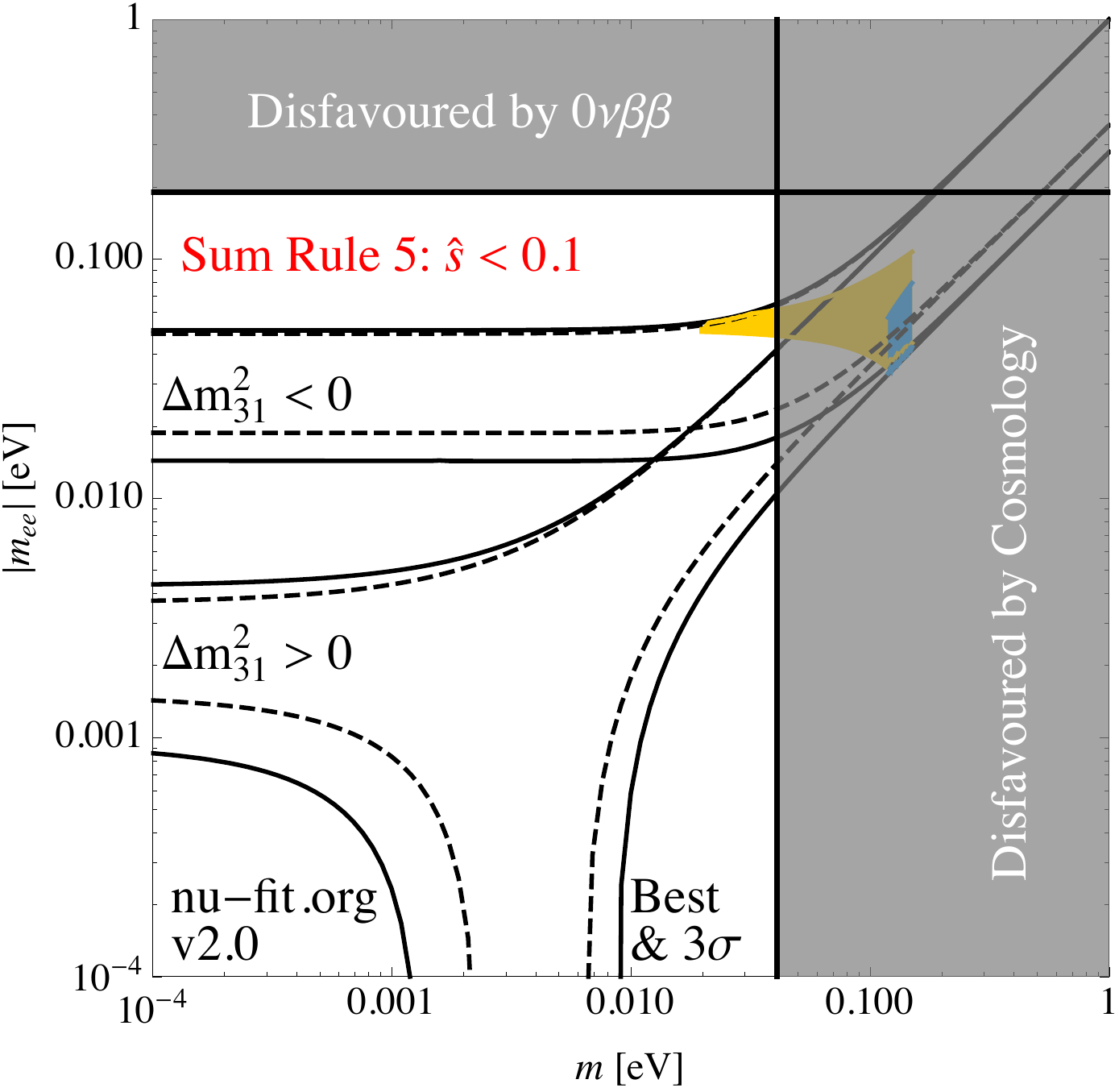} &
\includegraphics[width=5.4cm]{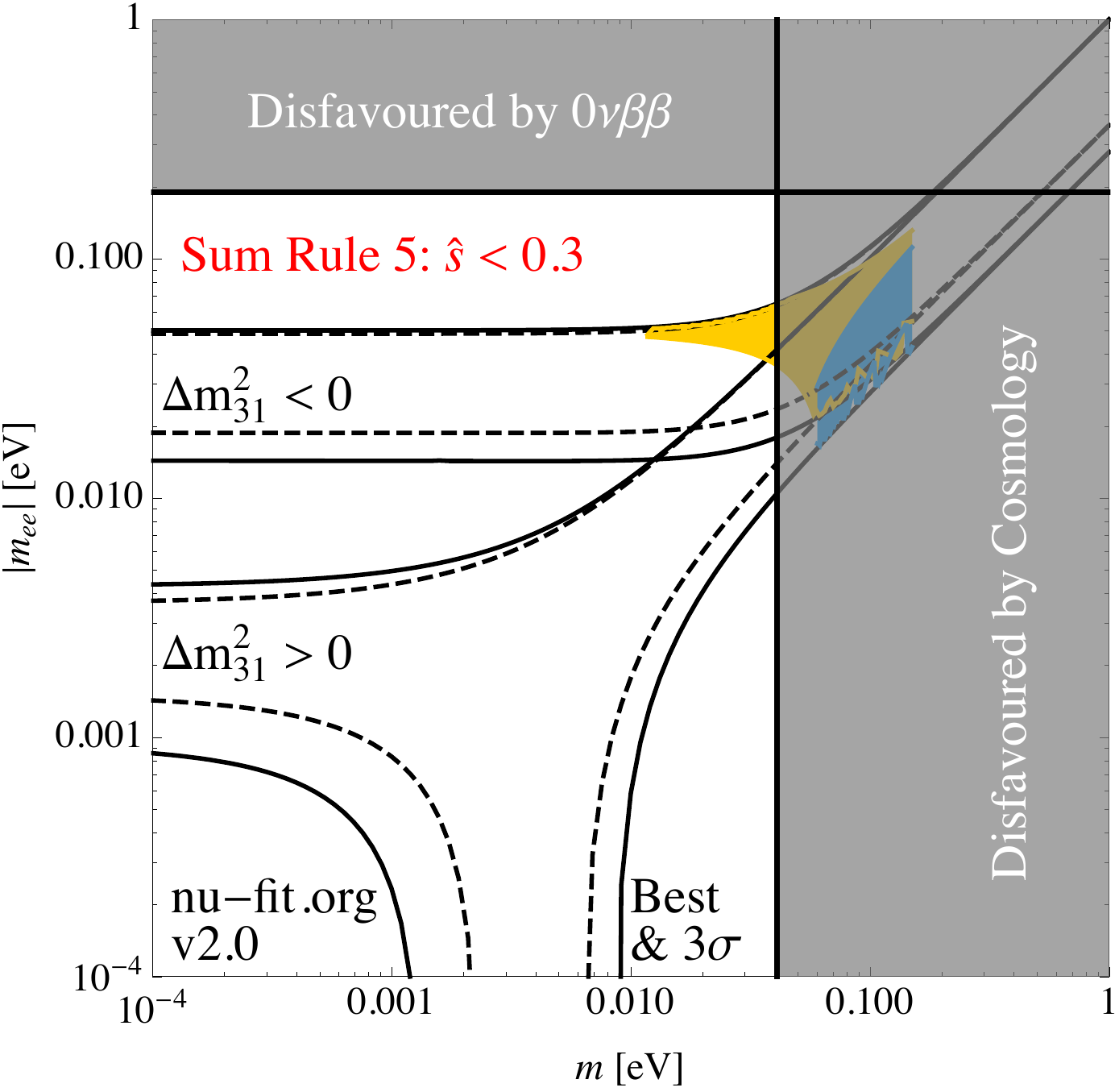}
\end{tabular}
\caption{\label{fig:SR5}Effective mass with SR~5 for $ \hat{s}=10^{-4}$, $0.1$, $0.3$.}
\end{figure}

\begin{figure}
\begin{tabular}{lll}
\hspace{-1cm}
\includegraphics[width=5.4cm]{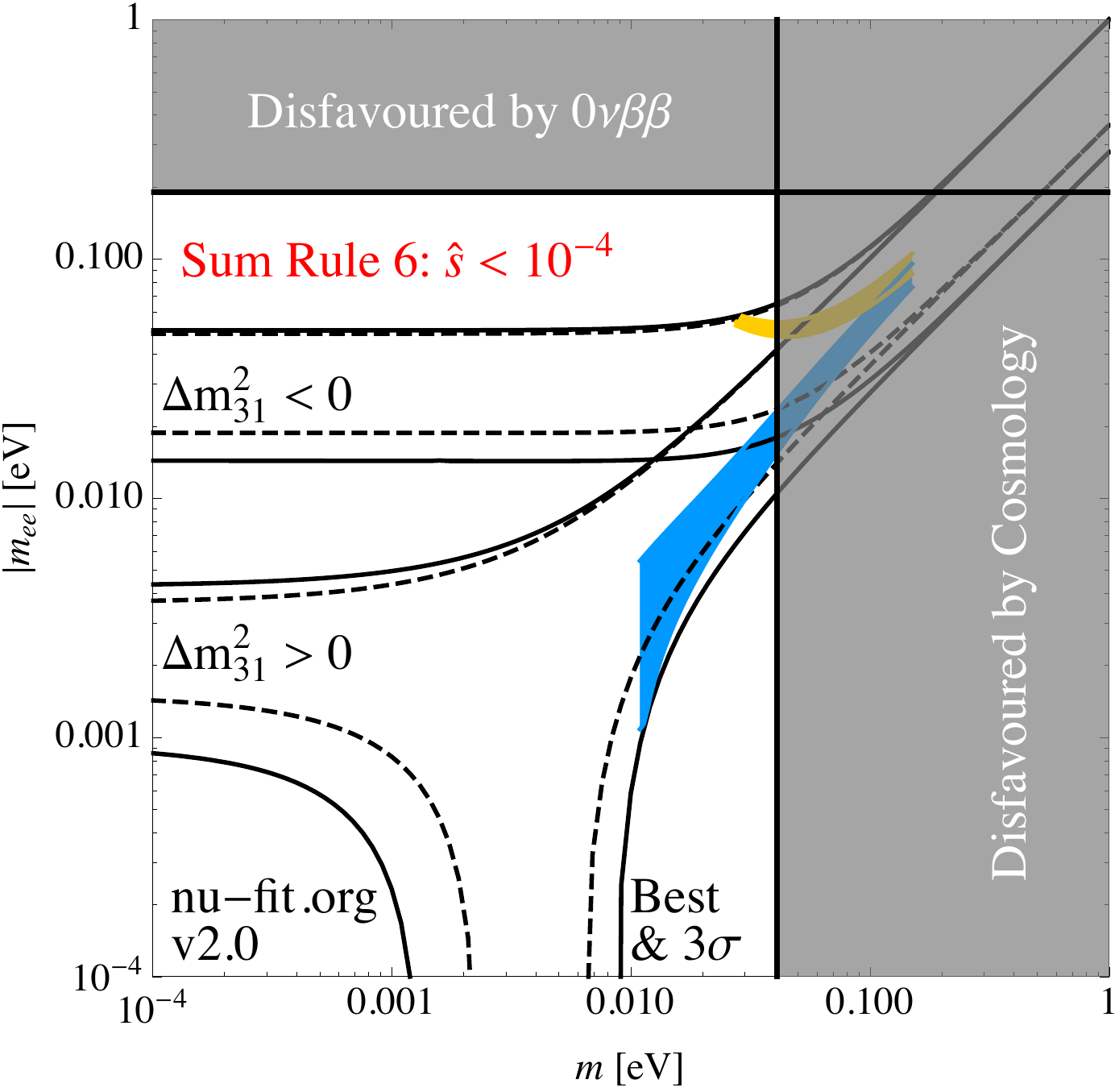} &
\includegraphics[width=5.4cm]{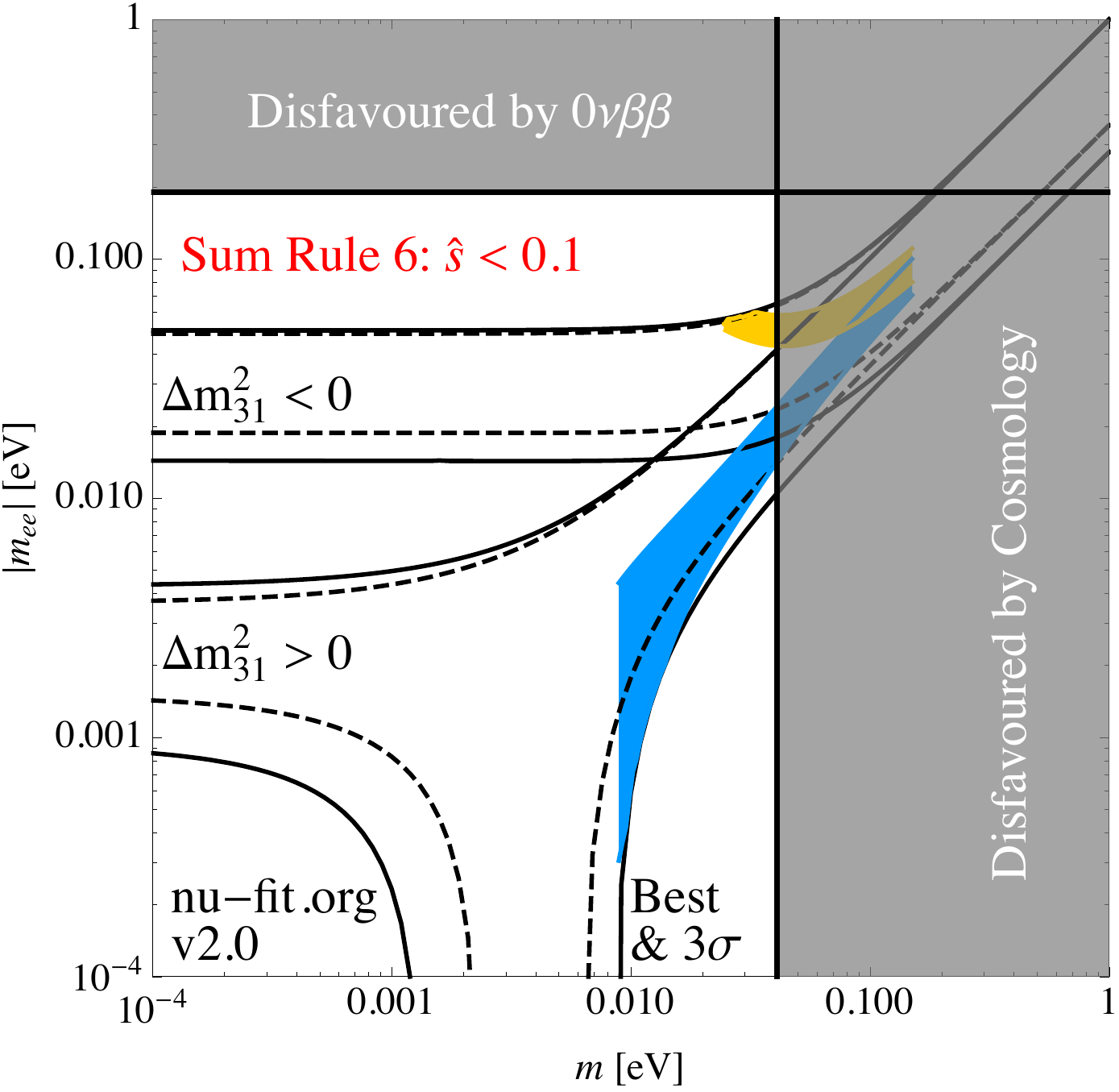} &
\includegraphics[width=5.4cm]{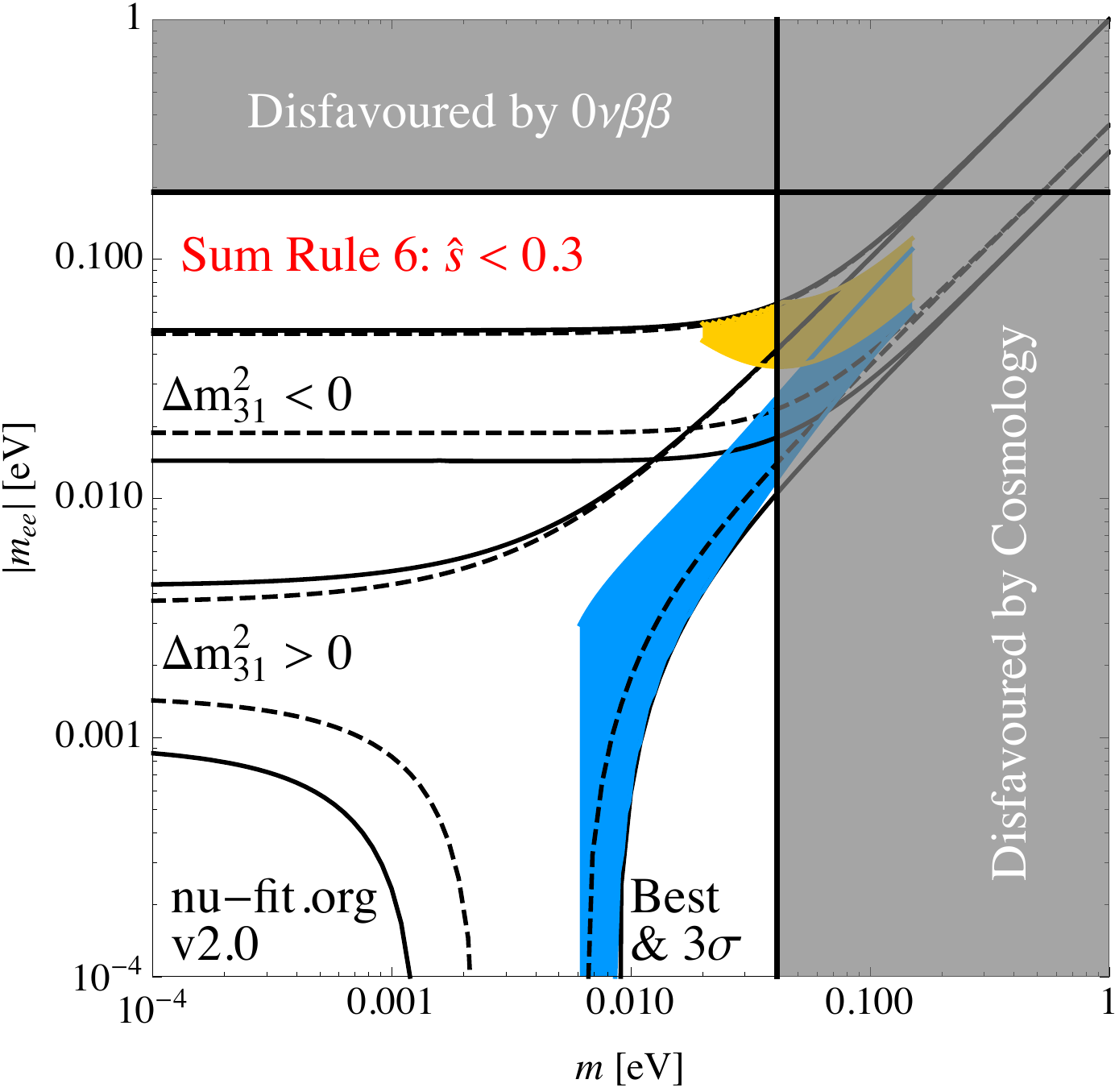}
\end{tabular}
\caption{\label{fig:SR6}Effective mass with SR~6 for $ \hat{s}=10^{-4}$, $0.1$, $0.3$.}
\end{figure}

\begin{figure}
\begin{tabular}{lll}
\hspace{-1cm}
\includegraphics[width=5.4cm]{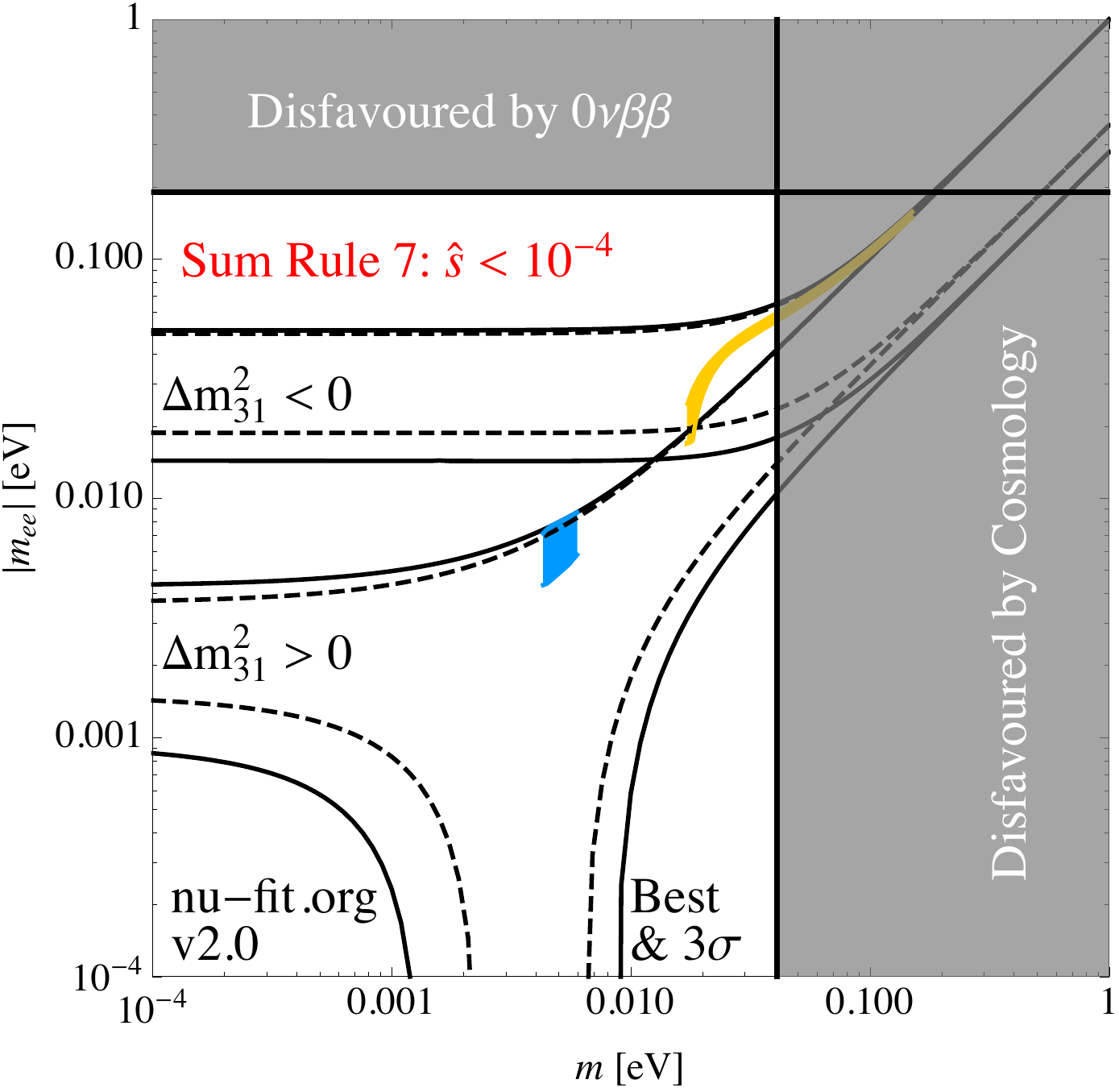} &
\includegraphics[width=5.4cm]{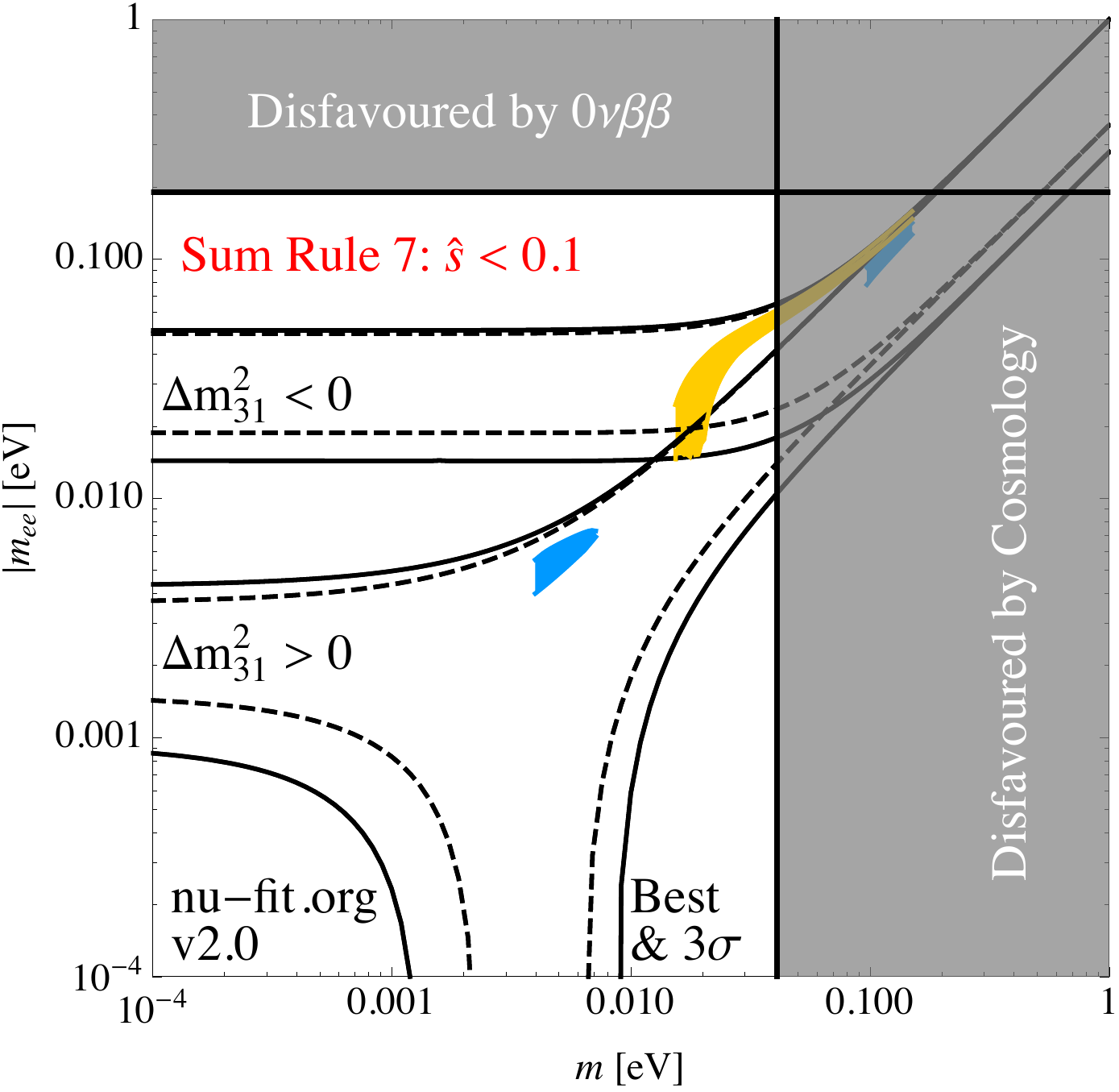} &
\includegraphics[width=5.4cm]{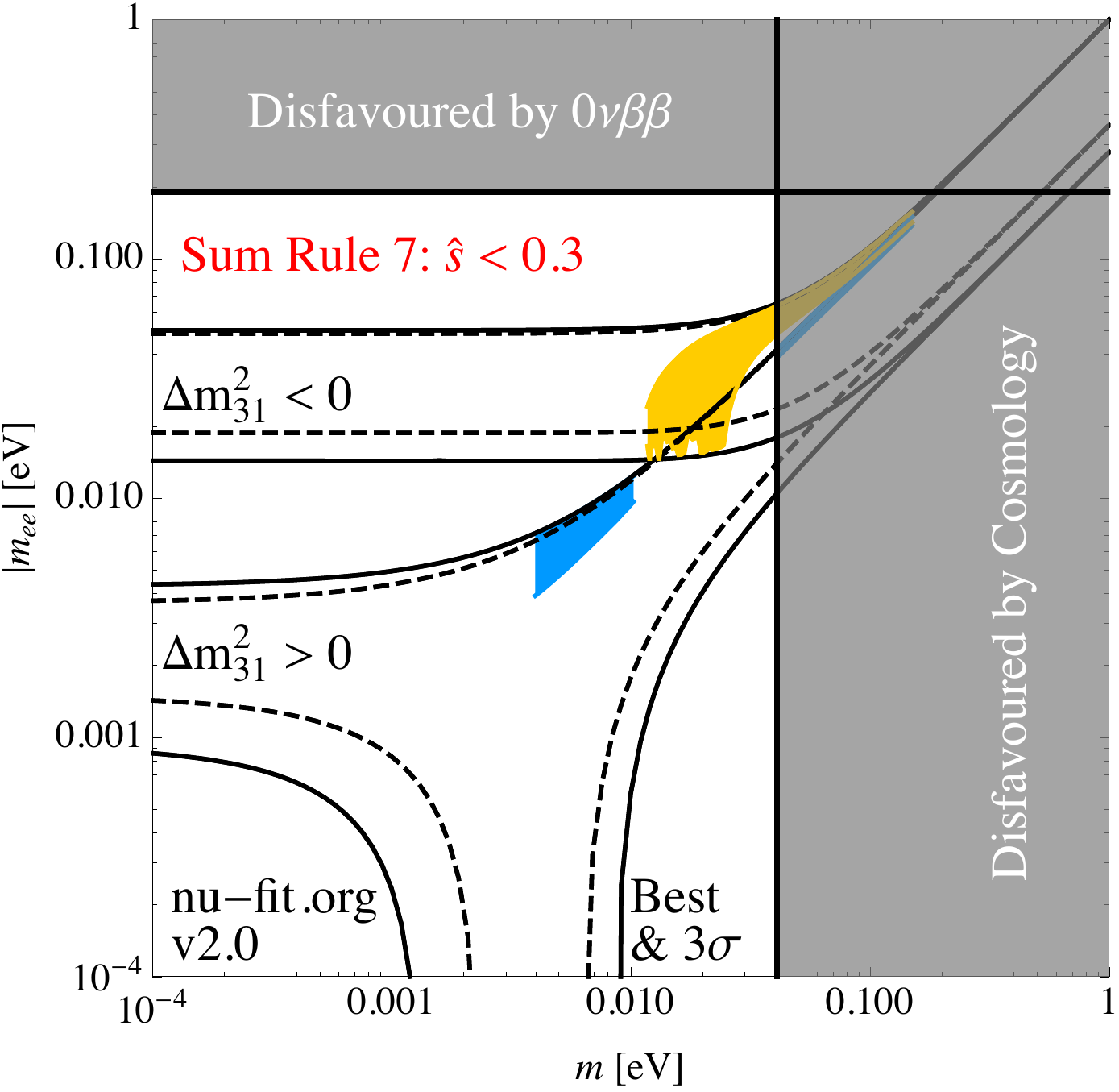}
\end{tabular}
\caption{\label{fig:SR7}Effective mass with SR~7 for $ \hat{s}=10^{-4}$, $0.1$, $0.3$.}
\end{figure}

\begin{figure}
\begin{tabular}{lll}
\hspace{-1cm}
\includegraphics[width=5.4cm]{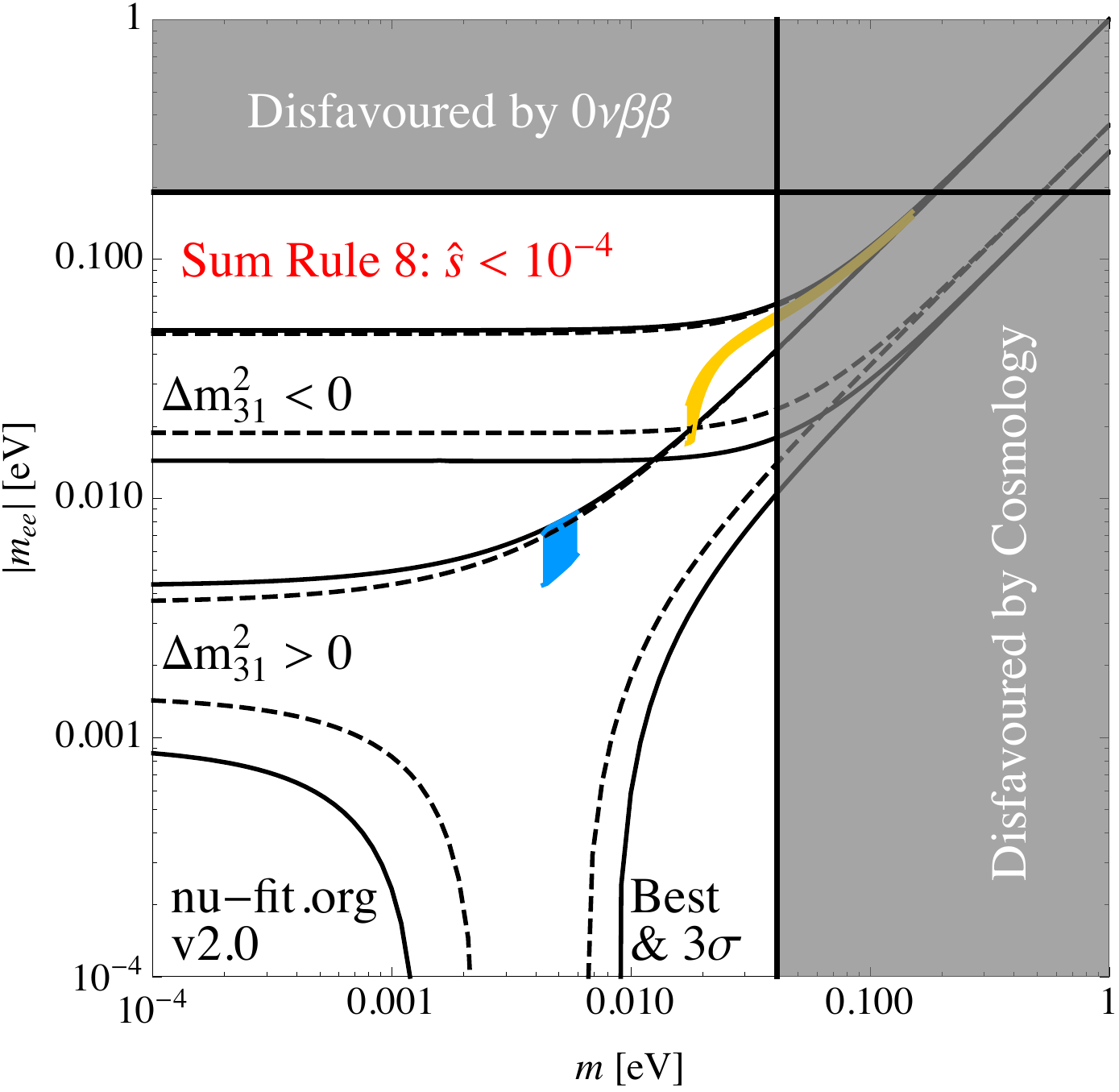} &
\includegraphics[width=5.4cm]{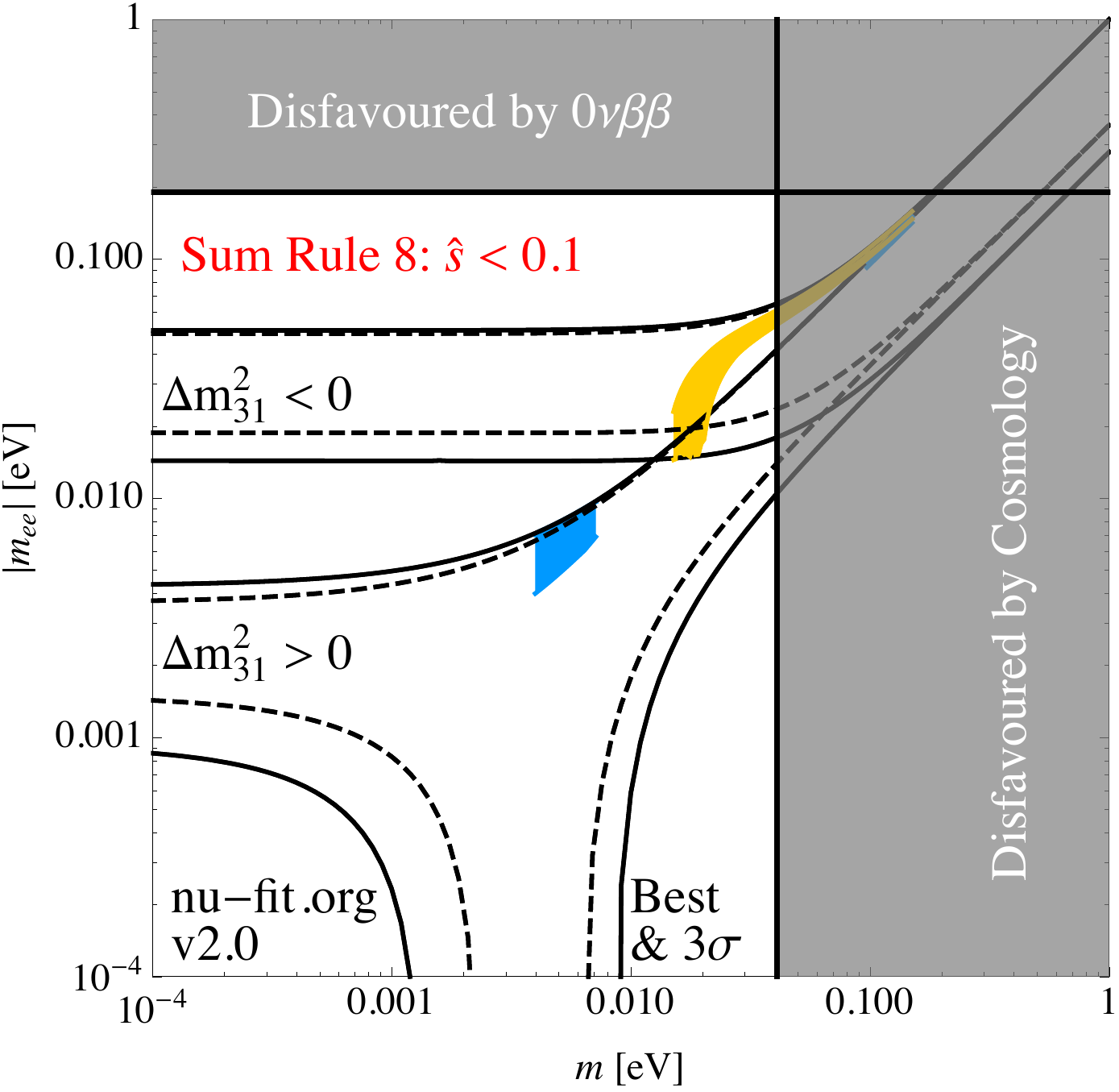} &
\includegraphics[width=5.4cm]{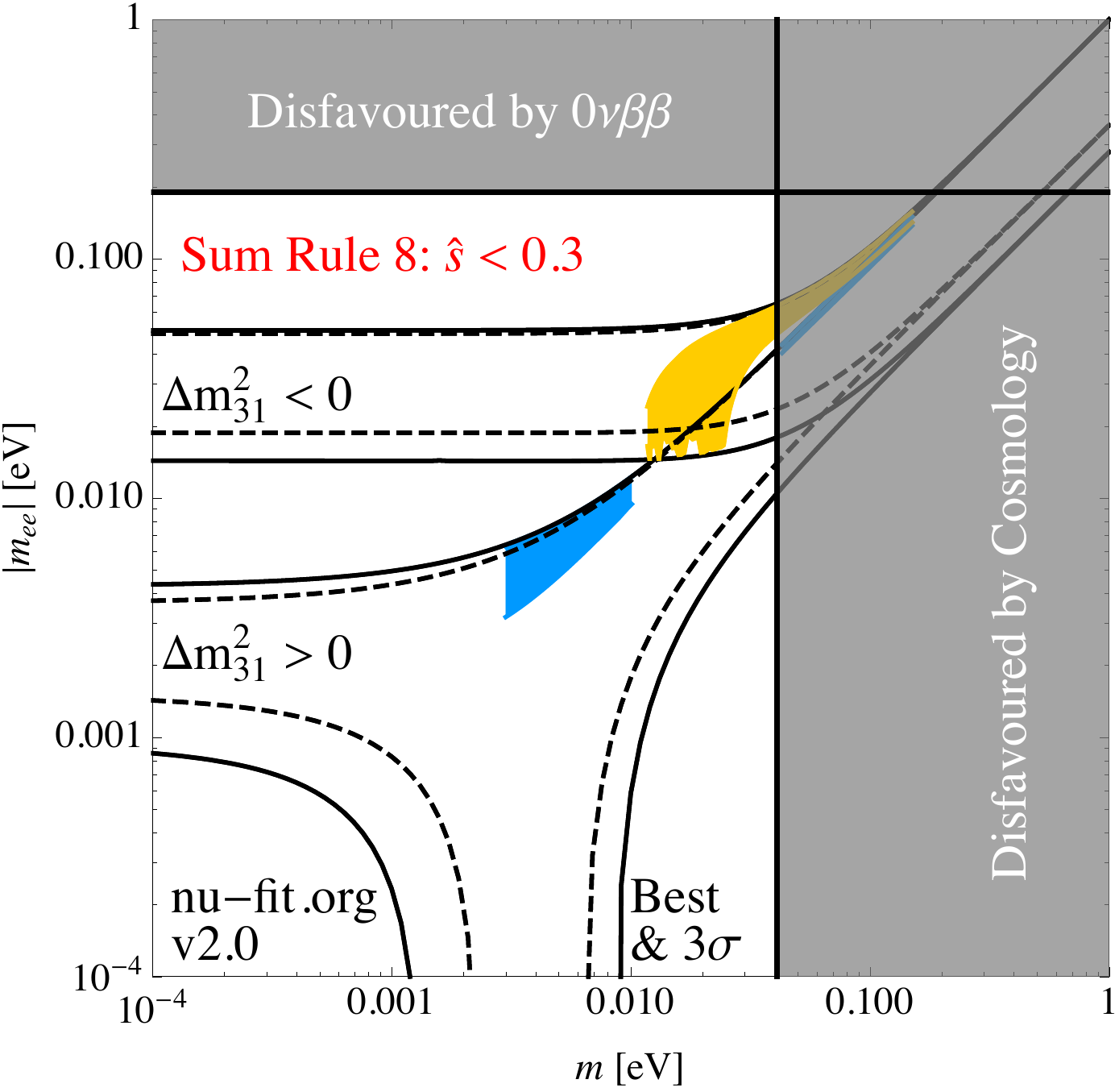}
\end{tabular}
\caption{\label{fig:SR8}Effective mass with SR~8 for $ \hat{s}=10^{-4}$, $0.1$, $0.3$.}
\end{figure}

\begin{figure}
\begin{tabular}{lll}
\hspace{-1cm}
\includegraphics[width=5.4cm]{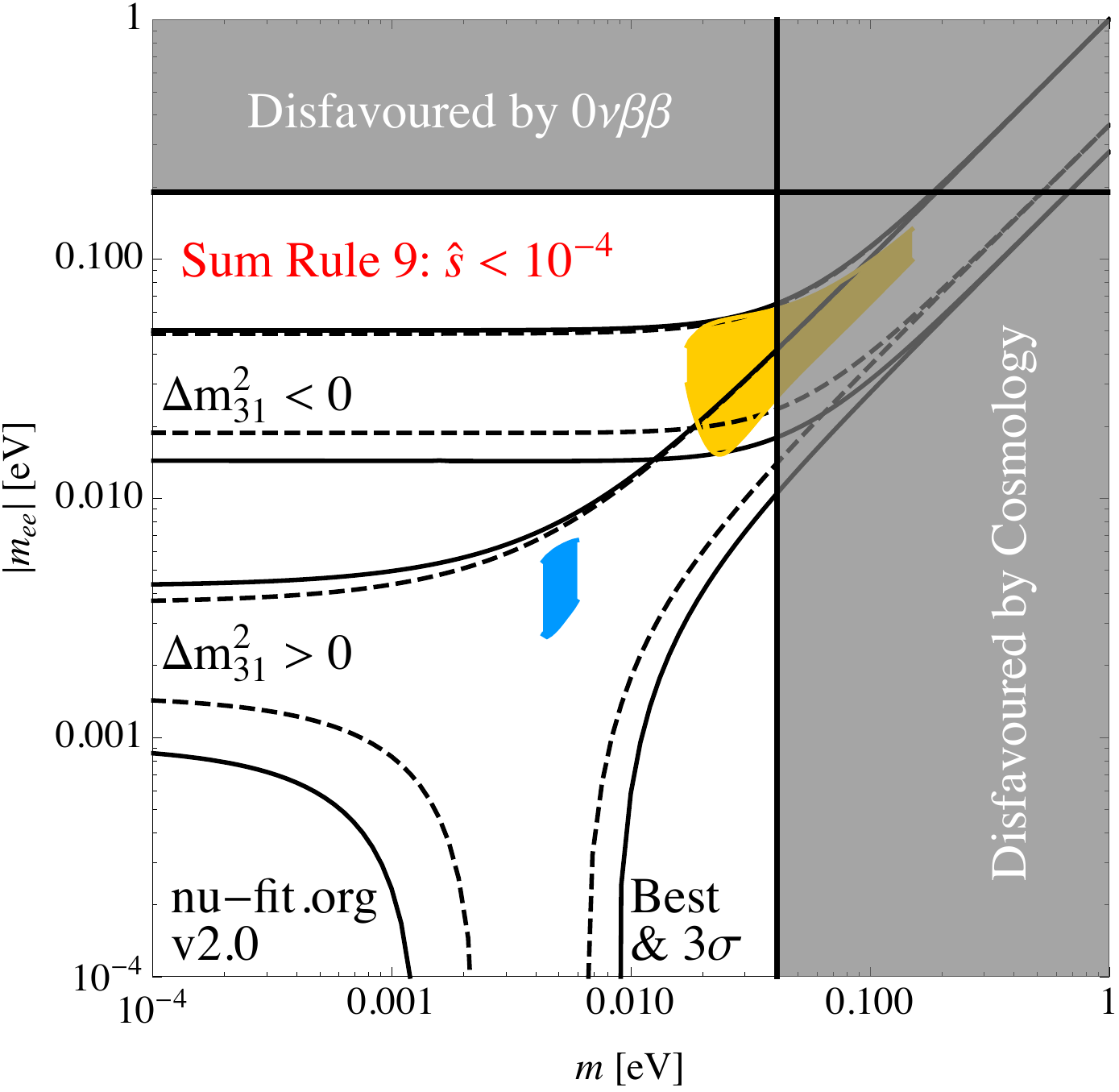} &
\includegraphics[width=5.4cm]{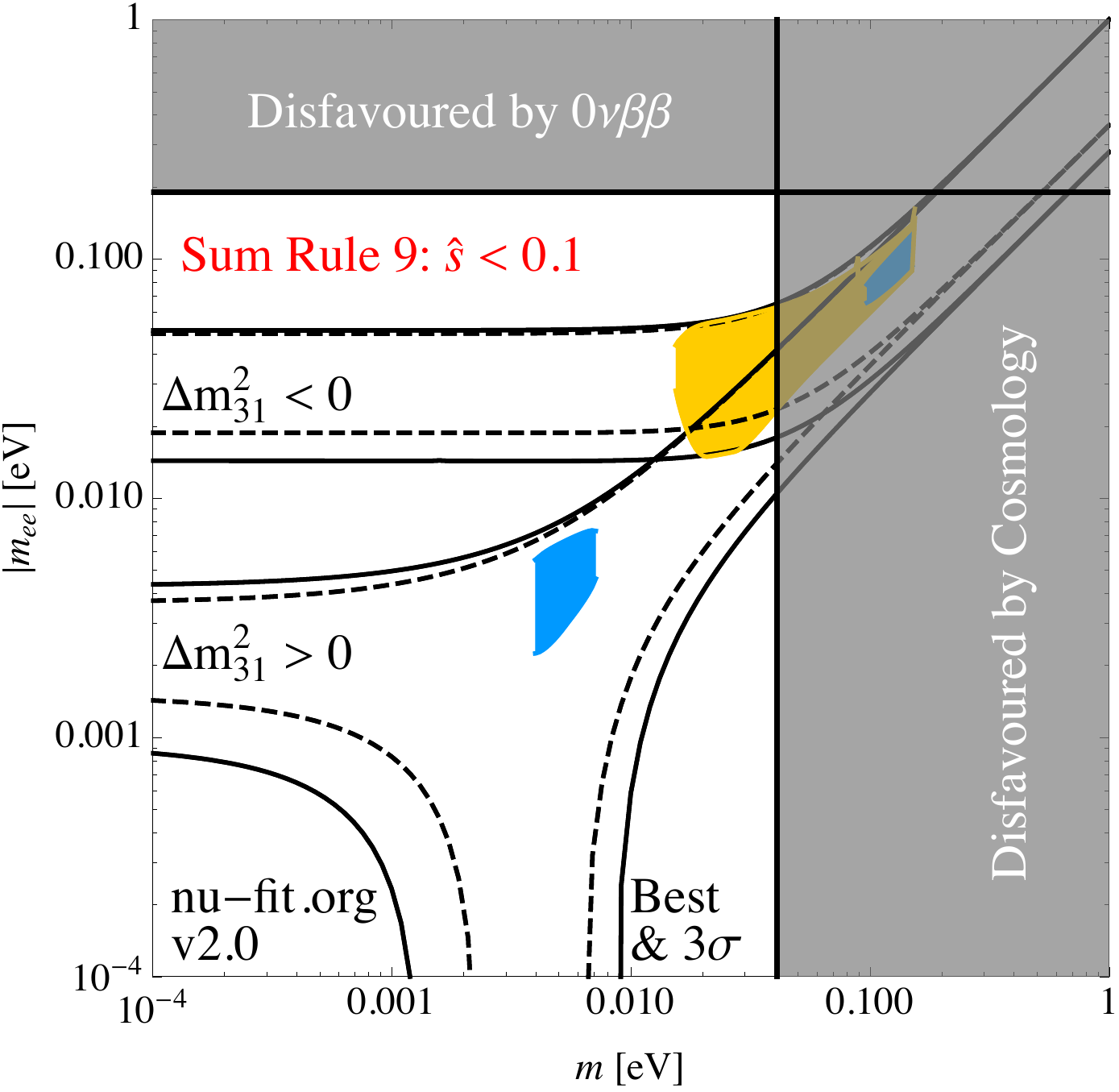} &
\includegraphics[width=5.4cm]{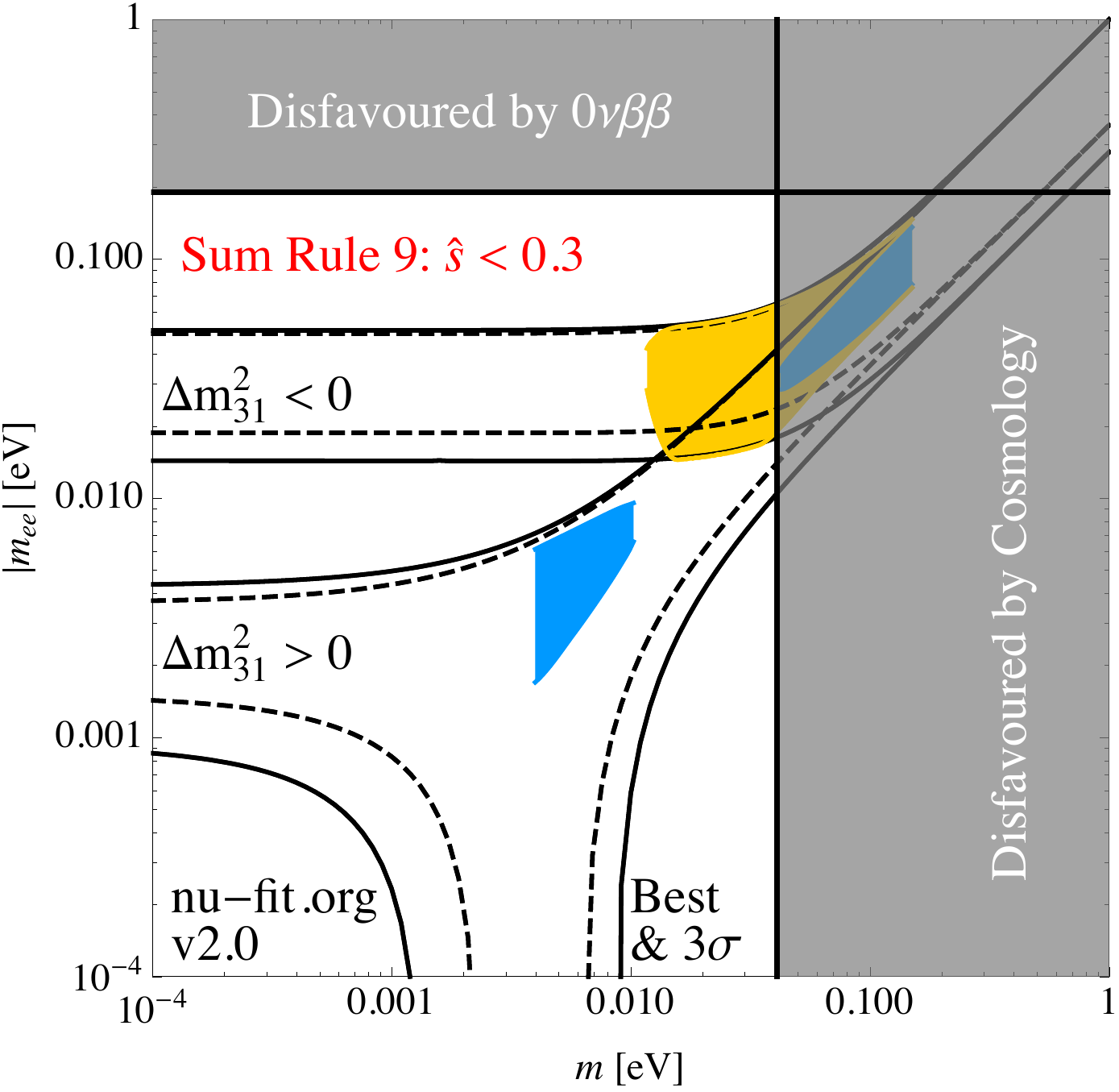}
\end{tabular}
\caption{\label{fig:SR9}Effective mass with SR~9 for $ \hat{s}=10^{-4}$, $0.1$, $0.3$.}
\end{figure}

\begin{figure}
\begin{tabular}{lll}
\hspace{-1cm}
\includegraphics[width=5.4cm]{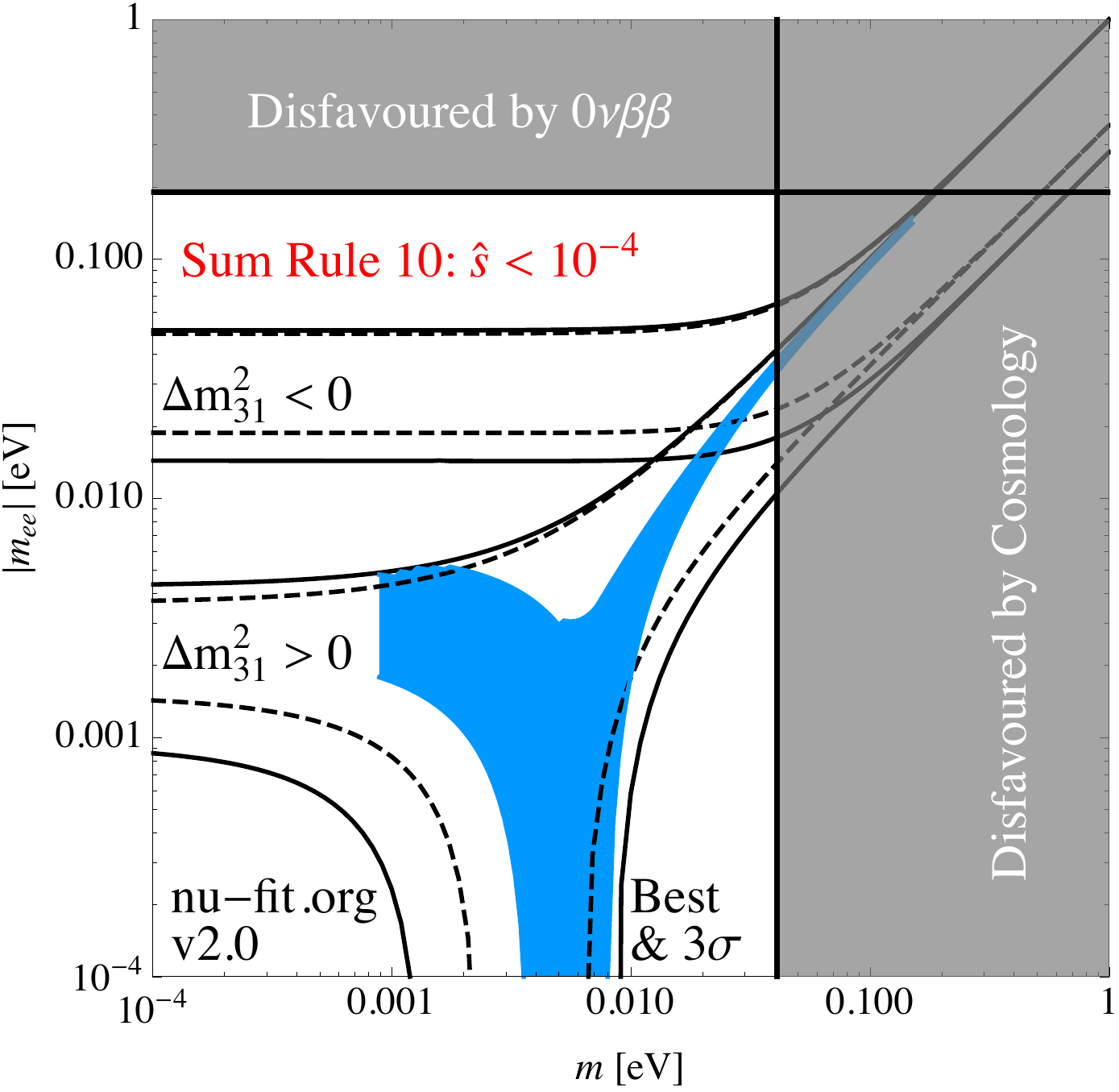} &
\includegraphics[width=5.4cm]{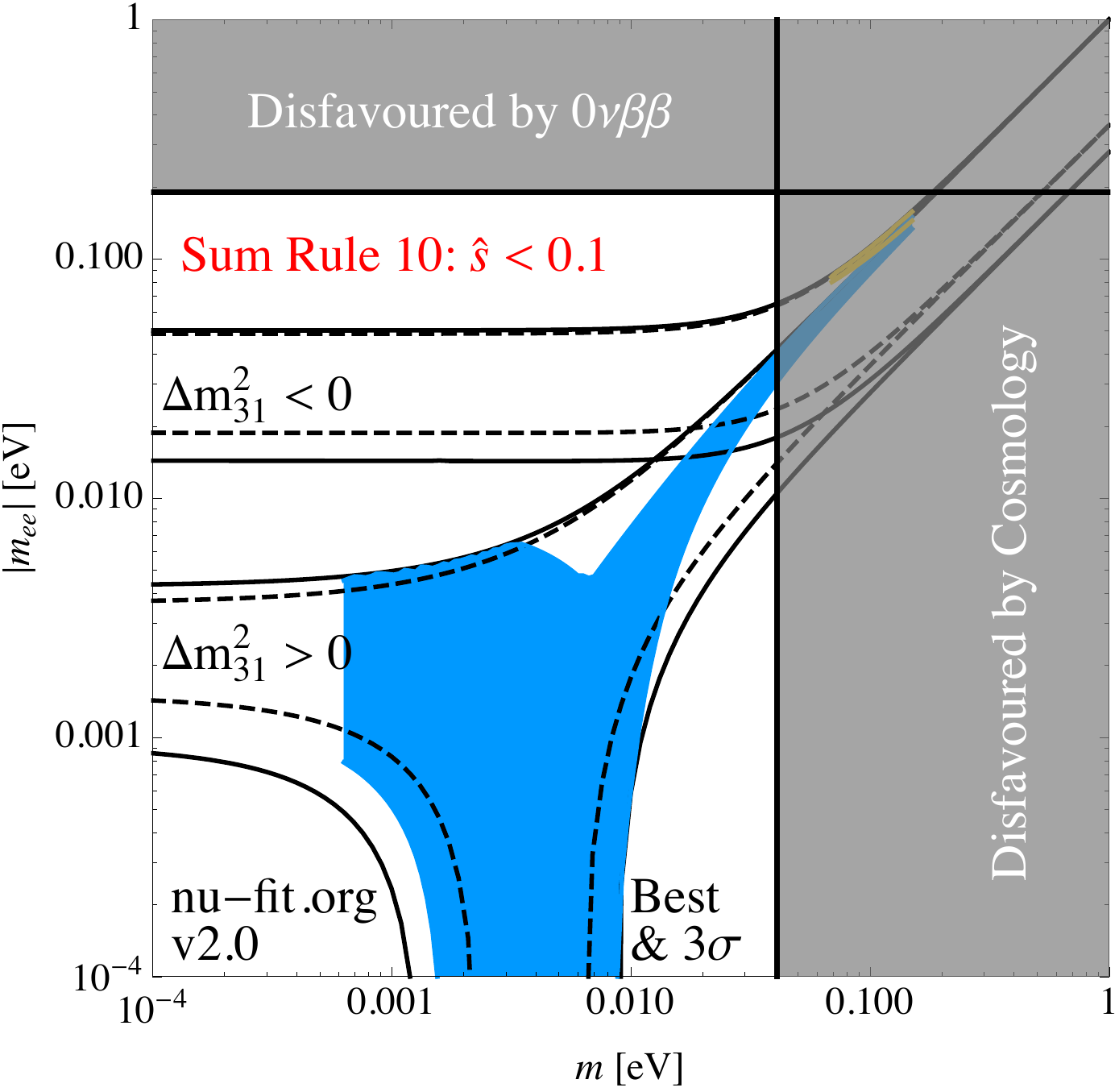} &
\includegraphics[width=5.4cm]{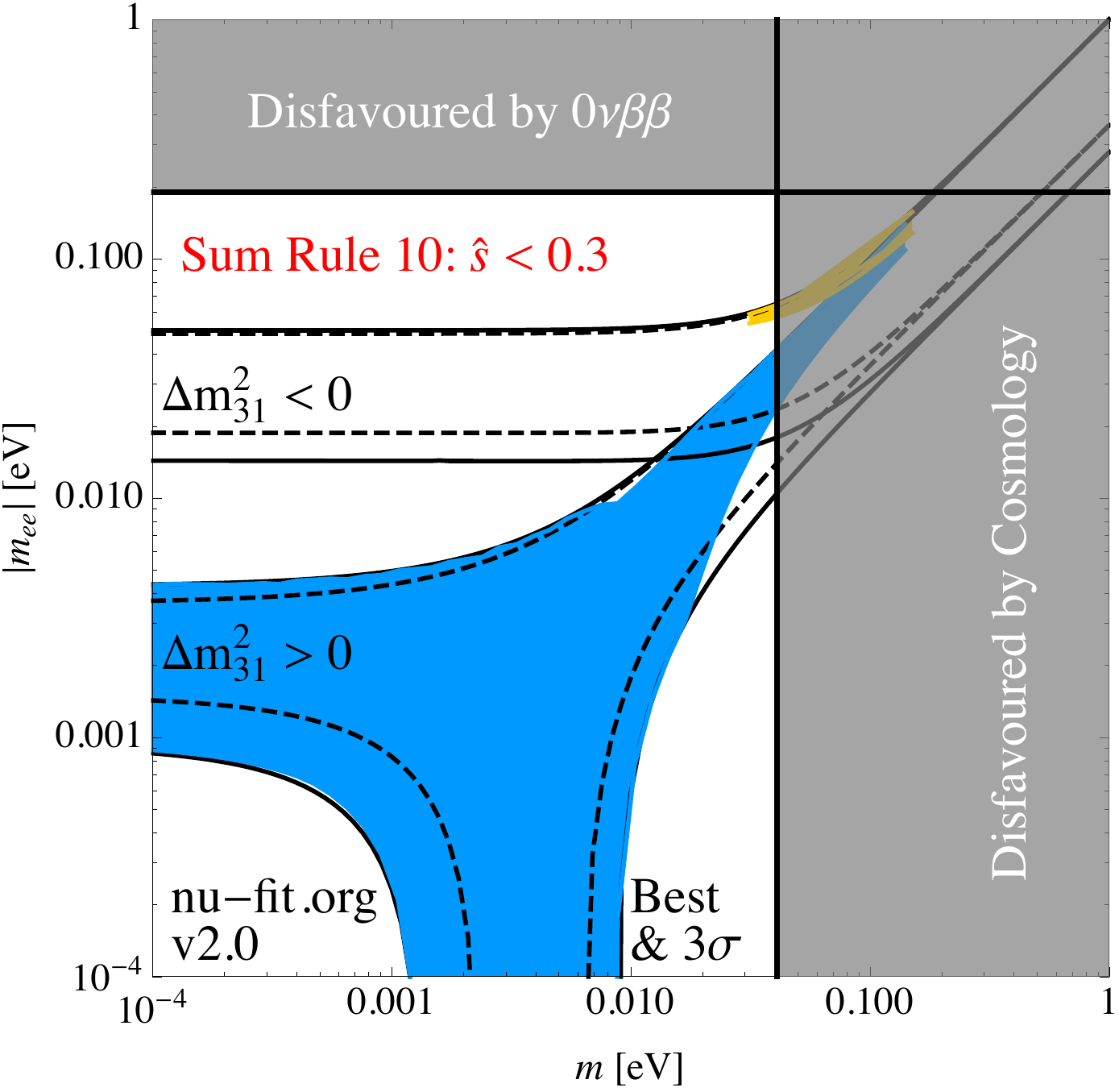}
\end{tabular}
\caption{\label{fig:SR10}Effective mass with SR~10 for $ \hat{s}=10^{-4}$, $0.1$, $0.3$.}
\end{figure}

\begin{figure}
\begin{tabular}{lll}
\hspace{-1cm}
\includegraphics[width=5.4cm]{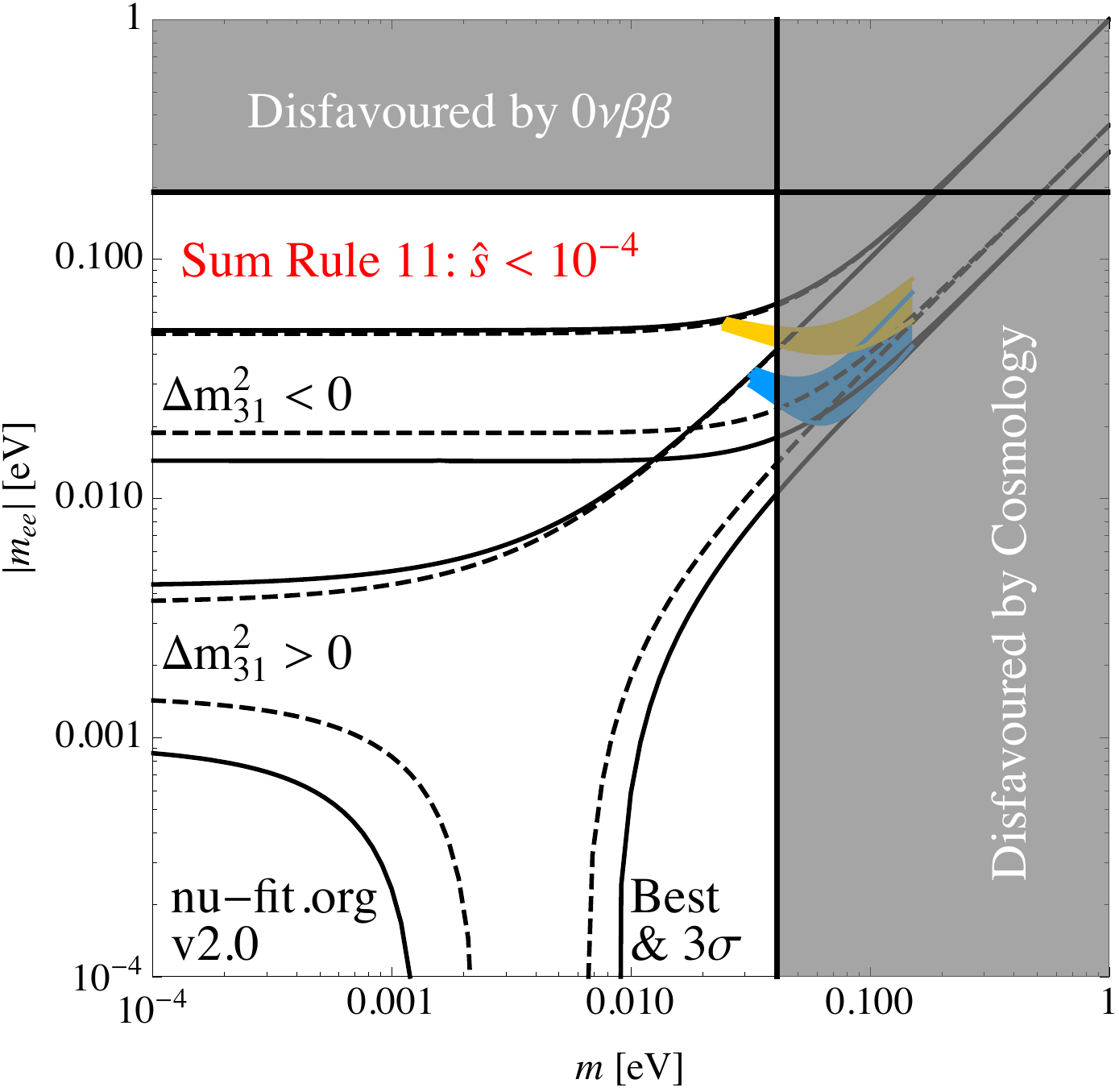} &
\includegraphics[width=5.4cm]{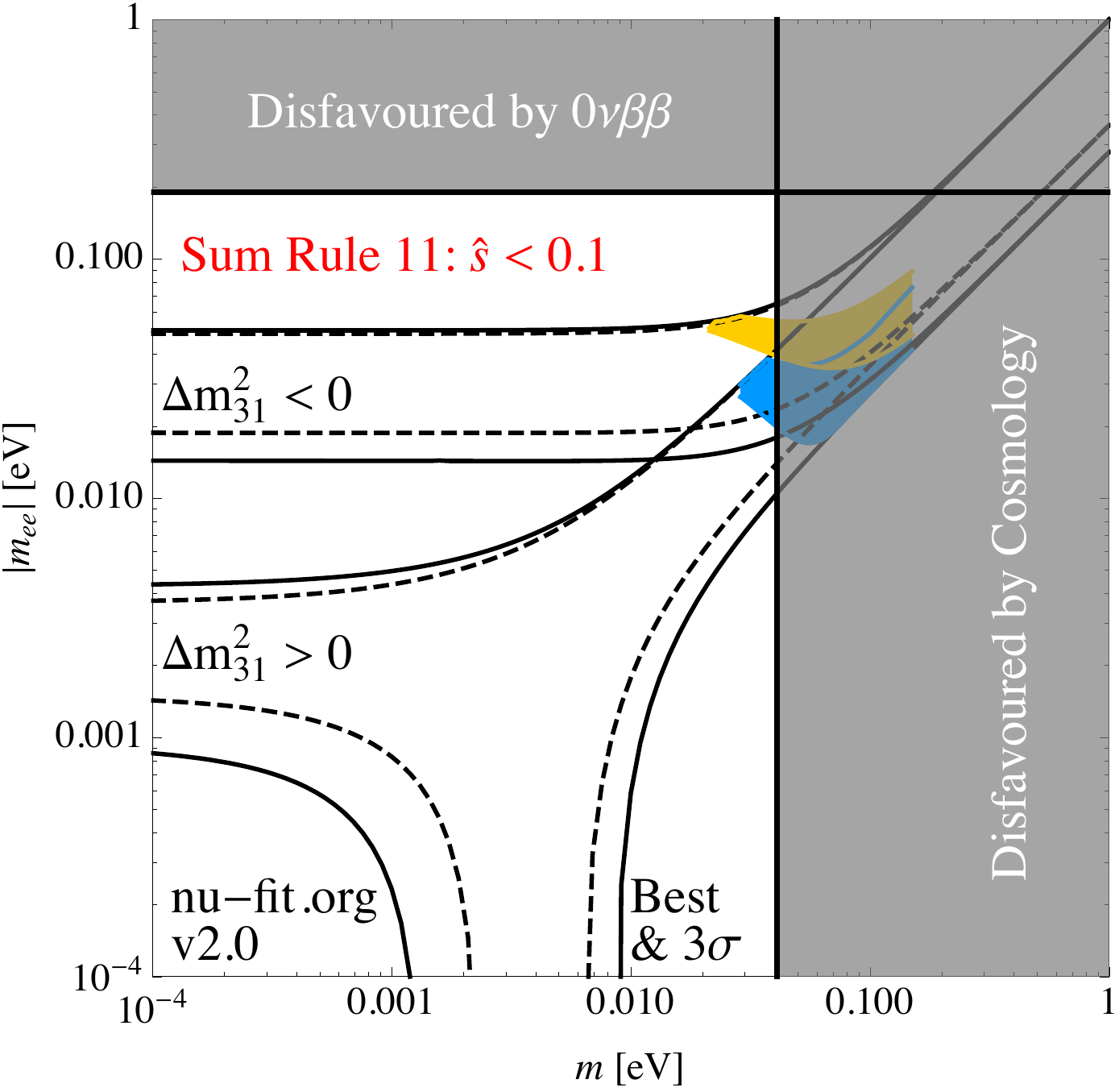} &
\includegraphics[width=5.4cm]{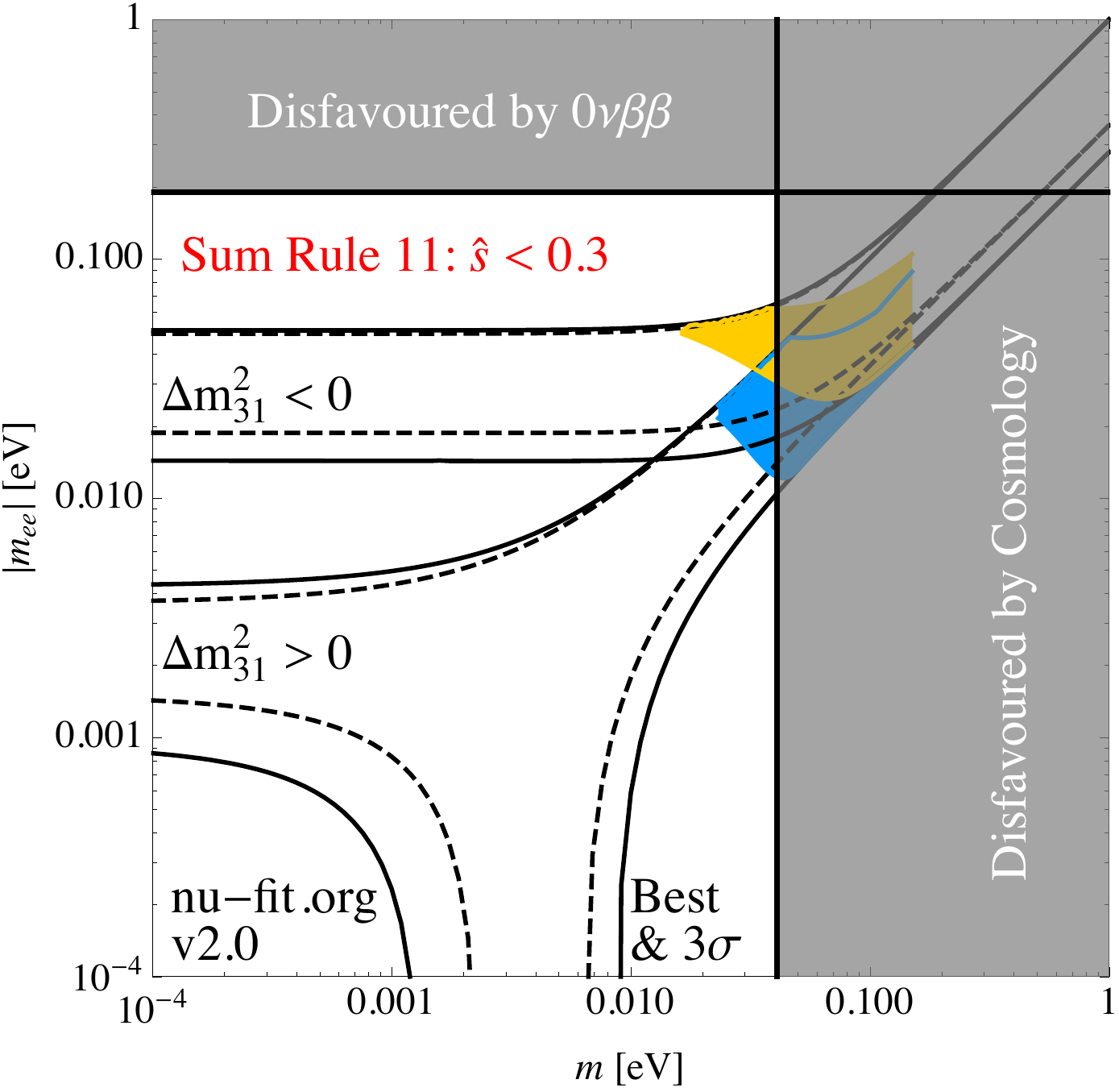}
\end{tabular}
\caption{\label{fig:SR11}Effective mass with SR~11 for $ \hat{s}=10^{-4}$, $0.1$, $0.3$.}
\end{figure}

\begin{figure}
\begin{tabular}{lll}
\hspace{-1cm}
\includegraphics[width=5.4cm]{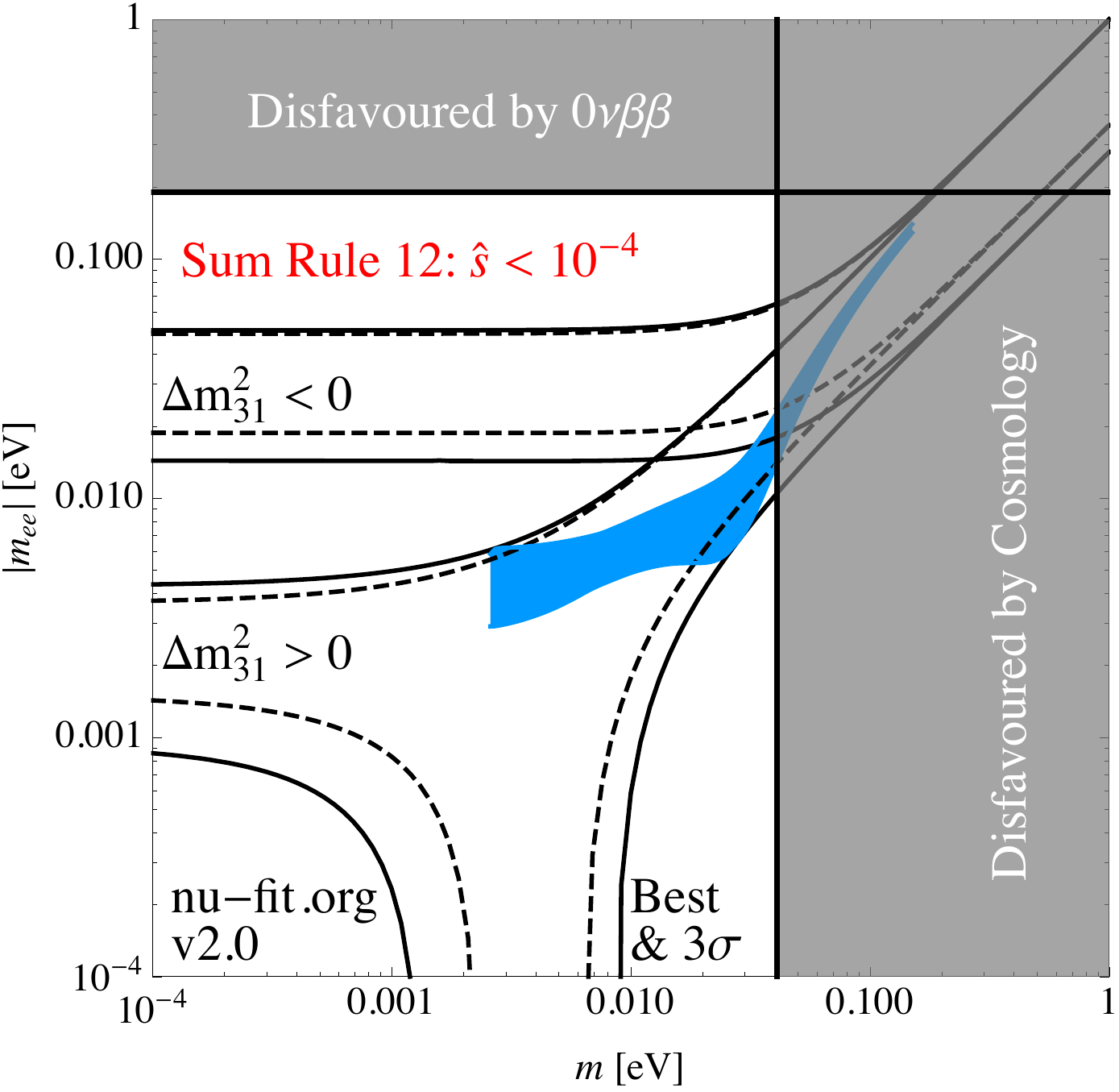} &
\includegraphics[width=5.4cm]{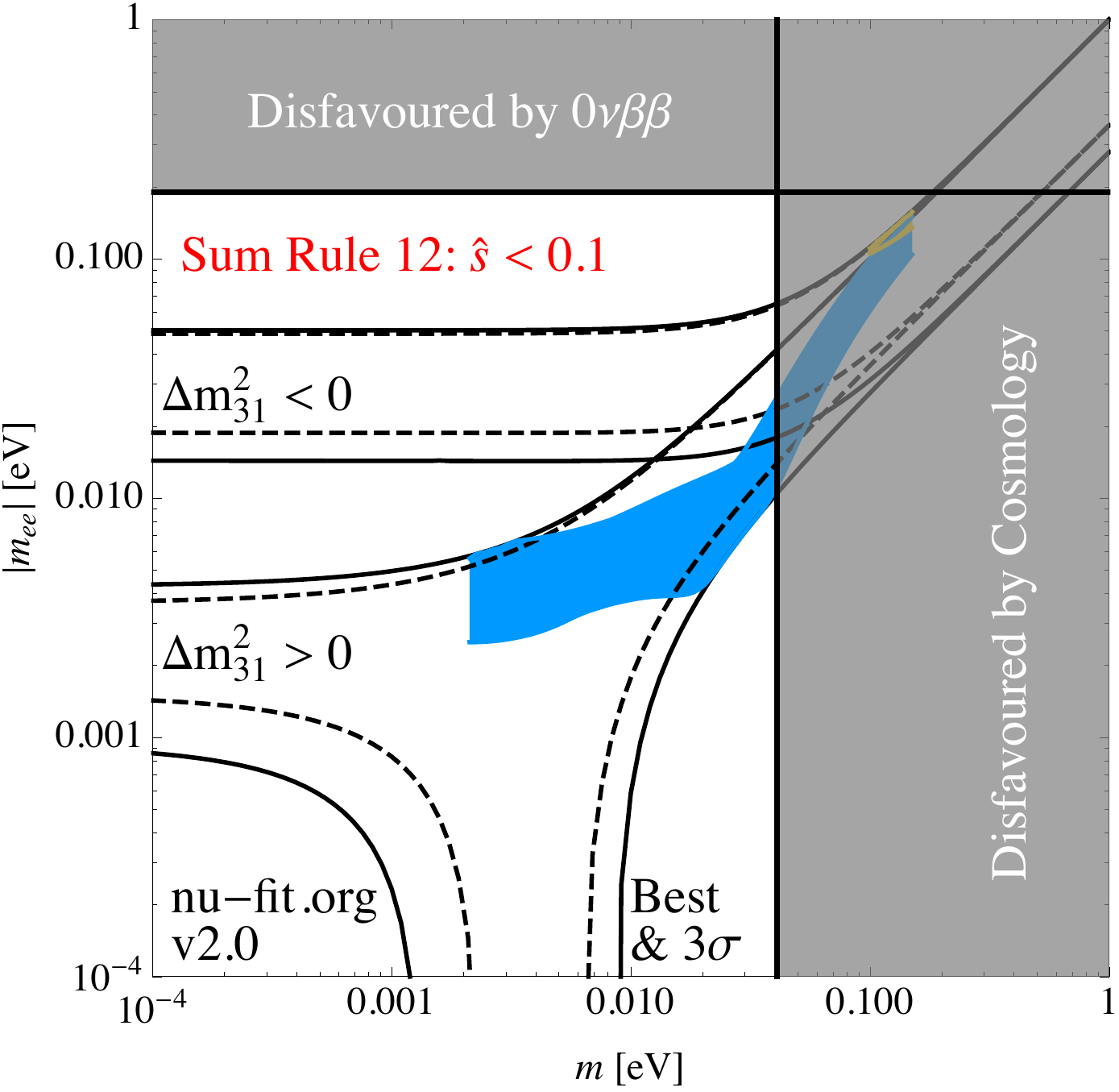} &
\includegraphics[width=5.4cm]{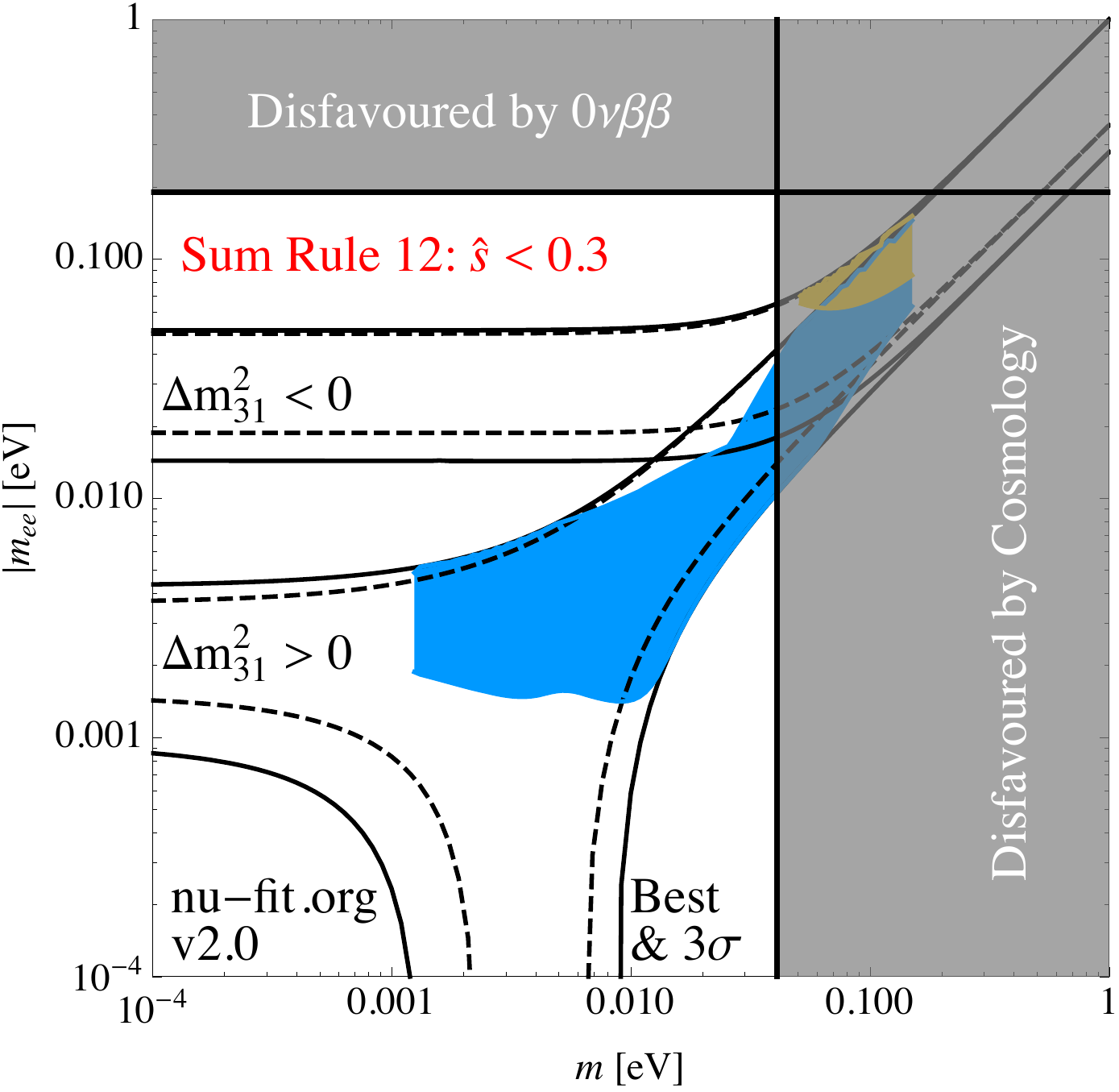}
\end{tabular}
\caption{\label{fig:SR12}Effective mass with SR~12 for $ \hat{s}=10^{-4}$, $0.1$, $0.3$.}
\end{figure}

Finally, our main predictions are those for the effective mass
\begin{align}
 |m_{ee}| = \left|m_{1} U_{e1}^{2}+m_{2} U_{e2}^{2}+m_{3} U_{e3}^{2}\right|=\left| m_{1}c_{12}^{2}c_{13}^{2}\text{e}^{-\text{i}\phi_{1}}+m_{2}s_{12}^{2}c_{13}^{2}\text{e}^{-\text{i}\phi_{2}}+m_{3}s_{13}^{2}\text{e}^{-2 \text{i} \delta}\right|
 \end{align}
  as probed in neutrinoless double beta decay, see, e.g.,\ Refs.~\cite{mee-references} for detailed discussions on this quantity. We have, for each SR, numerically scanned the parameter space and we have derived the allowed regions for both normal and inverted mass orderings, see Figs.~\ref{fig:SR1} to~\ref{fig:SR12}. For each SR, we have investigated the following three cases depending on the size of $ \hat s$, as defined in Eq.~\eqref{eq:dsh}:
\begin{enumerate}

\item $ \hat s < 10^{-4}$ (\emph{left panels}): This case basically means that the SR is taken to be exact. All plots perfectly match the unperturbed SR predictions, as presented in Refs.~\cite{King:2013psa,Gehrlein:2015ena}.

\item $ \hat s < 0.1$ (\emph{middle panels}): This corresponds to a considerable perturbation of each SR. In particular, a correction of this size will reveal that, in cases where one mass ordering is forbidden for an exact SR, this ordering will open up due to the corrections.

\item $ \hat s < 0.3$ (\emph{right panels}): Here, the SRs are even less exact. This case is more or less the limiting case of what can be described by the approach followed in this work, given that we ultimately rely on a perturbative expansion. This case should in particular include the RGE corrections, as discussed in Ref.~\cite{Gehrlein:2015ena}, as long as they can be described accurately as small perturbations.

\end{enumerate}

Let us discuss the results for the different SRs one by one, with a particular focus on how the allowed regions for the perturbed SRs compare to those derived from the RGE-corrected SRs (cf.\ Ref.~\cite{Gehrlein:2015ena}, in which reference Sec.~4.X always contains the plots and the discussion on SR~X). The parameters for the respective SRs can be read off from Tab.~\ref{tab:overview_SR}.

Starting with SR~1 in Fig.~\ref{fig:SR1}, it is immediately visible from the left panel that, indeed, a very small deviation from an exact SR, $ \hat s < 10^{-4}$, practically does not change the prediction compared to that of the exact SR. This observation will hold true for all SRs, as to be expected. If we increase $ \hat s$, it is visible that, for NO, the allowed regions visibly increase. In particular, the lowest allowed value for $m$ evolves as $(0.027, 0.023, 0.017)$~eV for $ \hat s < (10^{-4}, 0.1, 0.3)$, while at the same time the lowest possible value for the effective mass $|m_{ee}|$ changes as $(0.025, 0.022, 0.016)$~eV. In particular, with increasing $ \hat s$, the allowed range for the effective mass becomes even slightly larger than if only RGE corrections were included (cf.\ Sec.~4.1 in Ref.~\cite{Gehrlein:2015ena}), where the minimal value for the effective mass would be at $0.026$~eV.

For IO, in turn, while the allowed region also increased with increasing $ \hat s$, the main difference is that the area does not grow sufficiently large as to cover the complete allowed region obtained when taking into account RGE-corrections. As already hinted, this is a reflection of the RGE corrections not always staying within the perturbative range for $ \hat s$, which is what we are considering in this work.

As shown in~\cite{Gehrlein:2015ena}, we encounter a parametric enhancement for the RGE effects for a small mass scale, because the correction is proportional to the inverse mass scale. Since we avoid parametric enhancements of the corrections in our parametrisation from Eqs.~\eqref{eq:s} to~\eqref{eq:ds} we do not obtain the same result as in~\cite{Gehrlein:2015ena}. Nevertheless, given that even with RGE-corrections the change in the prediction was less than dramatic, in particular when taking into account the nuclear uncertainties, we can again conclude that the predictions of the SRs are relatively robust compared to corrections.
 
Let us press on and jump to SR~2, see Fig.~\ref{fig:SR2}. Starting with NO, the qualitative change with increasing $ \hat s$ is similar to that with increasingly strong RGE-corrections, however, the allowed regions are different (and possibly slightly larger for the RGE-corrections). For $ \hat s < (10^{-4}, 0.1, 0.3)$, the lowest $m$ evolves as $(0.015, 0.013, 0.009)$~eV and the lower bound on the effective mass $|m_{ee}|$ as $(0.014, 0.012, 0.008)$~eV (compared to $0.016$~eV and $0.015$~eV for the RGE corrections). For IO, something interesting happens. As is visible from the left panel of Fig.~\ref{fig:SR2}, this mass ordering is not allowed for an exact SR. As noted in Sec.~\ref{sec:forbidden}, this behaviour remained true for the RGE corrections. However, with a more general correction to the SR, as implied by an approximate SR, the IO starts opening up. While it is hardly visible for $ \hat s < 0.1$, it is easy to spot for $\hat s < 0.3$. However, while the otherwise forbidden ordering does in principle open up, the resulting predicted region unfortunately lies in the part of the plot that is in any case strongly disfavoured by cosmology, cf.\ grey rectangle on the right in the plots. Thus, in essence, the prediction of IO being forbidden does not change.

Similarly to the previous case, for SR~3 the allowed area for NO also broadens, while for IO a small regions opens up which is, however, strongly disfavoured, cf.\ Fig.~\ref{fig:SR3}. However, the difference is that, for large enough $ \hat{s}$, the allowed region for NO may enter the ``tube'' in which cancellations inside $|m_{ee}|$ to practically zero are possible. This implies that, for $ \hat s < (10^{-4}, 0.1, 0.3)$, while the smallest neutrino mass eigenvalue $m$ is only reduced as $(0.015, 0.013, 0.009)$~eV, the minimum value of the effective mass $|m_{ee}|$ changes more dramatically, $(0.0030, 0.0020, 0.00029)$~eV, compared to a considerably larger lowest value of $0.0036$~eV for RGE corrections only.

For SR~4, cf.\ Fig.~\ref{fig:SR4}, no dramatic changes are visible. The allowed region is a bit broader for strong RGE corrections, but the minimal value of the effective mass is in any case $0.015$~eV, the lowest value at all possible for IO. The minimal value for the smallest neutrino mass eigenvalue $m$ decreases more strongly than for RGE corrections only, however, such small values are in any case not accessible by experiments. For NO, a small area opens up which is allowed by the SR as such, but again it is located in the disfavoured region of the plot.
 
Sum rule~5 exhibits once more the characteristic behaviour, see Fig.~\ref{fig:SR5}. For IO, the allowed region broadens even more strongly than for RGE corrections only, leading to a smallest neutrino mass $m$ of $(0.024, 0.020, 0.012)$~eV and a minimum effective mass $|m_{ee}|$ of $(0.051, 0.050, 0.047)$~eV for $ \hat s < (10^{-4}, 0.1, 0.3)$. For NO, the small region opening up is again disfavoured.

Sum rule~6 -- see Fig.~\ref{fig:SR6} -- looks similar to SR~5 for IO and to SR~3 for NO, in both cases fully including the regions predicted by the RGE corrections. For $ \hat s < (10^{-4}, 0.1, 0.3)$, the smallest neutrino mass $m$ and the minimum effective mass $|m_{ee}|$ change as $(0.028, 0.025, 0.020)$~eV and $(0.053, 0.051, 0.045)$~eV, respectively, for IO. The corresponding values for NO are, in turn, $(0.011, 0.0090, 0.0061)$~eV and $(0.0011, 0.00031, 3.5\cdot 10^{-12})$~eV, where the latter value simply indicates that a full cancellation of the effective mass can happen.

Figs.~\ref{fig:SR7} and~\ref{fig:SR8} look nearly identical and quite interesting, too, as several changes happen. Starting with IO, the allowed region again broadens, however, while a new edgy corner had appeared for the RGE-corrections, see Sec.~4.7 for SR~7 in Ref.~\cite{Gehrlein:2015ena}, in the case of approximate SRs the band simply broadens. The lowest value possible for the effective mass quickly reaches its absolute minimum and then cannot change anymore, $(0.017, 0.015, 0.015)$~eV for $ \hat s < (10^{-4}, 0.1, 0.3)$, while the smallest neutrino mass $m$ varies as $(0.017, 0.015, 0.012)$~eV.

For NO, in turn, the changes are more dramatic. First of all, the allowed patch considerably grows for $ \hat s < (10^{-4}, 0.1, 0.3)$, such that the minimum [maximum] mass eigenvalue $m$ varies as $(0.0043, 0.0040, 0.0040)$~eV [$(0.0060, 0.0071, 0.010)$~eV], while the minimum [maximum] effective mass changes as $(0.0043, 0.0040, 0.0039)$~eV [$(0.0087, 0.0073, 0.012)$~eV]. Furthermore, for large enough mass $m$, a second (disjoint) allowed region opens up for NO in the quasidegenerate mass region for $ \hat s < (0.1, 0.3)$. However, again this new addition is located in that part of the parameter space that is strongly disfavoured by cosmology.

Sum rule~9, cf.\ Fig.~\ref{fig:SR9},  hardly change at all for RGE corrections only. And also for approximate SRs, hardly any significant change is visible for IO. Already for the exact SR, the entire range is allowed for $|m_{ee}|$, while the smallest mass $m$ changes slightly, $(0.017, 0.015, 0.012)$~eV, for $ \hat s < (10^{-4}, 0.1, 0.3)$. However, for NO,  the ranges for $m$ change as $(0.0043 ... 0.0060, 0.0040 ... 0.0071, 0.0040 ... 0.010)$~eV, while the ones for $|m_{ee}|$ evolve as $(0.0027 ... 0.0067, 0.0022 ... 0.0073, 0.0017 ... 0.0095)$~eV. Again, for NO, a small region opens up for large masses which are, however, disfavoured.

Coming to SR~10, the change in NO looks more dramatic than it actually is, given that in all cases the full range for the effective mass is allowed. Also the smallest neutrino mass $m$ is always below detectability. However, for IO -- which looks rather innocent at first sight -- a small region opens up that can actually be detectable! While IO is completely forbidden for $ \hat{s}=10^{-4}$ and only opens up in the disfavoured region for $\hat{s}=0.1$, for the largest value of $ \hat{s}=0.3$ the IO points start to penetrate the allowed region, predicting smallest values of $(m, |m_{ee}|) = (0.031, 0.054)$~eV.

For SR~11 both the NO and IO allowed band broaden to some extent for increasing $ \hat{s}$, see Fig.~\ref{fig:SR11}, and in particular they broaden more than if only the RGE corrections were taken into account. For IO, the smallest value of $m$ [of $|m_{ee}|$] varies as $(0.024, 0.021, 0.016)$~eV [as $(0.044, 0.039, 0.031)$~eV] for $ \hat s < (10^{-4}, 0.1, 0.3)$. For NO, the smallest value of $m$ varies as $(0.031, 0.029, 0.023)$~eV, while that of $|m_{ee}|$ varies as $(0.022, 0.017, 0.013)$~eV. As a comparison, the minimum values for $(m, |m_{ee}|)$ have been determined to be $(0.024, 0.042)$~eV [$(0.032, 0.021)$~eV] for IO [NO], cf.\ Sec.~4.11 in Ref.~\cite{Gehrlein:2015ena}.

Finally, for SR~12, the broadening for NO appears to be much stronger than in the case where only RGE corrections are taken into account. For $ \hat s < (10^{-4}, 0.1, 0.3)$, the smallest value for  $m$ [of $|m_{ee}|$] varies as $(0.0026, 0.0021, 0.0012)$~eV [as $(0.0029, 0.0025, 0.0014)$~eV] in the case of NO. For IO, again a small region opens up for larger $ \hat s$, but only in the disfavoured region.

\section{\label{sec:conc}Summary and conclusions}
 
 In this paper, we have used a perturbative approach to investigate how the predictions of neutrino mass sum rules change if the sum rules are not exact relations, but rather approximate to a given degree. After establishing a formalism to treat corrections to an exact sum rule, we show that the perturbations can be linked to the geometrical image of a ``nonperfect'' triangle. This illustration makes it relatively easy to understand the generality of our approach, as long as the corrections are small enough to be covered by a perturbative computation.
 
We then discuss several scenarios in which corrections to an exact sum rule can potentially arise. The three most generic frameworks are higher-order terms resulting from flavour symmetry breaking, corrections to the light neutrino masses arising from the charged lepton sector, and modifications of sum rules due to renormalisation group running. We have for each approach presented several analytic approximations, which serve as analytic estimates to be compared to the numerical computations.

The latter are our main results. We have, for each of the known mass sum rules, investigated the effect of perturbations to the exact formulas. As to be expected, as long as the correction terms are very small, we basically recover the result obtained from an exact sum rule. However, as we turn on the perturbations, we can see that the allowed regions are modified. The first type of modification is a simple increase of the allowed area in the parameter space. This is to be expected, since suddenly more parameter combinations are allowed, but in fact these broadenings are rather mild for most of the cases (at least so long as we stay in the perturbative regime). On the other hand, in several cases, qualitatively new predictions arise: depending on the sum rule, the exact formulas may forbid one of the neutrino mass orderings, which can be restored in the perturbed case. This could strongly alter the predictions, however, it turns out that in most cases (except for SR~10) the newly allowed regions are not very big and are practically excluded by the cosmological bounds on the neutrino mass. Finally, an interesting result is that the predictions for approximate sum rules do in many cases \emph{not} cover the regions which had been derived by us in an earlier work on the corrections from renormalisation group running. While this result may seem to come as a surprise at first sight, it is in fact easy to understand, since in some cases the corrections induced by running due to some parametric enhancement are larger than a $30\%$ correction, so that they are not covered by our formalism.

Summing up, we have treated the topic of neutrino mass sum rules in unprecedented generality, to the point that we may have delivered the final step to what can at all be said about neutrino mass sum rules from a phenomenological point of view. Using this work as well as previous ones, any model predicting a new sum rule can be analysed to the point that it can clearly be matched to the experimental results. We thus pass the ball to the experimentalists, who will hopefully be able to deliver further new bounds which allow us to constrain whole groups of flavour models in a reliable manner. This can push our understanding of the leptonic flavour sector to a new level.

\section*{Acknowledgements}

AM acknowledges partial support by the Micron Technology Foundation, Inc. AM furthermore acknowledges partial support by the European Union through the FP7 Marie Curie Actions ITN INVISIBLES (PITN-GA-2011-289442) and by the Horizon 2020 research and innovation programme under the Marie Sklodowska-Curie grant agreements No.~690575 (InvisiblesPlus RISE) and No.~674896 (Elusives ITN). MS would like to thank Stefan Antusch for useful discussions on parametrisations and acknowledges support  by 
BMBF under contract no.\ 05H12VKF.

\appendix
\section{\label{sec:parametrisations} Notes on parametrisations and phases} 

In this appendix we want to give a comprehensive derivation of the relations between the mixing parameters in the PMNS matrix and the neutrino and charged lepton mixing parameters.

First of all we parametrise the relevant matrices as  unitary $3\times 3$ matrices by 3~angles and 6~phases. One possible  parametrisation is~\cite{King:2002nf}:
\begin{align}
U=P_1 U_{23} U_{13}U_{12},
\label{eq:unphys}
\end{align}
where the $U_{ij}$ are 
\begin{equation}
U_{23}=
\begin{pmatrix}
1&0&0\\
0&c_{23}&s_{23}\text{e}^{-\text{i}\delta_{23}}\\
0&-s_{23}\text{e}^{\text{i}\delta_{23}}&c_{23}\\
\end{pmatrix}
\label{eq:U23},
\end{equation}
and analogous expressions for $U_{12}$ and $U_{13}$. We use the usual abbreviations $c_{ij}\equiv \cos\theta_{ij}$ and $s_{ij}\equiv \sin\theta_{ij}$. The matrix $P_1$ is a diagonal matrix which only contains phases:
\begin{align}
P_1&=
\begin{pmatrix}
\text{e}^{\text{i}\omega_1}&0&0\\
0&\text{e}^{\text{i}\omega_2}&0\\
0&0&\text{e}^{\text{i}\omega_3}\\
\end{pmatrix}.
\end{align}
The phase matrix $P_1$ can be removed by an additional charged lepton phase rotation to make the charged lepton masses real~\cite{King:2002nf}.
Hence we are left with only three phases $\delta_{ij}$ and
\begin{align}
U= U_{23} U_{13}U_{12}~. 
\label{eq:Udelta}
\end{align}

In the standard parametrisation of the PMNS matrix which contains the Dirac phases and the two Majorana phases we will also introduce an unphysical phase matrix $P_2$, which can be removed by a charged lepton phase rotation as
\begin{align}
P_2=
\begin{pmatrix}
\text{e}^{\text{i}\eta_1}&0&0\\
0&\text{e}^{\text{i}\eta_2}&0\\
0&0&\text{e}^{\text{i}\eta_3}\\
\end{pmatrix}~.
\end{align}
in the combination
\begin{align}
P_2 R_{23}U_{13}R_{12}P_0~.
\end{align}
 Where $\delta_{13}$ in $U_{13}$ is replaced with the Dirac phase $\delta$ and $P_0$ contains the Majorana phases:
\begin{align}
P_0=
\begin{pmatrix}
\text{e}^{-\text{i}\phi_1/2}&0&0\\
0&\text{e}^{-\text{i}\phi_2/2}&0\\
0&0&1\\
\end{pmatrix}~.
\end{align}
$R_{ij}$ are the Euler matrices which are of the form of Eq.~\eqref{eq:U23} but without any phases. The relation between the $\delta_{ij}$ in Eq.~\eqref{eq:unphys} and the phases in the matrix $P_0$ is~\cite{King:2002nf}:
\begin{align}
\phi_1&=-2(\delta_{12}+\delta_{23})~,
\label{eq:maj1}\\
\phi_2&=-2\delta_{23}~,\label{eq:maj2}\\
\delta&=\delta_{13}-\delta_{23}-\delta_{12}~.
\label{eq:delta}
\end{align}
Replacing the $\delta_{ij}$ in Eq.~\eqref{eq:unphys} with Eqs.~(\ref{eq:maj1}, \ref{eq:maj2}, \ref{eq:delta}) leads to 
\begin{align}
P_1 U_{23} U_{13}U_{12} \underbrace{=}_{\text{Eqs.~(\ref{eq:maj1},\ref{eq:maj2},\ref{eq:delta})}}P_2 R_{23}U_{13}R_{12}P_0~.
\end{align}
By comparing both sides of the equation we see that the phases in $P_1$ are related to the phases in $P_2$ as
\begin{align}
\omega_1&=\eta_1-\frac{\phi_1}{2}~,\label{eq:omega1}\\
\omega_2&=\eta_2-\frac{\phi_2}{2}~,\label{eq:omega2}\\
\omega_3&=\eta_3~.\label{eq:omega3}
\end{align}
In the following we will use the parametrisation in Eq.~\eqref{eq:unphys} for the mixing matrix of the neutrinos and the charged leptons. For the PMNS matrix we will replace the $\omega_i$ in  Eq.~\eqref{eq:unphys} by Eqs.~(\ref{eq:omega1}, \ref{eq:omega2}, \ref{eq:omega3}).

Our aim is to obtain expressions for the physical phases in the PMNS matrix (the Majorana phases and the Dirac phase) in terms of the neutrino and charged lepton mixing parameters. We therefore use the relation $U_{\text{PMNS}}=U_e^\dagger U_{\nu}$. From the elements in the first row the PMNS matrix we obtain:
\begin{align}
 c_{13}^{\text{PMNS}} c_{12}^{\text{PMNS}} \text{e}^{\text{i}(\eta_1-\phi_1/2)}&=c_{12}^e c_{12}^\nu c_{13}^e c_{13}^\nu 
 \text{e}^{-\text{i}(\omega_1^e-\omega_1^\nu)}+ \text{e}^{-\text{i}(\delta_{12}^e+\delta_{13}^e+\delta_{23}^e+\omega_3^e-\omega_3^\nu)} \nonumber \\
 & (\text{e}^{\text{i}(\delta_{12}^e+\delta_{23}^e)}\theta_{13}^e c_{12}^e-\text{e}^{\text{i}\delta_{13}^e} \theta_{23}^e s_{12}^e)(\text{e}^{\text{i}\delta_{13}^\nu}c_{12}^\nu c_{23}^\nu s_{13}^\nu-\text{e}^{\text{i}(\delta_{12}^e+\delta_{23}^\nu)} s_{12}^\nu s_{23}^\nu) \nonumber\\
 & +\text{e}^{-\text{i}(\delta_{12}^e+\delta_{23}^\nu+\omega_2^e-\omega_2^\nu)} s_{12}^e(\text{e}^{\text{i}(\delta_{12}^\nu+\delta_{23}^\nu)} c_{23}^\nu s_{12}^\nu+\text{e}^{\text{i}\delta_{13}^\nu}c_{12}^\nu s_{13}^\nu s_{23}^\nu, \label{eq:mpt_a1}\\
  s_{12}^{\text{PMNS}} c_{13}^{\text{PMNS}}\text{e}^{\text{i} (\eta_1-\phi_2/2)} &= \text{e}^{-\text{i}(\delta_{12}^e+\delta_{12}^\nu+\delta_{13}^e+\delta_{23}^e+\delta_{23}^\nu+\omega_1^e+\omega_2^e+\omega_3^e)}(\text{e}^{\text{i}(\delta_{12}^\nu+\delta_{23}^\nu+\omega_1^e)}c_{12}^\nu \nonumber\\  
 & (-\text{e}^{\text{i}(\delta_{13}^e+\delta_{23}^e+\omega_2^\nu+\omega_3^e)}c_{23}^\nu s_{12}^e +\text{e}^{\text{i}(\delta_{12}^e+\delta_{23}^e+\delta_{23}^\nu+\omega_2^e+\omega_3^\nu)}\theta_{13}^e c_{12}^e s_{23}^\nu \nonumber\\ 
& -\text{e}^{\text{i}(\delta_{13}^e+\delta_{23}^\nu+\omega_2^e+\omega_3^\nu)}\theta_{23}^e s_{12}^e s_{23}^\nu+s_{12}^\nu (\text{e}^{\text{i}(\delta_{12}^e+\delta_{13}^e+\delta_{23}^e+\delta_{23}^\nu+\omega_1^\nu+\omega_2^e \omega_3^e)}c_{12}^e c_{13}^\nu\nonumber\\
&+\text{e}^{\text{i}(\delta_{13}^e+\delta_{13}^\nu+\delta_{23}^\nu+\omega_1^e+\omega_2^e+\omega_3^\nu)} \theta_{13}^e c_{12}^e c_{23}^\nu s_{13}^\nu\nonumber\\
 & -\text{e}^{\text{i}(\delta_{13}^e+\delta_{13}^\nu+\delta_{23}^\nu+\omega_1^e+\omega_2^e+\omega_3^\nu)}\theta_{23}^e c_{23}^\nu s_{12}^e s_{13}^\nu\nonumber\\  
 & +\text{e}^{\text{i}(\delta_{13}^e+\delta_{13}^\nu+\delta_{23}^e+\omega_1^e+\omega_2^\nu+\omega_3^e)}s_{12}^e s_{13}^\nu s_{23}^\nu)), \label{eq:mpt_a2}\\
     s_{13}^{\text{PMNS}}\text{e}^{\text{i}( \eta_1-\delta)} &= \text{e}^{-\text{i}(\delta_{12}^e+\delta_ {13}^e+\delta_{13}^\nu+\delta_{23}^e+\delta_{23}^\nu+\omega_1^e+\omega_2^e+\omega_3^e)} \nonumber\\
    & (-\text{e}^{\text{i}(\delta_{12}^e+\delta_{13}^\nu+\delta_{23}^e   +\delta_{23}^\nu+\omega_1^e+\omega_2^e+\omega_3^\nu)}\theta_{13}^e c_{12}^e c_{13}^\nu c_{23}^\nu \nonumber\\
    & +\text{e}^{\text{i}(\delta_{13}^e+\delta_{13}^\nu+\delta_{23}^\nu+\omega_1^e+\omega_2^e+\omega_3^\nu)}\theta_{23}^e c_{13}^\nu c_{23}^\nu s_{12}^e \nonumber \\
   &  +\text{e}^{\text{i}(\delta_{12}^e+\delta_{ 13}^e+\delta_{23}^e+\delta_{23}^\nu+\omega_1^\nu+\omega_2^e+\omega_3^e)}c_{12}^e s_{13}^\nu \nonumber\\
    & -\text{e}^{\text{i}(\delta_{13}^e +\delta_{13}^\nu+\delta_{23}^e+\omega_1^e+\omega_2^\nu+\omega_3^e)} c_{13}^\nu s_{12}^e s_{23}^\nu)~.  \label{eq:mpt_a3}
\end{align}
These expressions are exact to leading order in $\theta_{13}^e$ and $\theta_{23}^e$. If we exploit the structure of the PMNS matrix we furthermore obtain:
\begin{multline}
c_{12}^{\text{PMNS}} \left(c_{13}^{\text{PMNS}}\right)^2 c_{23}^{\text{PMNS}} s_{13}^{\text{PMNS}} \left(s_{12}^{\text{PMNS}}s_{23}^{\text{PMNS}}\text{e}^{-\text{i}\delta}-c_{12}^{\text{PMNS}}c_{23}^{\text{PMNS}}s_{13}^{\text{PMNS}}\right)=\\
\left(U_{11}^{\text{PMNS}}\right)^*U_{13}^{\text{PMNS}}U_{31}^{\text{PMNS}} \left(U_{33}^{\text{PMNS}}\right)^*~.
\label{eq:delta1}
\end{multline}
Together with Eqs.~(\ref{eq:mpt_a1}, \ref{eq:mpt_a2}, \ref{eq:mpt_a3}, \ref{eq:delta1}) we obtain equations for $\delta$, $\eta_1$, and for the Majorana phases. The mixing angles in Eq.~\eqref{eq:delta1} can be expressed in terms of the leptonic mixing parameters using Eqs.~(\ref{eq:mpt_a1}, \ref{eq:mpt_a2}, \ref{eq:mpt_a3}). 

As a concrete example to employ the formalism to derive the expressions for the Majorana phases, we consider the $A_5\times SU(5)$ model proposed in~\cite{Gehrlein:2014wda,Gehrlein:2015dxa,Gehrlein:2015dza}. We have $~\theta_{13}^e\approx 0,~\theta_{23}^e\approx 0,~\theta_{12}^e\neq 0$ and in the neutrino sector we have Golden Ratio mixing with $\theta_{13}^\nu=0,~\theta_{23}^\nu=45^\circ$ and $\theta_{12}^\nu=\text{arctan}\left(\frac{2}{1+\sqrt{5}}\right)$. Since $\theta_{13}^e $ and $\theta_{23}^e$ are negligibly small, we will set their values to zero in the following. The phase $\delta_{12}^e $ in the charged lepton sector will be treated as a free parameter. For simplicity, we take the neutrino mass matrix to be real. The phases in the neutrino sector which lead to positive eigenvalues are then:
\begin{align}
\omega_1^\nu=\pi/2, ~
\omega_2^\nu=\pi,~
\omega_3^\nu=\pi/2,~
\delta_{12}^{\nu}=3\pi/2 ,~
\delta_{23}^\nu=3\pi/2.
\end{align}
With these parameters we obtain for Eqs.~(\ref{eq:mpt_a1}, \ref{eq:mpt_a2}, \ref{eq:mpt_a3}, \ref{eq:delta1},) to first order in $\theta_{12}^e$:
\begin{align}
c_{12}^{\text{PMNS}} \text{e}^{\text{i}(\eta_1-\phi_1/2)}&\approx \frac{\text{e}^{\text{i}\pi/2} (\sqrt{3+\sqrt{5}} + \theta_{12}^e  \text{e}^{-\text{i}\delta_{12}^e})}{\sqrt{5+\sqrt{5}}}~,\\
 s_{12}^{\text{PMNS}} \text{e}^{\text{i}(\eta_1-\phi_2/2)}&\approx -\frac{2}{\sqrt{10+2\sqrt{5}}}+ \frac{\theta_{12}^{e}\text{e}^{-\text{i}\delta_{12}^e }}{\sqrt{5-\sqrt{5}}}~,\\
 \theta_{13}^{\text{PMNS}}\text{e}^{\text{i}( \eta_1-\delta)} &\approx\frac{\text{e}^{\text{i}\pi/2}\theta_{12}^e \text{e}^{-\text{i}\delta_{12}^{e}}}{\sqrt{2}}~,\label{eq:dunsinn}\\
s_{12}^{\text{PMNS}}s_{23}^{\text{PMNS}}\text{e}^{-\text{i}\delta}&\approx \frac{-\sqrt{2}(15+7\sqrt{5})\text{e}^{-\text{i}\delta_{12}^e}+(5+2\sqrt{5})\theta_{12}^e \text{e}^{-2\text{i}\delta_{12}^e}+(20+9\sqrt{5})\theta_{12}^e}{4(5+2\sqrt{5})^{3/2}}~.
\end{align}
One might wonder if it is possible to take the limit $\theta_{12}^e\rightarrow 0$ in Eq.~\eqref{eq:dunsinn}  and to obtain a sensible result. This is not possible since a diagonal charged lepton mass matrix corresponds to $\theta_{13}^{\text{PMNS}}=0$ where $\delta$ is unphysical and in the derivation of Eq.~\eqref{eq:dunsinn} from Eq.~\eqref{eq:delta} we have to divide by $\theta_{13}^{\text{PMNS}}$. For a nondiagonal charged lepton mass matrix, we obtain for $\delta $ 
\begin{align}
\delta\approx \pi+\delta_{12}^e+\frac{\theta_{12}^e \sin(\delta_{12}^e)}{\sqrt{2}}~.
\end{align}
We can easily obtain that $\eta_1$ is 
\begin{align}
\eta_1\approx-\frac{\pi}{2}+\frac{1}{\sqrt{2}}\theta_{12}^e \sin(\delta_{12}^e)~.
\end{align}
Finally we get
\begin{align}
\phi_1&\approx\sqrt{3+\sqrt{5}}\theta_{12}^e \sin(\delta_{12}^e)~,\label{eq:a1we}\\
\phi_2&\approx\pi-\frac{\sqrt{5}-1}{\sqrt{2}}\theta_{12}^e\sin(\delta_{12}^e) ~. \label{eq:a2we}
\end{align} 

One might wonder if  it is necessary to include the unphysical phases in order to derive the expressions for Majorana phases. Indeed, the correct consideration of the unphysical phases is essential since taking the $\omega_i$ in Eq.~\eqref{eq:unphys} to zero which corresponds to $\eta_i=\phi_i/2$ for $i=1,~2$ changes the matrix element from which we extract the Majorana phases. For example, to obtain information about $\phi_2$, we consider the 1-2 element if we include the unphysical phases, but if we neglect the unphysical phases, we would have to consider the 2-3 element of the PMNS matrix. The dependence of these elements on the charged lepton mixing parameters differs in general. Even in the case of only a 1-2 mixing in the charged lepton we would miss the correct $\theta_{12}^e$ dependence of $\phi_2$ without the unphysical phases.

The formulas derived in~\cite{King:2002nf,Antusch:2005kw,Antusch:2008yc} assume that the unphysical phases have been already correctly taken into account. The reader has to be aware that these formulas therefore strictly apply to this case only. 

A phase matrix on the left side of the PMNS matrix can always be absorbed by an additional charged lepton phase rotation. For this reason, the phases $\eta_i$ do \emph{not} appear in physical observables.


\begin{thebibliography}{99}

\bibitem{Neutrino-Kin}
  C.~Kraus {\it et al.},
  Eur.\ Phys.\ J.\ C {\bf 40} (2005) 447
  [hep-ex/0412056];
  V.~M.~Lobashev {\it et al.},
  Phys.\ Lett.\ B {\bf 460} (1999) 227.
  
\bibitem{0nbb-Exp}
  \emph{See, for example}:
  C.~Alduino {\it et al.} [CUORE Collaboration],
  Phys.\ Rev.\ C {\bf 93} (2016) 045503
  [arXiv:1601.01334 [nucl-ex]];
  J.~B.~Albert {\it et al.} [EXO-200 Collaboration],
  Nature {\bf 510} (2014) 229
  [arXiv:1402.6956 [nucl-ex]];
  M.~Agostini {\it et al.} [GERDA Collaboration],
  Phys.\ Rev.\ Lett.\  {\bf 111} (2013) 122503
  [arXiv:1307.4720 [nucl-ex]];
  A.~Gando {\it et al.} [KamLAND-Zen Collaboration],
  arXiv:1605.02889 [hep-ex];
  R.~Arnold {\it et al.} [NEMO-3 Collaboration],
  Phys.\ Rev.\ D {\bf 92} (2015) 072011
  [arXiv:1506.05825 [hep-ex]].
  
\bibitem{Pagliaroli:2010ik}
  G.~Pagliaroli, F.~Rossi-Torres and F.~Vissani,
  Astropart.\ Phys.\  {\bf 33} (2010) 287
  [arXiv:1002.3349 [hep-ph]].

\bibitem{Cosmo-Nu}  
  P.~A.~R.~Ade {\it et al.} [Planck Collaboration],
  arXiv:1502.01589 [astro-ph.CO];
  E.~Di Valentino, E.~Giusarma, O.~Mena, A.~Melchiorri and J.~Silk,
  Phys.\ Rev.\ D {\bf 93} (2016) 083527
  [arXiv:1511.00975 [astro-ph.CO]];
  A.~J.~Cuesta, V.~Niro and L.~Verde,
  Phys.\ Dark Univ.\  {\bf 13} (2016) 77
  [arXiv:1511.05983 [astro-ph.CO]].
  
\bibitem{Gonzalez-Garcia:2014bfa}
  M.~C.~Gonzalez-Garcia, M.~Maltoni and T.~Schwetz,
  JHEP {\bf 1411} (2014) 052
  [arXiv:1409.5439 [hep-ph]].
  
\bibitem{flavour-reviews}
  G.~Altarelli and F.~Feruglio,
  Rev.\ Mod.\ Phys.\  {\bf 82} (2010) 2701
  [arXiv:1002.0211 [hep-ph]];
  W.~Grimus and P.~O.~Ludl,
  J.\ Phys.\ A {\bf 45} (2012) 233001
  [arXiv:1110.6376 [hep-ph]];
  S.~Morisi and J.~W.~F.~Valle,
  Fortsch.\ Phys.\  {\bf 61} (2013) 466
  [arXiv:1206.6678 [hep-ph]];
  S.~F.~King and C.~Luhn,
  Rept.\ Prog.\ Phys.\  {\bf 76} (2013) 056201
  [arXiv:1301.1340 [hep-ph]];
  S.~F.~King, A.~Merle, S.~Morisi, Y.~Shimizu and M.~Tanimoto,
  New J.\ Phys.\  {\bf 16} (2014) 045018
  [arXiv:1402.4271 [hep-ph]].
  
\bibitem{Haba:2000be}
  N.~Haba and H.~Murayama,
  Phys.\ Rev.\ D {\bf 63} (2001) 053010
  [hep-ph/0009174].
  
\bibitem{Adulpravitchai:2009re}
  A.~Adulpravitchai, M.~Lindner, A.~Merle and R.~N.~Mohapatra,
  Phys.\ Lett.\ B {\bf 680} (2009) 476
  [arXiv:0908.0470 [hep-ph]].
  
\bibitem{Mixing-SR}
  S.~Antusch, P.~Huber, S.~F.~King and T.~Schwetz,
  JHEP {\bf 0704} (2007) 060
  [hep-ph/0702286 [HEP-PH]];
  P.~Ballett, S.~F.~King, C.~Luhn, S.~Pascoli and M.~A.~Schmidt,
  Phys.\ Rev.\ D {\bf 89} (2014) 1,  016016
  [arXiv:1308.4314 [hep-ph]];
  P.~Ballett, S.~F.~King, C.~Luhn, S.~Pascoli and M.~A.~Schmidt,
  JHEP {\bf 1412} (2014) 122
  [arXiv:1410.7573 [hep-ph]];
   S.~Antusch and S.~F.~King,
   Phys.\ Lett.\ B {\bf 631} (2005) 42
   [hep-ph/0508044];
     S.~T.~Petcov,
  Nucl.\ Phys.\ B {\bf 892} (2015) 400
  [arXiv:1405.6006 [hep-ph]];   
  I.~Girardi, S.~T.~Petcov and A.~V.~Titov,
  Nucl.\ Phys.\ B {\bf 894} (2015) 733
  [arXiv:1410.8056 [hep-ph]];
  I.~Girardi, S.~T.~Petcov and A.~V.~Titov,
  Int.\ J.\ Mod.\ Phys.\ A {\bf 30} (2015) 13,  1530035
  [arXiv:1504.02402 [hep-ph]];
  I.~Girardi, S.~T.~Petcov and A.~V.~Titov,
  Eur.\ Phys.\ J.\ C {\bf 75} (2015) 7,  345
  [arXiv:1504.00658 [hep-ph]];
  I.~Girardi, S.~T.~Petcov, A.~J.~Stuart and A.~V.~Titov,
  Nucl.\ Phys.\ B {\bf 902} (2016) 1
  [arXiv:1509.02502 [hep-ph]];
    A.~D.~Hanlon, S.~F.~Ge and W.~W.~Repko,
   Phys.\ Lett.\ B {\bf 729} (2014) 185
   [arXiv:1308.6522 [hep-ph]];
   S.~F.~Ge, D.~A.~Dicus and W.~W.~Repko,
   Phys.\ Rev.\ Lett.\  {\bf 108} (2012) 041801
   [arXiv:1108.0964 [hep-ph]];
   S.~F.~Ge, D.~A.~Dicus and W.~W.~Repko,
   Phys.\ Lett.\ B {\bf 702} (2011) 220
   [arXiv:1104.0602 [hep-ph]];
   J.~Zhang and S.~Zhou,
   arXiv:1604.03039 [hep-ph].


  
\bibitem{Altarelli:2008bg}
  G.~Altarelli, F.~Feruglio and C.~Hagedorn,
  JHEP {\bf 0803} (2008) 052
  [arXiv:0802.0090 [hep-ph]].
  
\bibitem{Hirsch:2008rp}
  M.~Hirsch, S.~Morisi and J.~W.~F.~Valle,
  Phys.\ Rev.\ D {\bf 78} (2008) 093007
  [arXiv:0804.1521 [hep-ph]].
 
     
\bibitem{Bazzocchi:2009da}
  F.~Bazzocchi, L.~Merlo and S.~Morisi,
  Phys.\ Rev.\ D {\bf 80} (2009) 053003
  [arXiv:0902.2849 [hep-ph]].  

\bibitem{Altarelli:2009kr}
  G.~Altarelli and D.~Meloni,
  J.\ Phys.\ G {\bf 36} (2009) 085005
  [arXiv:0905.0620 [hep-ph]].
  
\bibitem{Chen:2009um}
  M.~C.~Chen and S.~F.~King,
  JHEP {\bf 0906} (2009) 072
  [arXiv:0903.0125 [hep-ph]].

\bibitem{Barry:2010yk}
  J.~Barry and W.~Rodejohann,
  Nucl.\ Phys.\ B {\bf 842} (2011) 33
  [arXiv:1007.5217 [hep-ph]].

  
\bibitem{SR11}
  L.~Dorame, D.~Meloni, S.~Morisi, E.~Peinado and J.~W.~F.~Valle,
  Nucl.\ Phys.\ B {\bf 861} (2012) 259
  [arXiv:1111.5614 [hep-ph]].
  
\bibitem{King:2013psa}
  S.~F.~King, A.~Merle and A.~J.~Stuart,
  JHEP {\bf 1312} (2013) 005
  [arXiv:1307.2901 [hep-ph]].
   
  
\bibitem{SR-GERDA}
  M.~Agostini, A.~Merle and K.~Zuber,
  Eur.\ Phys.\ J.\ C {\bf 76} (2016) 176
  [arXiv:1506.06133 [hep-ex]].
  
\bibitem{Cooper:2012bd}
  I.~K.~Cooper, S.~F.~King and A.~J.~Stuart,
  Nucl.\ Phys.\ B {\bf 875} (2013) 650
  [arXiv:1212.1066 [hep-ph]].
  
\bibitem{Gehrlein:2015ena}
  J.~Gehrlein, A.~Merle and M.~Spinrath,
  JHEP {\bf 1509} (2015) 066
  [arXiv:1506.06139 [hep-ph]].
  
\bibitem{pdg}
  K.~A.~Olive {\it et al.}  [Particle Data Group Collaboration],
  Chin.\ Phys.\ C {\bf 38} (2014) 090001.
  


  \bibitem{Gehrlein:2014wda} 
  J.~Gehrlein, J.~P.~Oppermann, D.~Sch\"afer and M.~Spinrath,
  Nucl.\ Phys.\ B {\bf 890} (2014) 539
  [arXiv:1410.2057 [hep-ph]];
  

\bibitem{seesaw}
  P.~Minkowski,
  Phys.\ Lett.\ B {\bf 67} (1977) 421;
  P.~Ramond,
  hep-ph/9809459;
  T.~Yanagida,
  Conf.\ Proc.\ C {\bf 7902131} (1979) 95;
  M.~Gell-Mann, P.~Ramond and R.~Slansky,
  Conf.\ Proc.\ C {\bf 790927} (1979) 315
  [arXiv:1306.4669 [hep-th]];
  S.~L.~Glashow,
  NATO Sci.\ Ser.\ B {\bf 59} (1980) 687;
  R.~N.~Mohapatra and G.~Senjanovic,
  Phys.\ Rev.\ Lett.\  {\bf 44} (1980) 912;
  J.~Schechter and J.~W.~F.~Valle,
  Phys.\ Rev.\ D {\bf 22} (1980) 2227.


 
  
\bibitem{Ding:2010pc}
  G.~J.~Ding,
  Nucl.\ Phys.\ B {\bf 846} (2011) 394
  [arXiv:1006.4800 [hep-ph]].
  
\bibitem{Ma:2005sha}
  E.~Ma,
  Phys.\ Rev.\ D {\bf 72} (2005) 037301
  [hep-ph/0505209].
  
\bibitem{Ma:2006wm}
  E.~Ma,
  Mod.\ Phys.\ Lett.\ A {\bf 21} (2006) 2931
  [hep-ph/0607190].
  
\bibitem{Honda:2008rs}
  M.~Honda and M.~Tanimoto,
  Prog.\ Theor.\ Phys.\  {\bf 119} (2008) 583
  [arXiv:0801.0181 [hep-ph]].
  
\bibitem{Brahmachari:2008fn}
  B.~Brahmachari, S.~Choubey and M.~Mitra,
  Phys.\ Rev.\ D {\bf 77} (2008) 073008
  [Phys.\ Rev.\ D {\bf 77} (2008) 119901]
  [arXiv:0801.3554 [hep-ph]].
  
\bibitem{Kang:2015xfa}
  S.~K.~Kang and M.~Tanimoto,
  Phys.\ Rev.\ D {\bf 91} (2015) 073010
  [arXiv:1501.07428 [hep-ph]].
  
\bibitem{SR1}
  F.~Bazzocchi, L.~Merlo and S.~Morisi,
  Nucl.\ Phys.\ B {\bf 816} (2009) 204
  [arXiv:0901.2086 [hep-ph]];
  L.~L.~Everett and A.~J.~Stuart,
  Phys.\ Rev.\ D {\bf 79} (2009) 085005
  [arXiv:0812.1057 [hep-ph]];
  M.~S.~Boucenna, S.~Morisi, E.~Peinado, Y.~Shimizu and J.~W.~F.~Valle,
  Phys.\ Rev.\ D {\bf 86} (2012) 073008
  [arXiv:1204.4733 [hep-ph]].
  
\bibitem{SR2}
  R.~N.~Mohapatra and C.~C.~Nishi,
  Phys.\ Rev.\ D {\bf 86} (2012) 073007
  [arXiv:1208.2875 [hep-ph]]. 

\bibitem{SR3}
  G.~Altarelli and F.~Feruglio,
  Nucl.\ Phys.\ B {\bf 720} (2005) 64
  [hep-ph/0504165];
  G.~Altarelli, F.~Feruglio and Y.~Lin,
  Nucl.\ Phys.\ B {\bf 775} (2007) 31
  [hep-ph/0610165];
  E.~Ma,
  Mod.\ Phys.\ Lett.\ A {\bf 22} (2007) 101
  [hep-ph/0610342];
  F.~Bazzocchi, S.~Kaneko and S.~Morisi,
  JHEP {\bf 0803} (2008) 063
  [arXiv:0707.3032 [hep-ph]];
  F.~Bazzocchi, S.~Morisi and M.~Picariello,
  Phys.\ Lett.\ B {\bf 659} (2008) 628
  [arXiv:0710.2928 [hep-ph]];
  Y.~Lin,
  Nucl.\ Phys.\ B {\bf 813} (2009) 91
  [arXiv:0804.2867 [hep-ph]];
  E.~Ma,
  Mod.\ Phys.\ Lett.\ A {\bf 25} (2010) 2215
  [arXiv:0908.3165 [hep-ph]];
  P.~Ciafaloni, M.~Picariello, A.~Urbano and E.~Torrente-Lujan,
  Phys.\ Rev.\ D {\bf 81} (2010) 016004
  [arXiv:0909.2553 [hep-ph]];
  F.~Bazzocchi and S.~Morisi,
  Phys.\ Rev.\ D {\bf 80} (2009) 096005
  [arXiv:0811.0345 [hep-ph]];
  F.~Feruglio, C.~Hagedorn and R.~Ziegler,
  Eur.\ Phys.\ J.\ C {\bf 74} (2014) 2753
  [arXiv:1303.7178 [hep-ph]];
  M.~C.~Chen and K.~T.~Mahanthappa,
  Phys.\ Lett.\ B {\bf 652} (2007) 34
  [arXiv:0705.0714 [hep-ph]];
  G.~J.~Ding,
  Phys.\ Rev.\ D {\bf 78} (2008) 036011
  [arXiv:0803.2278 [hep-ph]];
  M.~C.~Chen and K.~T.~Mahanthappa,
  Phys.\ Lett.\ B {\bf 681} (2009) 444
  [arXiv:0904.1721 [hep-ph]];
  F.~Feruglio, C.~Hagedorn, Y.~Lin and L.~Merlo,
  Nucl.\ Phys.\ B {\bf 775} (2007) 120
   [Nucl.\ Phys.\  {\bf 836} (2010) 127]
  [hep-ph/0702194];
  L.~Merlo, S.~Rigolin and B.~Zaldivar,
  JHEP {\bf 1111} (2011) 047
  [arXiv:1108.1795 [hep-ph]];
  C.~Luhn, K.~M.~Parattu and A.~Wingerter,
  JHEP {\bf 1212} (2012) 096
  [arXiv:1210.1197 [hep-ph]];
  T.~Fukuyama, H.~Sugiyama and K.~Tsumura,
  Phys.\ Rev.\ D {\bf 82} (2010) 036004
  [arXiv:1005.5338 [hep-ph]].
  
\bibitem{Altarelli:2005yx}
  G.~Altarelli and F.~Feruglio,
  Nucl.\ Phys.\ B {\bf 741} (2006) 215
  [hep-ph/0512103].
  
  
\bibitem{Chen:2009gy}
  M.~C.~Chen, K.~T.~Mahanthappa and F.~Yu,
  Phys.\ Rev.\ D {\bf 81} (2010) 036004
  [arXiv:0907.3963 [hep-ph]].

\bibitem{SR4}
  G.~J.~Ding and Y.~L.~Zhou,
  Nucl.\ Phys.\ B {\bf 876} (2013) 418
  [arXiv:1304.2645 [hep-ph]];
  M.~Lindner, A.~Merle and V.~Niro,
  JCAP {\bf 1101} (2011) 034
  [JCAP {\bf 1407} (2014) E01]
  [arXiv:1011.4950 [hep-ph]].

\bibitem{SR5}
  K.~Hashimoto and H.~Okada,
  arXiv:1110.3640 [hep-ph].

\bibitem{SR6}
  G.~J.~Ding, L.~L.~Everett and A.~J.~Stuart,
  Nucl.\ Phys.\ B {\bf 857} (2012) 219
  [arXiv:1110.1688 [hep-ph]].

\bibitem{SR7}
  S.~Morisi, M.~Picariello and E.~Torrente-Lujan,
  Phys.\ Rev.\ D {\bf 75} (2007) 075015
  [hep-ph/0702034];
  B.~Adhikary and A.~Ghosal,
  Phys.\ Rev.\ D {\bf 78} (2008) 073007
  [arXiv:0803.3582 [hep-ph]];
  Y.~Lin,
  Nucl.\ Phys.\ B {\bf 824} (2010) 95
  [arXiv:0905.3534 [hep-ph]];
  C.~Csaki, C.~Delaunay, C.~Grojean and Y.~Grossman,
  JHEP {\bf 0810} (2008) 055
  [arXiv:0806.0356 [hep-ph]];
  C.~Hagedorn, E.~Molinaro and S.~T.~Petcov,
  JHEP {\bf 0909} (2009) 115
  [arXiv:0908.0240 [hep-ph]];
  T.~J.~Burrows and S.~F.~King,
  Nucl.\ Phys.\ B {\bf 835} (2010) 174
  [arXiv:0909.1433 [hep-ph]];
  G.~J.~Ding and J.~F.~Liu,
  JHEP {\bf 1005} (2010) 029
  [arXiv:0911.4799 [hep-ph]];
  M.~Mitra,
  JHEP {\bf 1011} (2010) 026
  [arXiv:0912.5291 [hep-ph]];
  F.~del Aguila, A.~Carmona and J.~Santiago,
  JHEP {\bf 1008} (2010) 127
  [arXiv:1001.5151 [hep-ph]];
  T.~J.~Burrows and S.~F.~King,
  Nucl.\ Phys.\ B {\bf 842} (2011) 107
  [arXiv:1007.2310 [hep-ph]];
  Y.~H.~Ahn and P.~Gondolo,
  Phys.\ Rev.\ D {\bf 91} (2015) 1,  013007
  [arXiv:1402.0150 [hep-ph]];
  B.~Karmakar and A.~Sil,
  Phys.\ Rev.\ D {\bf 91} (2015) 013004
  [arXiv:1407.5826 [hep-ph]];
  Y.~H.~Ahn,
  Phys.\ Rev.\ D {\bf 91} (2015) 5,  056005
  [arXiv:1410.1634 [hep-ph]].

\bibitem{SR8}
  X.~G.~He, Y.~Y.~Keum and R.~R.~Volkas,
  JHEP {\bf 0604} (2006) 039
  [hep-ph/0601001];
  J.~Berger and Y.~Grossman,
  JHEP {\bf 1002} (2010) 071
  [arXiv:0910.4392 [hep-ph]];
  A.~Kadosh and E.~Pallante,
  JHEP {\bf 1008} (2010) 115
  [arXiv:1004.0321 [hep-ph]];
  L.~Lavoura, S.~Morisi and J.~W.~F.~Valle,
  JHEP {\bf 1302} (2013) 118
  [arXiv:1205.3442 [hep-ph]].

\bibitem{SR9}
  S.~F.~King, C.~Luhn and A.~J.~Stuart,
  Nucl.\ Phys.\ B {\bf 867} (2013) 203
  [arXiv:1207.5741 [hep-ph]].
  
\bibitem{SR10}
    A.~Adulpravitchai, M.~Lindner and A.~Merle,
  Phys.\ Rev.\ D {\bf 80} (2009) 055031
  [arXiv:0907.2147 [hep-ph]].

\bibitem{SR12}
  L.~Dorame, S.~Morisi, E.~Peinado, J.~W.~F.~Valle and A.~D.~Rojas,
  Phys.\ Rev.\ D {\bf 86} (2012) 056001
  [arXiv:1203.0155 [hep-ph]].
    
\bibitem{Antusch:2003kp}
  S.~Antusch, J.~Kersten, M.~Lindner and M.~Ratz,
 Nucl.\ Phys.\ B {\bf 674} (2003) 401
  [hep-ph/0305273]. 
      
\bibitem{mee-references}
 M.~Lindner, A.~Merle, W.~Rodejohann,
 Phys.\ Rev.\ D {\bf 73} (2006) 053005
 [hep-ph/0512143];
 A.~Merle, W.~Rodejohann,
 Phys.\ Rev.\ D {\bf 73} (2006) 073012
 [hep-ph/0603111].

\bibitem{Gehrlein:2015dxa}
  J.~Gehrlein, S.~T.~Petcov, M.~Spinrath and X.~Zhang,
  Nucl.\ Phys.\ B {\bf 896} (2015) 311
   [Nucl.\ Phys.\ B {\bf 899} (2015) 617]
  [arXiv:1502.00110 [hep-ph]].
  
  
\bibitem{Gehrlein:2015dza}
  J.~Gehrlein, S.~T.~Petcov, M.~Spinrath and X.~Zhang,
  Nucl.\ Phys.\ B {\bf 899} (2015) 617
  [arXiv:1508.07930 [hep-ph]].
  
\bibitem{Antusch:2005kw}
  S.~Antusch and S.~F.~King,
  Phys.\ Lett.\ B {\bf 631} (2005) 42
  [hep-ph/0508044].

\bibitem{King:2002nf}
  S.~F.~King,
  JHEP {\bf 0209} (2002) 011
  [hep-ph/0204360].
  
\bibitem{Antusch:2008yc}
  S.~Antusch, S.~F.~King and M.~Malinsky,
  Nucl.\ Phys.\ B {\bf 820} (2009) 32
  [arXiv:0810.3863 [hep-ph]].

 
\end{thebibliography}
\end{document}